\documentclass[useAMS,usenatbib]{mn2e}
\usepackage{txfonts}
\usepackage{graphicx}
\usepackage{verbatim}
\usepackage{rotating}
\usepackage[flushleft]{threeparttable}
\usepackage{color}

\newcommand\finetilde{{\raise.17ex\hbox{$\scriptstyle\sim$}}} 

\def\msun{\,{\rm M_\odot}}

\title[EPTA Limits]{European Pulsar Timing Array Limits On An Isotropic Stochastic Gravitational-Wave Background}
\author[L. Lentati et al.]{\parbox{\textwidth}{L. Lentati$^{1}$\thanks{E-mail:
ltl21@cam.ac.uk}, S.~R. Taylor$^{2, 3}$, C. M. F. Mingarelli$^{4,5, 6}$, A. Sesana$^{6, 7}$, S.~A.~Sanidas$^{8,9}$, A.~Vecchio$^{6}$, R. N. Caballero$^{5}$, K.~J.~Lee$^{10, 5}$, R. van Haasteren$^{3}$, S. Babak$^{7}$, C. G. Bassa$^{11,9}$, P. Brem$^{7}$, M. Burgay$^{12}$,  D. J. Champion$^{5}$, I. Cognard$^{13, 14}$, G. Desvignes$^{5}$, J. R. Gair$^{2}$, L. Guillemot$^{13,14}$, J. W. T. Hessels$^{11, 8}$, G. H. Janssen$^{11,9}$, R. Karuppusamy$^{5}$, M. Kramer$^{5, 9}$, A. Lassus$^{5,15}$, P. Lazarus$^{5}$,   K. Liu$^{5}$, S. Os{\l}owski$^{17, 5}$, D. Perrodin$^{12}$, A. Petiteau$^{15}$,  A. Possenti$^{12}$, M. B. Purver$^{9}$, P. A. Rosado$^{18,19}$, R. Smits$^{11}$, B. Stappers$^{9}$, G. Theureau$^{13,14,16}$, C. Tiburzi$^{20,12}$,   J. P. W. Verbiest$^{17,5}$}\vspace{0.4cm}\\ %
$^{1}$ Astrophysics Group, Cavendish Laboratory, JJ Thomson Avenue,  Cambridge, CB3 0HE, UK\\
$^{2}$ Institute of Astronomy, University of Cambridge, Madingley Road, Cambridge CB3 0HA \\
$^{3}$ Jet Propulsion Laboratory, California Institute of Technology, Pasadena, California 91109, USA\\
$^{4}$ TAPIR (Theoretical Astrophysics), California Institute of Technology, Pasadena, California 91125, USA \\
$^{5}$ Max-Planck-Institut f{\"u}r Radioastronomie, Auf dem H{\"u}gel 69, D-53121 Bonn, Germany \\
$^{6}$ School of Physics and Astronomy, University of Birmingham, Edgbaston, Birmingham B15 2TT, United Kingdom\\
$^{7}$ Max-Planck-Institut f{\"u}r Gravitationsphysik, Albert Einstein Institut, Am M{\"u}hlenberg 1, 14476 Golm, Germany \\
$^{8}$ Anton Pannekoek Institute for Astronomy, University of Amsterdam, Science Park 904, 1098 XH Amsterdam, The Netherlands \\
$^{9}$ Jodrell Bank Centre for Astrophysics, University of Manchester, Manchester, M13 9PL, United Kingdom\\
$^{10}$  Kavli institute for astronomy and astrophysics,Peking University, Beijing 100871,P.R.China \\
$^{11}$ ASTRON, the Netherlands Institute for Radio Astronomy, Postbus 2, 7990 AA, Dwingeloo, The Netherlands \\
$^{12}$ INAF - Osservatorio Astronomico di Cagliari, via della Scienza 5, I-09047 Selargius (CA), Italy \\
$^{13}$ Laboratoire de Physique et Chimie de l'Environnement et de l'Espace LPC2E CNRS-Universit{\'e} d'Orl{\'e}ans, F-45071 Orl{\'e}ans, France \\
$^{14}$ Station de radioastronomie de Nan{\c c}ay, Observatoire de Paris, CNRS/INSU F-18330 Nan{\c c}ay, France \\
$^{15}$ Universit\'e Paris-Diderot-Paris7 APC - UFR de Physique, Batiment Condorcet ,10 rue Alice Domont et L\'eonie Duquet 75205 PARIS CEDEX 13, France \\
$^{16}$ Laboratoire Univers et Th{\'e}ories LUTh, Observatoire de Paris, CNRS/INSU, Université Paris Diderot, 5 place Jules Janssen, 92190 Meudon, France \\
$^{17}$ Fakult\"at f\"ur Physik, Universit\"at Bielefeld, Postfach 100131, 33501 Bielefeld, Germany \\
$^{18}$ Centre for Astrophysics \& Supercomputing, Swinburne University of Technology, PO Box 218, Hawthorn VIC 3122, Australia \\
$^{19}$ Max Planck Institute for Gravitational Physics, Albert Einstein Institute, Callinstra\ss e 38, 30167, Hanover, Germany\\
$^{20}$ Dipartimento di Fisica - Universit\'a di Cagliari, Cittadella Universitaria, I-09042 Monserrato (CA), Italy}

\begin{document}

\maketitle

\label{firstpage}

\begin{abstract}
We present new limits on an isotropic stochastic gravitational-wave background (GWB) using a six pulsar dataset spanning 18 yr of observations from the 2015 European Pulsar Timing Array data release.  Performing a Bayesian analysis, we fit simultaneously for the intrinsic noise parameters for each pulsar, along with common correlated signals including clock, and Solar System ephemeris errors, obtaining a robust 95$\%$ upper limit on the dimensionless strain amplitude $A$ of the background of $A<3.0\times 10^{-15}$ at a reference frequency of $1\mathrm{yr^{-1}}$ and a spectral index of $13/3$, corresponding to a background from inspiralling super-massive black hole binaries,  constraining the GW energy density to $\Omega_\mathrm{gw}(f)h^2 < 1.1\times10^{-9}$ at 2.8 nHz.  We also present limits on the correlated power spectrum at a series of discrete frequencies, and show that our sensitivity to a fiducial isotropic GWB is highest at a frequency of $\sim 5\times10^{-9}$~Hz. Finally we discuss the implications of our analysis for the astrophysics of supermassive black hole binaries, and present 95$\%$ upper limits on the string tension, $G\mu/c^2$,  characterising a background produced by a cosmic string network for a set of possible scenarios, and for a stochastic relic GWB. For a Nambu-Goto field theory cosmic string network, we set a limit $G\mu/c^2<1.3\times10^{-7}$, identical to that set by the {\it Planck} Collaboration, when combining {\it Planck} and high-$\ell$ Cosmic Microwave Background data from other experiments. For a stochastic relic background we set a limit of $\Omega^\mathrm{relic}_\mathrm{gw}(f)h^2<1.2 \times10^{-9}$, a factor of 9 improvement over the most stringent limits previously set by a pulsar timing array.
\end{abstract}


\section{Introduction}
The first evidence for gravitational-waves (GWs) was originally obtained through the timing of the binary pulsar B1913+16.  The observed decrease in the orbital period of this system was found to be completely consistent with that predicted by general relativity, if the energy loss was due solely to the emission of gravitational radiation \citep{1989ApJ...345..434T}.  Despite  a decrease of only 2.3ms over the course of 30~yr, by exploiting the high precision with which the time of arrival (TOA) of electromagnetic radiation from pulsars can be measured, deviations from general relativity have been constrained by this system to be less than 0.3$\%$ \citep{2010ApJ...722.1030W}. 

Since then, observations of the double-pulsar, PSR J0737$-$3039, have provided even greater constraints, placing limits on deviations from general relativity of less than  0.05\% (\cite{2006Sci...314...97K}, Kramer et al. in prep.).  It is this extraordinary precision that also makes pulsar timing one possible route towards the {\it direct} detection of GWs, which remains a key goal in experimental astrophysics.

For a detailed review of pulsar timing we refer to \cite{2012hpa..book.....L}.  In general, one computes the difference between the expected arrival time of a pulse, given by a pulsar's timing model which characterises the properties of the pulsar's orbital motion, as well as its timing properties such as its spin frequency, and the actual arrival time. The residuals from this fit then carry physical information about the unmodelled effects in the pulse propagation, including those due to GWs \citep[e.g.][]{1978SvA....22...36S,1979ApJ...234.1100D}.

Individual pulsars have, for several decades, been used to set limits on the amplitude of gravitational radiation from a range of sources \citep[e.g.][]{1994ApJ...428..713K}. However, by using a collection of millisecond pulsars, known as a pulsar timing array \citep[PTA,][]{1990ApJ...361..300F}, one can both increase the signal-to-noise ratio of the effect of gravitational radiation in the timing residuals, and use the expected form for the cross correlation of the signal between pulsars in the array to discriminate between the GW signal of interest, and other sources of noise in the data, such as the intrinsic spin-noise due to rotational irregularities \citep[e.g.][]{2010ApJ...725.1607S}, or delays in the pulse arrival time due to propagation through the interstellar medium \citep[e.g.][]{2013MNRAS.429.2161K}.  In the specific case of an isotropic stochastic gravitational-wave background (GWB), which is the focus of this paper, this correlation is  known as the `Hellings-Downs' curve \citep{1983ApJ...265L..39H}, and is only a function of the angular separation of pairs of pulsars in the array.

The lowest frequency to which a particular pulsar timing dataset will be sensitive is set by the total observing span for that dataset.  Sensitivity to frequencies lower than this is significantly decreased due to the necessity of fitting a quadratic function in the pulsar timing model describing its spin down.  PTA datasets are now entering the regime where observations span decades, and as such are most sensitive to GWs in the range $10^{-9} - 10^{-8}$ Hz.  The primary GW sources in this band are thought to be supermassive black hole binaries (SMBHBs) \citep{1995ApJ...446..543R,2003ApJ...583..616J,2003ApJ...590..691W,2004ApJ...611..623S,2008MNRAS.390..192S}, however other sources such as cosmic strings \citep[see, e.g.][]{1981PhLB..107...47V,1994csot.book.....V} or relics from inflation \citep[see, e.g.][]{2005PhyU...48.1235G} have also been suggested.

The formation of SMBHBs is a direct consequence of the hierarchical structure formation paradigm. There is strong evidence that SMBHs are common in the nuclei of nearby galaxies \citep[see][and references therein]{2013ARA&A..51..511K}. The fact that many distant galaxies harbour active nuclei for a short period of their life implies that they were also common in the past. In $\Lambda$-Cold Dark Matter ($\Lambda$-CDM) cosmology models galaxies merge frequently \citep{1993MNRAS.262..627L}. During a galaxy merger the SMBHs harboured in the galactic nuclei will sink to the center of the merger remnant, eventually forming a SMBHB \citep{1980Natur.287..307B}. As a consequence the Universe should contain a potentially large number of gradually in-spiralling SMBHBs.   The incoherent superposition of GWs from these binaries is expected to form an isotropic stochastic GWB. Deviations from isotropy, however, such as from a small number of bright nearby sources, could result in individually resolvable systems \citep{2011MNRAS.414.3251L}, and an anisotropic distribution of power across the sky \citep{2013PhRvD..88f2005M, 2013PhRvD..88h4001T, 2014PhRvD..90h2001G}. These latter situations are the subject of two companion papers \citep{2015PhRvL.115d1101T, 2015arXiv150902165B} here we focus on the possibility of detecting a stochastic isotropic GWB, and we will discuss the implications of our findings for the astrophysics of SMBHBs, cosmic strings, and relics from inflation.

An isotropic, stochastic GWB of cosmological or astrophysical origin can be described in terms of its GW energy density content $\rho_\mathrm{gw}$ per unit logarithmic frequency, divided by the critical energy density, $\rho_c$, to close the Universe:
\begin{equation}
  \label{eq:omegagw}
  \Omega_{\mathrm{gw}}(f)=\frac{1}{\rho_c}\frac{\mathrm{d} \rho_{\mathrm{gw}}}{\mathrm{d}\ln f} =\frac{2\pi^2}{3H_0^2}f^2 h^2_c(f).
\end{equation}
Here, $f$ is the GW frequency, $\rho_c=3H_0^2/8\pi$ is the critical energy density required to close the Universe, $H_0=100~h$~km~s$^{-1}$~Mpc$^{-1}$ is the Hubble expansion rate, with $h$ the dimensionless  Hubble parameter, and $\rho_{\mathrm{gw}}$ is the total energy density in GWs \citep{1999PhRvD..59j2001A,2000PhR...331..283M}. 

Typically the `characteristic strain', $h_c(f)$,  associated with a GWB energy density $\Omega_{\mathrm{gw}}(f)$ is parametrised as a single power-law for several backgrounds of interest:
\begin{equation}
h_c=A\left(\frac{f}{\mathrm{yr}^{-1}}\right)^{\alpha},
  \label{hcA}
\end{equation}
where $A$ is the strain amplitude at a characteristic frequency of 1yr$^{-1}$, and $\alpha$ describes the slope of the spectrum. Finally, $h_c$ is directly related to the observable quantity induced by a GWB in our timing residuals, the one-sided power spectral density, $S(f)$, given by:
\begin{equation}
  S(f) = \frac{1}{12\pi^2}\frac{1}{f^3}h_c(f)^2 = \frac{A^2}{12\pi^2}\left(\frac{f}{\rm{yr}^{-1}}\right)^{-\gamma}\rm{yr}^3,
  \label{Sh}
\end{equation}
where $\gamma\equiv3-2\alpha$.  Note that unless explicitly stated otherwise, henceforth when referring to spectral indices we will be referring to the quantity $\gamma$. 

The expected spectral index varies depending on the source of the stochastic background.  For a GWB resulting from inspiraling SMBHBs the characteristic strain is approximately $h_c(f)\propto f^{-2/3}$ \citep{1995ApJ...446..543R,2003ApJ...583..616J,2003ApJ...590..691W,2004ApJ...611..623S}, or equivalently, $\gamma=13/3$, whereas primordial background contributions or cosmic strings are expected to have power-law indices of $\gamma=5$ \citep{2005PhyU...48.1235G}, and $\gamma=16/3$ \citep{2010PhRvD..81j4028O,2005PhRvD..71f3510D} respectively. However, for cosmic strings in particular, a single spectral index is not expected to accurately describe the spectrum in the PTA frequency band \citep{2012PhRvD..85l2003}.

\begin{figure*}
  \begin{center}$
    \begin{array}{c}
      \includegraphics[width=150mm]{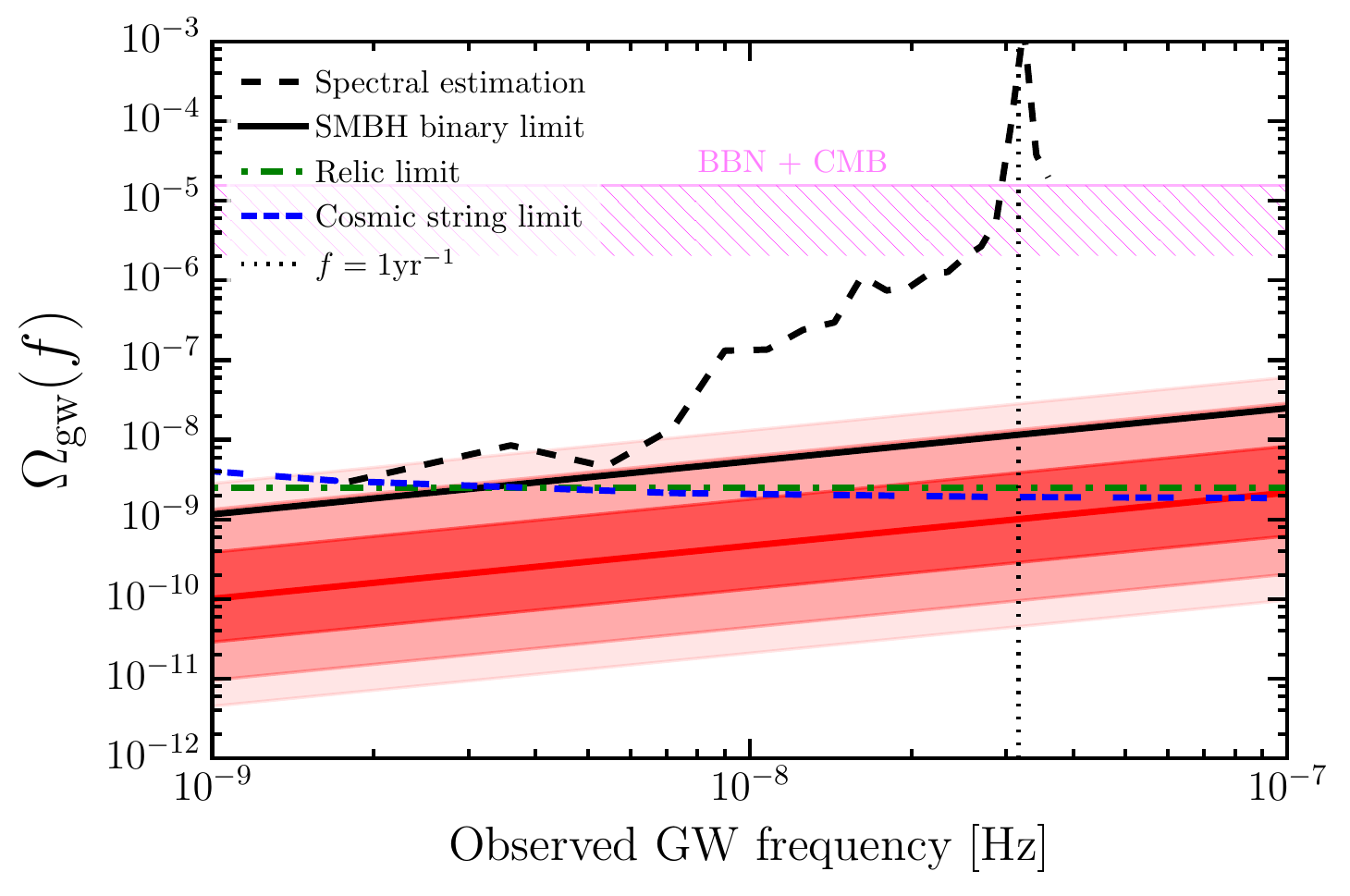}  \\
    \end{array}$
  \end{center}
  \caption{Summary of key results from the analysis of a 6 pulsar dataset from the 2015 EPTA data release (D15).  Results are presented in terms of $\Omega_{\mathrm{gw}}(f)$ as a function of GW frequency, with $H_0 = 70 \mathrm{km~s^{-1}~Mpc^{-1}}$.  We indicate the 95$\%$ upper limits on the amplitude of a correlated GWB assuming a power law model with a spectral index of $\gamma=13/3$ (solid black line; Section \ref{Section:Results}) and for a more general analysis where the power is determined simultaneously at a set of discrete frequencies (dashed line) as discussed in Section \ref{sec:bayesian}.  The red shaded areas represent the central 68\%, 95\%, and 99.7\% confidence interval of the predicted GWB amplitude according to \protect\cite{2013MNRAS.433L...1S} under the assumptions that a SMBHB evolves purely due to gravitational radiation reaction and binaries are circular (See Section \ref{Section:implicationsSMBHB} for more details).  Only about 5$\%$ of the distribution is excluded, meaning that our limit does not place significant restrictions on the cosmic SMBHB population.  We also indicate 95\% upper limits obtained for a stochastic relic background (green dash-dotted line; Section \ref{sec:relicGWs}), and for cosmic string network backgrounds (blue triple-dashed line; Section \ref{Section:implicationsStrings}). The cosmic string limit plotted corresponds to a fiducial model for a population of cosmic strings, with the following parameters: string tension $G\mu/c^2=10^{-7}$, the birth-scale of loops relative to the horizon $\alpha_{\rm cs}=1.6\times10^{-6}$, spectral index $q=4/3$, cut-off on the number of emission harmonics $n_*=1$ and intercommutation probability $p=1$.  Finally we indicate recent constraints placed by CMB \protect\citep{2012PhRvD..85l3002S}, and BBN  \citep{1997rggr.conf..373A, 2000PhR...331..283M, 2015arXiv150201589P} observations.}
  \label{figure:MoneyPlot}
\end{figure*}

A multitude of experiments have set limits on the amplitude of the stochastic GWB, either at a reference frequency as is done for PTAs \citep{2013Sci...342..334S} and ground-based interferometers~\citep{2014PhRvL.113w1101A}, or by reporting a value for GW energy density integrated over all frequencies as is done by Big Bang Nucleosynthesis measurements, e.g. \citep{2005APh....23..313C} and Cosmic Microwave Background (CMB) measurements \citep{2006PhRvL..97b1301S,2012PhRvD..85l3002S}. As such, an upper limit on the stochastic GWB reported in terms of either $\Omega_{\mathrm{gw}}(f)h^2$, or $\Omega_{\mathrm{gw}}(f)$ for a specified value of $h$ provides a clear way to report our limits.

In the last few years the European PTA (EPTA), Parkes PTA (PPTA), and the North American NanoHertz Observatory for Gravitational waves (NANOGrav) have placed 95$\%$ upper limits on the amplitude of a stochastic GWB at a reference frequency of $1\mathrm{yr^{-1}}$ of 6$\times 10^{-15}$ \citep{2011MNRAS.414.3117V}, 2.4$\times 10^{-15}$ \citep{2013Sci...342..334S}, and 7$\times 10^{-15}$ \citep{2013ApJ...762...94D}  respectively.  While many of the same pulsars are used by all the PTAs, and both the EPTA and PPTA have similar total observing spans, all of these limits have been placed using different datasets, and different methodologies. As such, these similarly constraining limits should not be seen as redundant, but rather as complementary. For example, the first EPTA limit used Bayesian analysis methods, producing an upper limit while simultaneously fitting for the intrinsic timing noise of the pulsars.  Subsequent limits have used simulations to obtain conservative upper bounds consistent with the data, or made use of frequentist methods, fixing the noise at values derived from analysis of the individual pulsars.   Naturally a simultaneous analysis of the intrinsic properties of the pulsars with the GWB is the preferred method, and we will show explicitly in Section \ref{Section:Results} that fixing the noise properties of the individual pulsars can lead to an erroneously stringent limit on the amplitude of a GWB in the pulsar timing data.  The three PTA projects also work together as the International PTA \citep[IPTA;][]{2010CQGra..27h4013H}, where all three datasets are combined in order to produce ever more robust and constraining limits on the GWB, with the eventual goal of making a first detection.

In this work we make use of the Bayesian methods presented in \cite{2013PhRvD..87j4021L} (henceforth L13),  which allows us to greatly extend what is computationally feasible for a Bayesian analysis of pulsar timing data. In particular,  while we obtain upper limits on the amplitude of a GWB using a simple, two parameter power law as in \cite{2011MNRAS.414.3117V}, we can also make use of a much more general model, enabling us to place robust limits on the correlated power spectrum at discrete frequencies.  We can also include additional sources of common noise in our analysis simultaneously with the GWB, such as those that could be expected from errors in the Solar System ephemeris, or in the reference time standard used to measure the TOAs of the pulses. Finally we also take two approaches to parameterising the spatial correlations between pulsars, without having to assume anything about the form it might take.  This spatial correlation is the `smoking gun' of a signal from a GWB, and so the ability to extract it directly from the data is crucial for the credibility of any future detections from pulsar timing data.

The key results of our analysis, compared to current theoretical predictions for a range of models of stochastic background and indirect limits in the PTA range are summarised in Fig. \ref{figure:MoneyPlot}.  In Section \ref{Section:Models} we describe the deterministic and stochastic models that we included in this analysis.  In Section \ref{Section:AnalysisMethods} we discuss the implementation of these methods in our Bayesian and frequentist frameworks.  In Section \ref{Section:Dataset} we introduce the EPTA dataset adopted for the analysis (Desvignes et al. in prep.), and in Section \ref{Section:Results} we present the results obtained from our analysis. The implications of our findings for SMBHB astrophysics, cosmic strings, and relics from inflation are discussed in Section \ref{Section:Discussion}, and finally, in Section \ref{Section:Conclusions} we summarise and discuss future prospects.

This research is the result of the common effort to directly detect gravitational-waves using pulsar timing, known as the EPTA \citep[EPTA;][]{2013CQGra..30v4009K} \footnote{www.epta.eu.org/}.


\section{Signal and noise models}
\label{Section:Models}

The search for a stochastic GWB in pulsar timing data requires the estimation of a correlated signal of common origin in the pulse TOAs recorded for the different pulsars in the array. The difficulty lies in the intrinsic weakness of the signal and the presence of a range of effects -- both deterministic and stochastic -- that conspire to mask the signal of interest. At the heart of our analysis methods is the variance-covariance matrix
\begin{equation}
\Psi_{IJ}[i,j] = \langle d_I[i] d_J[j] \rangle\,,
\label{e:varcov}
\end{equation}
that describes the expectation value of the correlation between TOA $i$ from pulsar $I$, with a TOA $j$ from pulsar $J$. In the following description  upper-case latin indices $I, J, \dots$ identify pulsars, and lower case latin indices $i, j, \dots$ are short hand notation for the TOAs $t_i, t_j, \dots$. Eq.~(\ref{e:varcov}) depends on the unknown parameters that describe the model adopted to describe the data and enter the likelihood function in the Bayesian analysis, and the optimal statistic in the frequentist approach.

For any pulsar we adopt a model for the observed pulse TOAs, which we denote $\mathbf{d}$, that results from a number of contributions and physical effects according to:
\begin{equation}
\label{Eq:TotalSignal}
\mathbf{d} = \mathbf{\tau}^\mathbf{TM} + \mathrm{\tau}^\mathbf{WN} + \mathrm{\tau}^\mathbf{SN} + \mathrm{\tau}^\mathbf{DM}  + \mathrm{\tau}^\mathbf{CN} + \mathrm{\tau}^\mathbf{GW}\,.
\end{equation}
In Eq.~(\ref{Eq:TotalSignal}) we have:

\begin{itemize}
\item $\mathbf{\tau}^\mathbf{TM}$, the deterministic model that characterises the pulsar's astrometric properties, such as position and proper motion, as well as its timing properties, such as spin period, and additional orbital parameters if the pulsar is in a binary.
\item $\mathrm{\tau}^\mathbf{WN}$, the stochastic contribution due to the combination of instrumental thermal noise, and intrinsic pulsar white noise. 
\item $\mathrm{\tau}^\mathbf{SN}$, the stochastic contribution due to red spin-noise.
\item $\mathrm{\tau}^\mathbf{DM}$, the stochastic contribution due to changes in the dispersion of radio pulses traveling through the interstellar medium. 
\item $\mathrm{\tau}^\mathbf{CN}$, the stochastic contribution due to `common noise',  present across all pulsars in the timing array (described in Sec. \ref{Section:AdditionalNoise}), as could be expected from errors in the Solar System ephemeris, or in the reference time standard used to measure the TOAs of the pulses.
\item $\mathrm{\tau}^\mathbf{GW}$, the stochastic contribution due to a GWB.
\end{itemize}

Our model assumes that all stochastic contributions are zero mean random Gaussian processes. Each of the contributions just described depends on a number of unknown parameters that need to be \emph{simultaneously} estimated in the analysis. %
While all these elements, which we set out in detail below, will be present in the Bayesian analysis described in Section \ref{Section:Bayes}, we do not incorporate the common pulsar noise terms in the frequentist optimal-statistic analysis described in Section \ref{Section:Frequentist} as this approach by design interprets all cross-correlated power as originating from a stochastic GWB. 

\subsection{The timing model}

The first contribution to the total signal model that we must consider is the deterministic effect due to the intrinsic evolution of the `pulsar clock', encapsulated by the pulsar's timing ephemeris. We identify with $\mathbf{\epsilon}_I$ the $m$-dimensional parameter vector for pulsar $I$ that describes the relevant set of timing model parameters, and denote as $\mathbf{\tau}(\mathbf{\epsilon})$ the set of arrival times determined by the adopted model and specific value of the parameters. We use \textsc{Tempo2}  \citep{2006MNRAS.369..655H, 2006MNRAS.372.1549E} to construct a weighted least-squares fit, in which the stochastic contributions have been determined from a Bayesian analysis of the individual pulsars using the  \textsc{TempoNest} plugin \citep{2014MNRAS.437.3004L}.  We can define the set of `post-fit' residuals that results from subtracting the predicted TOA for each pulse at the Solar System Barycenter from our observed TOAs as:
\begin{equation}
\mathbf{d}_{\mathrm{post}} = \mathbf{d} - \mathbf{\tau}(\mathbf{\epsilon}).
\end{equation}
In everything that follows, rather than use the full non-linear timing model we consider an initial estimate of the $m$ timing model parameters $\mathbf{\epsilon_{\mathrm{0}}}$, and construct a linear approximation to that model such that any deviations from those initial estimates are encapsulated using the $m$ parameters $\mathbf{\delta\epsilon}$ such that:

\begin{equation}
\delta\epsilon_i = \epsilon_i - \epsilon_{\mathrm{0}i}.
\end{equation}
Therefore, we can express the change in the post-fit residuals that results from the deviation in the timing model parameters $\mathbf{\delta\epsilon}$ as:
\begin{equation}
\mathbf{\delta t} = \mathbf{d}_{\mathrm{post}} -  \mathbfss{M}\mathbf{\delta\epsilon},
\end{equation}
where $\mathbfss{M}$ is the $N_{\mathrm{d}}\times m$ `design matrix' which describes the dependence of the timing residuals on the model parameters.

When we perform our Bayesian GWB analysis, we will marginalise analytically over the linear timing model, as described in Section \ref{Section:Bayes}.  When performing this marginalisation the matrix $\mathbfss{M}$ is numerically unstable. To remedy this issue we follow the same process as in \cite{2014PhRvD..90j4012V} and take the SVD of $\mathbfss{M}$, to form the set of matrices $\mathbfss{U}\mathbfss{S}\mathbfss{V}^T$. Here $ \mathbfss{U}$ is an $N_d\times N_d$  matrix, which we can divide into two components:
\begin{equation}
\mathbf{U} = \big(\mathbfss{G}^\mathrm{C}, \mathbfss{G}\big),
\end{equation}
where $\mathbfss{G}$ is a $N_d\times(N_d-m)$ matrix, which can be thought of as a projection matrix \citep{2013MNRAS.428.1147V}, and $\mathbfss{G}^\mathrm{C}$ is the $N_d\times m$ complement.  $\mathbfss{G}^\mathrm{C}$ represents a set of orthonormal basis vectors that contain the same information as $\mathbf{M}$ but is stable numerically.  We therefore replace $\mathbfss{M}$ with $\mathbfss{G}^\mathrm{C}$ in the subsequent analysis.

\subsection{White noise}
\label{Section:White}

We next consider the contribution to the total signal model that results from a stochastic white noise component, $\mathrm{\tau}^\mathbf{WN}$. This noise component is usually divided into two components, and this is the model that we adopt in our analysis:

\begin{itemize}
  \item For a given pulsar $I$, each TOA has an associated error bar, $\sigma_{(I,i)}$, the size of which will vary across a set of observations.  We can introduce an extra free parameter, referred to as EFAC, to account for possible mis-calibration of this radiometer noise.  The EFAC parameter therefore acts as a multiplier for all the TOA error bars for a given pulsar, observed with a particular `system' (i.e. a unique combination of telescope, recording system and receiver). \\
 \item A second white noise component is also used to represent some additional source of time independent noise, which we call EQUAD, and adds in quadrature to the TOA error bar.  In principle this parameter represents something physical about the pulsar, for example, contributions from the high frequency tail of the pulsar's red spin-noise power spectrum, or jitter noise that results from the time averaging of a finite number of single pulses to form each TOA \citep[see e.g.][]{1985ApJS...59..343C, 2011MNRAS.417.2916L, 2014MNRAS.443.1463S}. While this term should be independent of the observing system used to generate a given TOA, differences in the integration times between TOAs for different observing epochs can muddy this physical interpretation.
\end{itemize}
We can therefore modify the uncertainty $\sigma_{(I,i)}$, defining $\hat{\sigma}_{(I,i)}$ such that the statistical description is:
\begin{equation}
\langle \tau^{\mathrm{WN}}_I[i] \tau^{\mathrm{WN}}_J[j] \rangle = \delta_{IJ} \delta_{ij} \hat{\sigma}_{(I,i)}^2
\end{equation}
where
\begin{equation}
\hat{\sigma}_{(I,i)}^2 = (\alpha_{(I,i)}\sigma_{(I,i)})^2 + \beta_{(I,i)}^2
\end{equation}
where $\alpha$ and $\beta$ represent the EFAC and EQUAD parameters applied to TOA $i$ for pulsar $I$ respectively. In Section \ref{Section:Dataset} we list the number of different observing systems per pulsar used in the analysis presented in this paper.

\subsection{Spin-noise}

Individual pulsars are known to sometimes suffer from `spin-noise', which is observed in the pulsar's residuals as a red noise process. This is a particularly important noise source, as most models for a stochastic GWB predict that this too will induce a red spectrum signal in the timing residuals. The spin-noise component is specific to each individual pulsar, and is uncorrelated between pulsars in the timing array. The statistical properties of the spin-noise signal are therefore given by:

\begin{equation}
\langle \tau^{\mathrm{SN}}_I[i] \tau^{\mathrm{SN}}_J[j] \rangle = \delta_{IJ} C^{\mathrm{SN}}_{(I,i,j)},
\end{equation}
where the matrix element $C^{\mathrm{SN}}_{(I,i,j)}$ denotes the covariance in the spin-noise signal between residuals $i, j$ for pulsar $I$. In order to construct the matrix $\mathbfss{C}^{\mathrm{SN}}$, we will use the time-frequency method described in L13, which we will summarise below. 

We begin by writing the spin-noise component of the stochastic signal as
\begin{equation}
\mathrm{\tau}^{\mathbf{SN}} = \mathbfss{F}^{\mathbf{\mathrm{SN}}}\mathbf{a_{\mathrm{SN}}}
\end{equation}
where the matrix $\mathbfss{F}^{\mathbf{\mathrm{SN}}}$ denotes the Fourier transform such that for signal frequency $\nu$ and time $t$ we will have both:

\begin{equation}
\label{Eq:FMatrix}
F^{\mathbf{\mathrm{SN}}}(\nu,t) = \sin\left(2\pi\nu t\right),
\end{equation}
and an equivalent cosine term, and $\mathbf{a_{\mathrm{SN}}}$ are the set of free parameters that describe the amplitude of the sine and cosine components at each frequency. 

We include in our model the set of frequencies with values $n/T$, where $T$ the longest period to be included in the model and the number of frequencies to be sampled is $n_{\mathrm{SN}}$. In our analysis presented in Section \ref{Section:Dataset} we take $T$ to be $\sim 18~\mathrm{yr}$, which is the total observing span across all the pulsars in our dataset, and we take $n_{\mathrm{SN}} = 50$ such that we include in our model periods up to $\sim 130~\mathrm{days}$ which is sufficient to describe the stochastic signals present in the data (Caballero et al. in prep.).  For typical PTA data \cite{2012MNRAS.423.2642L} and \cite{2013MNRAS.428.1147V} showed that taking $T$ to be the longest time baseline in the dataset is sufficient to accurately describe the expected long-term variations present in the data, as the quadratic term present in the timing model significantly diminishes our sensitivities to periods longer than this in the data.

The covariance matrix of the spin-noise coefficients $\mathbf{a^{\mathbf{\mathrm{SN}}}}$ between pulsars $I,J$ at model frequencies $i,j$, which we denote $\mathbf{\Psi^{\mathbf{\mathrm{SN}}}_{(I,J)}}$  will be diagonal, with components:

\begin{equation}
\label{Eq:BPrior}
\Psi^{\mathbf{\mathrm{SN}}}_{(I,J,i,j)} = \left< a^{\mathbf{\mathrm{SN}}}_{(I,i)}a^{\mathbf{\mathrm{SN}}}_{(J,j)}\right> = \varphi^{\mathbf{\mathrm{SN}}}_{I,i}\delta_{ij}\delta_{IJ},
\end{equation}
where the set of coefficients $\mathbf{\varphi}^{\mathbf{\mathrm{SN}}}_{I}$ represent the theoretical power spectrum of the spin-noise signal present in pulsar $I$.  In our analysis of the dataset presented in Section \ref{Section:Dataset} we assume that this intrinsic spin-noise can be well described by a 2-parameter power law model in frequency, given by:

\begin{equation}
\label{Eq:RedPowerLaw}
\varphi^\mathrm{SN}(\nu, A_{\mathrm{SN}}, \gamma_{\mathrm{SN}}) = \frac{A_{\mathrm{SN}}^2}{12\pi^2}\left(\frac{1}{1\mathrm{yr}}\right)^{-3} \frac{\nu^{-\gamma_{\mathrm{SN}}}}{T},
\end{equation}
with $A_{\mathrm{SN}}$ and $\gamma_{\mathrm{SN}}$ the amplitude and spectral index of the power law.

We note that as discussed in L13, whilst Eq.~(\ref{Eq:BPrior}) states that the spin-noise model components are orthogonal to one another, this does not mean that we assume they are orthogonal in the time domain where they are sampled, and it can be shown that this non-orthogonality is accounted for within the likelihood \citep{2015MNRAS.446.1170V}.  The covariance matrix $\mathbfss{C}^{\mathrm{SN}}_{I}$ for the red noise signal present in the data alone can then be written:

\begin{equation}
\label{Eq:RedCMatrix}
\mathbfss{C}^{\mathrm{SN}}_I = \mathbfss{N}_I^{-1} - \mathbfss{N}_I^{-1}\mathbfss{F}^{\mathbf{\mathrm{SN}}}_I\left[(\mathbfss{F}^{\mathbf{\mathrm{SN}}}_I)^T\mathbfss{N}_I^{-1}\mathbfss{F}^{\mathbf{\mathrm{SN}}}_I + (\Psi^{\mathbf{\mathrm{SN}}})^{-1}\right]^{-1}(\mathbfss{F}^{\mathbf{\mathrm{SN}}}_I)^T\mathbfss{N}_I^{-1},
\end{equation}
with $\mathbfss{N}_I$ the diagonal matrix containing the TOA uncertainties, such that $N_{(I,i,j)} = \hat{\sigma}_{(I,i)}^2\delta_{ij}$.

\subsection{Dispersion measure variations}

The plasma located in the interstellar medium (ISM) can result in delays in the propagation of the pulse signal between the pulsar and the observatory.  Variations in the column-density of this plasma along the line of sight to the pulsar can appear as a red noise signal in the timing residuals.

Unlike other red noise signals however, the severity of the observed dispersion measure (DM) variations is dependent upon the observing frequency, and as such we can use this additional information to isolate the component of the red noise that results from this effect.

In particular, the group delay $t_g(\nu_o)$ at an observed frequency $\nu_o$ is given by the relation:
\begin{equation}
t_g(\nu_o) = DM/(K\nu_o^2)
\end{equation}
where the dispersion constant $K$ is defined to be:
\begin{equation}
K \equiv 2.41 \times 10^{-16}~\mathrm{Hz^{-2}~cm^{-3}~pc~s^{-1}}
\end{equation}
and the DM is defined as the integral of the electron density $n_e$ from the Earth to the pulsar:
\begin{equation}
DM = \int_0^L n_e \mathrm{d}l\,.
\end{equation}

While many different approaches to performing DM correction exist (e.g. \cite{2014MNRAS.441.2831L,2013MNRAS.429.2161K}), in our analysis we use the methods described in L13.  DM corrections can then be included in the analysis as an additional set of stochastic parameters in a similar manner to the intrinsic spin-noise. Further details on the DM variations present in the EPTA dataset, including comparisons between different models, will be presented in a seperate paper (Janssen et al. in prep.). In our analysis, as for the spin-noise, we assume a 2-parameter power law model, with an equivalent form to Eq.~(\ref{Eq:RedPowerLaw}), however we omit the factor $12\pi^2$ for the DM variations.

We first define a vector $\mathbf{D}$ of length $N_d$ for a given pulsar as:

\begin{equation}
D_i = 1/(K\nu^2_{(o,i)})
\end{equation}
for observation $i$ with observing frequency $\nu_{(o,i)}$.

We then make a change to Eq.~(\ref{Eq:FMatrix}) such that our DM Fourier modes are described by:

\begin{equation}
F^{\mathrm{DM}}(\nu,t_i) = \sin\left(2\pi\nu t_i\right)D_i
\end{equation}
and an equivalent cosine term, where the set of frequencies to be included is defined in the same way as for the red spin-noise, such that we choose the number of  frequencies, $n_{\mathrm{DM}}$, to also be 50. Unlike when modelling the red spin-noise, where the quadratic terms in the timing model that accounts for pulsar spin-down acts as a proxy to the low frequency ($\nu < 1/T$) fluctuations in our data, we are still sensitive to the low frequency power in the DM signal.  As such these terms must be accounted for either by explicitly including these low frequencies in the model, or by including a quadratic term in DM to act as a proxy, defined as: 

\begin{equation}
\label{Eq:DMQuad}
 Q^{\mathrm{DM}}(t_i)= q_0 t_iD_i + q_1 t_i^2D_i
\end{equation}
with $q_{0,1}$ free parameters to be fit for, and $t_i$ the barycentric arrival time for TOA $i$.  This can be achieved most simply by adding the timing model parameters $DM1$ and $DM2$ into the pulsar timing model, which are equivalent to $q_0$ and $q_1$ in Eq.~(\ref{Eq:DMQuad}), and this is the approach we take in our analysis here.

As for the spin-noise component we can then write down the time domain signal for our DM variations as:

\begin{equation}
\mathrm{\tau}^{\mathbf{DM}} = \mathbfss{F}^{\mathbf{\mathrm{DM}}}\mathbf{a_{\mathrm{DM}}},
\end{equation}
with $\mathbf{a_{\mathrm{DM}}}$ the set of free parameters that describe the amplitude of the sine and cosine components at each frequency.

The covariance matrix of the coefficients $\mathbf{a^{\mathbf{\mathrm{DM}}}}$ between pulsars $I,J$ at model frequencies $i,j$, which we denote $\mathbf{\Psi^{\mathbf{\mathrm{DM}}}_{(I,J)}}$  is then equivalent to the spin-noise matrix in Eq.~(\ref{Eq:BPrior}), and we can similarily construct the covariance matrix for the signal, $\mathrm{\tau}^{\mathbf{DM}}$, as in Eq.~(\ref{Eq:RedCMatrix}).

\subsection{Combining model terms}
\label{Section:ModelComb}

In order to simplify notation from this point forwards, for each pulsar $I$ we combine the matrices $\mathbfss{G}^\mathrm{C}_{I}$, $\mathbfss{F}^{\mathbf{\mathrm{SN}}}_I$ and $\mathbfss{F}^{\mathbf{\mathrm{DM}}}_I$ into a single, $N_{d,I}\times(m_I + 2n_{\mathrm{SN}} +2n_{\mathrm{DM}})$ matrix, where $N_{d,I}$ is the number of TOAs in pulsar $I$, $m_I$ is the number of timing model parameters, and the factor 2 in front of both $n_{\mathrm{SN}}$ and $n_{\mathrm{DM}}$ accounts for the sine and cosine terms included for each model frequency.  We denote this combined matrix $\mathbfss{T}_I$, such that:

\begin{equation}
\mathbfss{T}_I = \left(\mathbfss{G}^\mathrm{C}_{I}, \mathbfss{F}^{\mathbf{\mathrm{SN}}}_I, \mathbfss{F}^{\mathbf{\mathrm{DM}}}_I\right),
\end{equation}
and similarily we append the vectors $\mathbf{\delta\epsilon,}_I$,  $\mathbf{a_{\mathrm{SN,}}}_I$, and  $\mathbf{a_{\mathrm{DM,}}}_I$ to form the single vector $\mathbf{b}_I$.  In this way we can write our complete signal model for a single pulsar $I$ as:

\begin{equation}
\mathbf{\tau_I} = \mathbfss{T}_I\mathbf{b_I}.
\end{equation}

We can then construct the block diagonal matrix $\mathbfss{T}$ such that each block is given by the matrix $\mathbfss{T}_I$ for each pulsar $I$, and finally append the set of vectors $\mathbf{b}_I$ for all pulsars to form the complete vector of signal coefficients $\mathbf{b}$.  In this way the concatenated signal model as described thus far for all pulsars, which we denote here as $\mathbf{\tau}$,  can be written simply:

\begin{equation}
\mathbf{\tau} = \mathbfss{T}\mathbf{b}.
\end{equation}

\subsection{Common noise}
\label{Section:AdditionalNoise}

In Tiburzi (2015 PhD Thesis) and Tiburzi et al. (2015, in prep.) it was shown that additional sources of noise which are common to all pulsars in the PTA can be highly correlated with the quadrupole signature of a stochastic GWB.  If these sources of noise are present in our dataset, we will become less sensitive to a GWB if we do not include them in our model.  Therefore,  in order to ensure that our analysis remains robust to the presence of such signals, we will include in our model the 3 most likely sources of additional common noise:

\begin{description}
  \item[1:]  A common, uncorrelated noise term. This allows us to account for the possibility that all the millisecond pulsars in our dataset suffer from a similar, potentially steep, red noise process, as discussed in \cite{2010ApJ...725.1607S}. \\

  \item[2:]  A clock error. \cite{2012MNRAS.427.2780H} showed that a PTA is sensitive to errors in the time standard used to measure the arrival times of pulses. Errors in this time standard would result in a monopole signal being present in all pulsars in the dataset.  \\

  \item[3:]  An error in the Solar System ephemeris. \cite{2010ApJ...720L.201C} demonstrated that any error in the planet masses, or any unmodelled Solar System
 bodies will result in an error in our determining the barycentric time of arrival of the pulses.  This leads to a dipole correlation being induced in the timing residuals.\\
\end{description}

We note that there are other possible sources of common correlated noise in a PTA dataset beyond the three listed above.  In Section \ref{Section:Correlation} we will describe models that allow us to fit for a correlated signal, where the form of the correlation is unknown, and is described by free parameters in our analysis.  In principle one could then simultaneously fit for both a GWB, and this additional more general signal.  While this would significantly decrease our sensitivity to the GWB it would ensure that our analysis remained robust to the existence of unknown correlated signals in the data.  More optimally, one could perform an evidence comparison between a model that includes a GWB, and a model that includes a signal with an arbitrary correlation between pulsars in the PTA, in order to test which model the data supports.

A common, uncorrelated noise term can be trivially included by adding the model power spectrum to the diagonal of the elements of the matrix $\bmath{\Psi}$ that correspond to the intrinsic red noise, such that we have:

\begin{equation}
\label{Eq:UCPrior}
\Psi^{\mathbf{\mathrm{SN}}}_{(I,J,i,j)} =  \varphi^{\mathbf{\mathrm{SN}}}_{I,i}\delta_{ij}\delta_{IJ} + \varphi^{\mathbf{\mathrm{UC}}}_{i}\delta_{ij}\delta_{IJ},
\end{equation}
where the set of coefficients $\mathbf{\varphi}^{\mathbf{\mathrm{UC}}}$ represent the theoretical power spectrum  of the common uncorrelated signal, which is the same for all pulsars in the array.

\begin{figure}
\begin{center}$
\begin{array}{c}
\hspace{-1cm}
\includegraphics[width=90mm]{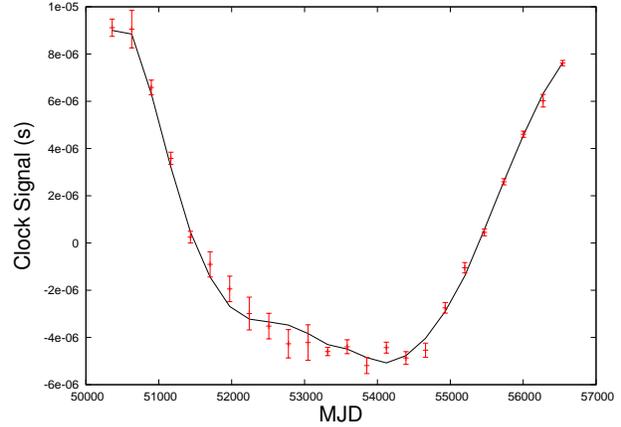} \\
\end{array}$
\end{center}
\caption{Simulated clock error used used in our analysis (black line) after subtracting the maxmium likelihood timing models from the joint analysis, and the time averaged maximum likelihood clock signal with 1$\sigma$ uncertainties (red points with error bars).  We find the recovered signal is consistent with the injected signal across the whole dataspan.}
\label{Fig:Clockplot}
\end{figure}

In order to include a clock error within the framework described thus far, we append to our matrix $\mathbf{T}$ an additional set of matrices -- one for each pulsar in the array -- each of them identical to the matrix $\mathbfss{F}^{\mathrm{SN}}_I$, given by Eq.~(\ref{Eq:FMatrix}), for the corresponding pulsar $I$.  Each of these matrices is multiplied by the same set of signal coefficients $\mathbf{a}_{\mathrm{clk}}$, which are appended to the vector of coefficients $\mathbf{b}$, representing a single signal being fit to all pulsars simultaneously.  We use the same number of frequencies in the model for the clock error as for the intrinsic spin-noise, and assume a 2-parameter power law model for the power spectrum, which we denote $\varphi^{\mathbf{\mathrm{clk}}}$, as in Eq.~(\ref{Eq:RedPowerLaw}).  From this we construct the covariance matrix $\mathbf{\Psi}^{\mathbf{\mathrm{clk}}}$ which we define:

\begin{equation}
\Psi^{\mathbf{\mathrm{clk}}}_{(i,j)} = \left< a^{\mathbf{\mathrm{clk}}}_{i}a^{\mathbf{\mathrm{clk}}}_{j}\right> = \varphi^{\mathbf{\mathrm{clk}}}_{i}\delta_{ij},
\end{equation}
the elements of which can be appended to the total covariance matrix for the signal coefficients $\mathbf{\Psi}$. We stress that modelling the clock signal in this way ensures that we correctly account for both the uneven time spans, and unequal weighting of the individual pulsars. Additionally, because we fit for the timing model simultaneously with the clock signal, the uncertainty in the low frequency variations of the signal are factored into the analysis appropriately.  We show this in a simple simulation in which we use the time sampling from our dataset described in Section \ref{Section:Dataset}, and include a clock error consistent with 10 times the difference between the TAI and BIPM2013 time standards, and white noise consistent with the TOA uncertainties in our dataset.  In Fig.~(\ref{Fig:Clockplot}) we show the clock signal used in our simulation after the maximum likelihood timing model has been subtracted from the joint analysis (black line), and the time averaged maximum likelihood recovered  clock signal with 1$\sigma$ uncertainties (red points with error bars).  The uncertainties in the clock error vary by a factor $\sim$ 9 across the dataset, as different pulsars contribute different amounts to the constraints.  We find the recovered signal is consistent with the injected signal across the whole dataspan.

Finally, in order to model an error in the Solar System ephemeris, we can define an error signal $\mathbf{e}$, which will be observed in any pulsar $I$ as the dot product between this error vector, and the position vector of the pulsar $\mathbf{k}_I$, such that the induced residual as a function of time, $\mathbf{\tau}^{\mathrm{eph}}_I$ will be given by:

\begin{equation}
\mathbf{\tau}^{\mathrm{eph}}_I = \mathbf{e}\cdot\mathbf{k}_I.
\end{equation}
We can incorporate this effect into our analysis by defining a set of basis vectors separately for each of the 3 components of $\mathbf{e}$, similarly to Eq \ref{Eq:FMatrix}.  For example, the component in the $x$ direction for pulsar $I$ will have basis vectors:

\begin{equation}
\mathbfss{F}^{\mathrm{eph,x}}_I = \mathbfss{F}^{\mathrm{SN}}_Ik_{(I,x)},
\end{equation}
such that the signal induced in the pulsar will be given by:

\begin{equation}
\mathbf{\tau}^{\mathrm{eph,x}}_I = \mathbfss{F}^{\mathrm{eph,x}}_I\mathbf{a}_{\mathrm{(eph,x)}},
\end{equation}
with $\mathbf{a}_{\mathrm{(eph,x)}}$ the set of signal coefficients to be fit for.  This model term is incorporated into our analysis in exactly the same way as for the clock error, with the basis vectors $\mathbfss{F}^{\mathrm{eph}}$ for the 3 components appended to the total matrix $\mathbfss{T}$, the 3 sets of signal coefficients appended to the vector $\mathbf{b}$, and the diagonal covariance matrix $\mathbf{\Psi}^{\mathbf{\mathrm{eph}}}$ constructed from the power law model appended to the matrix $\mathbf{\Psi}$.

While this parametrisation does not constitute a physical model of the Solar System dynamics, it allows us to incorporate our uncertainty regarding possible errors in the Solar System ephemeris, such as errors in the mass measurements of a number of planets or the effects of unknown Solar System bodies.  Given the dominant source of error in the Solar System ephemeris is likely to come from errors in the masses of planets such as Saturn, it could be advantageous to include these parameters explicitly in our model.  In our analysis presented in Section \ref{Section:Results} we opt for the more conservative approach, and use the general model described here to model such errors.

Once again we include the same number of frequencies in the model as for the spin-noise model, and parameterise the power spectrum for each of the 3 components, (x,y,z), of the error vector $\mathbf{e}$ with a separate 2 parameter power law, as  in Eq.~(\ref{Eq:RedPowerLaw}).

\subsection{Gravitational-wave background}
\label{Section:Correlation}

When dealing with a signal from a stochastic GWB, it is advantageous to include the cross correlated signal between the pulsars on the sky.  We do this by using the overlap reduction function -- a dimensionless
function which quantifies the response of the pulsars to the
stochastic GWB. For isotropic stochastic GWBs, when the pulsars are separated from the Earth and from each other by many GW wavelengths \citep[i.e., in the short-wavelength approximation, cf][]{2014PhRvD..90f2011M}, this is also known as the Hellings-Downs curve \citep{1983ApJ...265L..39H}:

\begin{eqnarray}
\Gamma(\zeta_{IJ})=\frac{3}{8}\left[1+\frac{\cos\zeta_{IJ}}{3}+4(1-\cos\zeta_{IJ})\ln\left(\sin\frac{\zeta_{IJ}}{2}\right) \right](1+\delta_{IJ}) \, .
\end{eqnarray}
Here $\zeta_{IJ}$ is the angle between the pulsars $I$ and $J$ on the sky and $\Gamma(\zeta_{IJ})$ is the overlap reduction function, which represents the expected correlation between the TOAs given an isotropic stochastic GWB, and the $\delta_{IJ}$
term accounts for the pulsar term for the autocorrelation. With this addition, our covariance matrix for the Fourier coefficients becomes

\begin{eqnarray}
\label{Eq:FreqMatrix}
\Psi^{\mathbf{\mathrm{SN}}}_{I,J,i,j} = \varphi^{\mathbf{\mathrm{SN}}}_{I,i}\delta_{ij}\delta_{IJ} + \varphi^{\mathbf{\mathrm{UC}}}_{i}\delta_{ij}\delta_{IJ} + \Gamma(\zeta_{IJ}) \varphi^{\mathbf{\mathrm{GWB}}}_{i}\delta_{ij}.
\end{eqnarray}

In our analysis presented in Section \ref{Section:sgwblimits} we define $\varphi^{\mathbf{\mathrm{GWB}}}$ using both the 2-parameter power law model given in Eq.~(\ref{Eq:RedPowerLaw}), and also take a more general approach, where the power at each frequency included in the model is a free parameter in the analysis. In this case we define $\varphi^{\mathbf{\mathrm{GWB}}}$ simply as:

\begin{equation}
\varphi^{\mathbf{\mathrm{GWB}}}_i = \rho^2_i
\end{equation}
where we fit for the set of parameters $\rho$, and use a prior that is uniform in the amplitude $\rho$.

If we do not want to assume the isotropic (Hellings-Downs) overlap reduction function as the description of the correlations between pulsars in our dataset, we can instead fit for its shape.  In Section \ref{Section:Results}, we will do this in two ways: firstly fitting directly for the correlation coefficient between each pulsar, $\Gamma(\zeta_{IJ}) $, and secondly using a set of four Chebyshev polynomials, where we fit for the coefficients $c_{1..4}$ parameterised such that, defining $x=(\zeta_{IJ} - \pi/2)/(\pi/2)    $ we will have:

\begin{equation}
\Gamma(x)  = c_1 + c_2x + c_3(2x^2 - 1) +c_4(4x^3 - 3x)\, .
\label{e:gamma_Chebyshev}
\end{equation}

\section{Analysis Methods}
\label{Section:AnalysisMethods}

While the majority of the results presented in Section \ref{Section:Results} have been obtained using a Bayesian approach, we also employ a frequentist maximum-likelihood estimator of the GWB strain-spectrum amplitude as a consistency check.  In the following sections we outline the key elements of both these approaches to aid further discussion.

\subsection{Bayesian approach}
\label{Section:Bayes}

\subsubsection{General remarks}
Bayesian Inference provides a consistent approach to the estimation of a set of parameters $\Theta$ in a model or hypothesis $\mathcal{H}$ given the data, $D$.  Bayes' theorem states that:

\begin{equation}
\mathrm{Pr}(\Theta \mid D, \mathcal{H}) = \frac{\mathrm{Pr}(D\mid \Theta, \mathcal{H})\mathrm{Pr}(\Theta \mid \mathcal{H})}{\mathrm{Pr}(D \mid \mathcal{H})},
\end{equation}
where $\mathrm{Pr}(\Theta \mid D, \mathcal{H}) \equiv \mathrm{Pr}(\Theta)$ is the posterior probability distribution of the parameters,  $\mathrm{Pr}(D\mid \Theta, \mathcal{H}) \equiv L(\Theta)$ is the likelihood,\\ $\mathrm{Pr}(\Theta\mid\mathcal{H}) \equiv \pi(\Theta)$ is the prior probability distribution, and \\ $\mathrm{Pr}(D\mid\mathcal{H}) \equiv Z$ is the Bayesian Evidence.

In parameter estimation, the normalizing evidence factor is usually ignored, since it is independent of the parameters $\Theta$.   Inferences are therefore obtained by taking samples from the (unnormalised) posterior using, for example, standard Markov chain Monte Carlo (MCMC) sampling methods.

In contrast to parameter estimation, for model selection the evidence takes the central role and is simply the factor required to normalise the posterior over $\Theta$:

\begin{equation}
Z = \int L(\Theta)\pi(\Theta) \mathrm{d}^n\Theta,
\label{eq:Evidence}
\end{equation}
where $n$ is the dimensionality of the parameter space.

As the average of the likelihood over the prior, the evidence is larger for a model if more of its parameter space is likely and smaller for a model where large areas of its parameter space have low likelihood values, even if the likelihood function is very highly peaked.  Thus, the evidence automatically implements Occam's razor: a simpler theory with a compact parameter space will have a larger evidence than a more complicated one, unless the latter is significantly better at explaining the data.

The question of model selection between two models $\mathcal{H}_0$ and $\mathcal{H}_1$ can then be decided by comparing their respective posterior probabilities, given the observed data set $D$, via the posterior odds ratio $R$:

\begin{equation}
R= \frac{P(\mathcal{H}_1\mid D)}{P(\mathcal{H}_0\mid D)} = \frac{P(D \mid \mathcal{H}_1)P(\mathcal{H}_1)}{P(D\mid \mathcal{H}_0)P(\mathcal{H}_0)} = \frac{Z_1}{Z_0}\frac{P(\mathcal{H}_1)}{P(\mathcal{H}_0)},
\label{Eq:Rval}
\end{equation}
where $P(\mathcal{H}_1)/P(\mathcal{H}_0)$ is the {\it a priori} probability ratio for the two models, which can often be set to unity but occasionally requires further consideration.

The posterior odds ratio then allows us to obtain the probability of one model compared with the other, simply as:

\begin{equation}
P = \frac{R}{1+R}.
\end{equation}

\subsubsection{MultiNest}

The nested sampling approach \citep{2004AIPC..735..395S}  is a Monte-Carlo method targeted at the efficient calculation of the evidence, but also produces posterior inferences as a by-product.  In \cite{2008MNRAS.384..449F} and \cite{2009MNRAS.398.1601F} this nested sampling framework was built upon with the introduction of the \textsc{MultiNest} algorithm, which provides an efficient means of sampling from posteriors that may contain multiple modes and/or large (curving) degeneracies.  Since its release \textsc{MultiNest} has been used successfully in a wide range of astrophysical problems, from detecting the Sunyaev-Zel'dovich effect in galaxy clusters \citep{2012MNRAS.419.2921A}, to inferring the properties of a potential stochastic GWB in PTA data in a mock data challenge (L13).

In higher dimensions ($\gtrsim$ 50), the sampling efficiency of \textsc{MultiNest} begins to decrease significantly.    To help alleviate this problem, \textsc{MultiNest}  includes a `constant efficiency' mode, which ensures that the sampling efficiency meets some user set target.  This, however comes at the expense of less accurate evidence values.
Recently, the \textsc{MultiNest} algorithm has been updated to include the concept of importance nested sampling (INS; \citealt{2013arXiv1301.6450C})  which provides a solution to this problem.  Details can be found in \cite{2013arXiv1306.2144F}, but the key difference is that, where with normal nested sampling the rejected points play no further role in the sampling process, INS uses every point sampled to contribute towards the evidence calculation.  One outcome of this approach is that even when running in constant efficiency mode the evidence calculated is reliable even in higher ($\sim$ 50) dimensional problems.
In pulsar timing analysis, and especially when determining the properties of a correlated signal between pulsars, we will often have to deal with models that can contain $> 40$ parameters.  As such, the ability to run in constant efficiency mode whilst still obtaining accurate values for the evidence when these higher dimensional problems arise is crucial in order to perform reliable model selection.

All the analyses presented in Section \ref{Section:Results} are performed using INS, running in constant efficiency mode, with 5000 live points and an efficiency of $1\%$.

\subsubsection{Likelihood function}

Equivalent to the approach described in L13, we can write the joint probability density of:
\begin{enumerate}
  \item the linear parameters $\mathbf{b}$, which describe variations in the determinstic timing model and the signal realisations for the red noise and DM variations for each pulsar, and the common noise terms.
  \item the stochastic parameters, ($\mathbf{\alpha}$, $\mathbf{\beta}$) that describe the intrinsic white noise properties for each pulsar,
  \item the power-spectrum hyper-parameters that define the spin-noise and DM variation power laws, and the spectra of the common noise terms such as the stochastic GWB, which we collectively refer to as $\mathbf{\Theta}$,

\end{enumerate}
as:
\begin{eqnarray}
\label{Eq:Prob}
\mathrm{Pr}(\mathbf{b}, \mathbf{\alpha}, \mathbf{\beta}, \mathbf{\Theta},  \;|\; \mathbf{\delta t}) \; &\propto& \; \mathrm{Pr}(\mathbf{\delta t} | \mathbf{\alpha}, \mathbf{\beta}, \mathbf{b}) \; \\\ \nonumber
&\times & \mathrm{Pr}(\mathbf{b} | \mathbf{\Theta}) \; \mathrm{Pr}(\mathbf{\Theta})\mathrm{Pr}(\mathbf{\alpha}, \mathbf{\beta})\mathrm{Pr}(\mathbf{b}). \nonumber
\end{eqnarray}

In our analysis we simply use priors that are uniform in all the model parameters, so we can write the conditional distributions that make up Eq.~(\ref{Eq:Prob}) as:

\begin{eqnarray}
\label{Eq:ProbTime}
\mathrm{Pr}(\mathbf{\delta t} | \mathbf{\alpha}, \mathbf{\beta}, \mathbf{b}) \; &\propto& \; \frac{1}{\sqrt{\mathrm{det}(\mathbf{N})}}
 \exp\left[-\frac{1}{2}(\mathbf{\delta t} -  \mathbfss{T}\mathbf{b})^T\mathbf{N}^{-1}\right. \nonumber \\
 &\times&\left.(\mathbf{\delta t} -  \mathbfss{T}\mathbf{b})\right],
\end{eqnarray}
and:
\begin{equation}
\label{Eq:ProbFreq}
 \mathrm{Pr}(\mathbf{b} | \mathbf{\Theta}) \; \propto \; \frac{1}{\sqrt{\mathrm{det}\mathbf{\Psi}}} \exp\left[-\frac{1}{2}\mathbf{b}^{T}\mathbf{\Psi}^{-1}\mathbf{b}\right].
\end{equation}

We can now marginalise over all linear parameters $\mathbf{b}$ analytically in order to find the posterior for the remaining parameters alone. 

Defining $\mathbf{\Sigma}$ as  $(\mathbf{T}^T\mathbf{N}^{-1}\mathbf{T} + \mathbf{\Psi}^{-1})$, and $\bar{\mathbf{b}}$ as $\mathbf{T}^T\mathbf{N}^{-1}\mathbf{\delta t}$ our marginalised posterior for the stochastic parameters $\mathbf{\alpha}, \mathbf{\beta}, \mathbf{\Theta}$ alone is given by:

\begin{eqnarray}
\label{Eq:Margin}
\mathrm{Pr}(\mathbf{\alpha}, \mathbf{\beta}, \mathbf{\Theta} | \mathbf{\delta t}) &\propto& \frac{\mathrm{det} \left(\mathbf{\Sigma}\right)^{-\frac{1}{2}}}{\sqrt{\mathrm{det} \left(\mathbf{\Psi}\right)~\mathrm{det}\left(\mathbf{N}\right)}} \\
&\times&\exp\left[-\frac{1}{2}\left(\mathbf{\delta t}^T\mathbf{N}^{-1} \mathbf{\delta t} - \bar{\mathbf{b}}^T\mathbf{\Sigma}^{-1}\bar{\mathbf{b}}\right)\right]. \nonumber
\end{eqnarray}

\subsection{Frequentist techniques}
\label{Section:Frequentist}

As a consistency check of our Bayesian method, we also employ a weak-signal regime maximum-likelihood estimator of the GWB strain-spectrum amplitude, known as the \textit{optimal-statistic} \citep{2009PhRvD..79h4030A,2013CQGra..30v4015S,2014arXiv1410.8256C}. It also maximises the signal-to-noise ratio (S/N) in this regime, reproducing the results of an optimally-filtered cross-correlation search without explicitly introducing a filter function. 

The form of this statistic is
\begin{equation}
\hat{A}^2 = \frac{\sum_{IJ}\delta\mathbf{t}_I^{\rm T}\mathbf{P}_I^{-1}\tilde{\mathbf{S}}_{IJ}\mathbf{P}_J^{-1}\delta\mathbf{t}_J}{\sum_{IJ}{\rm tr}\left[\mathbf{P}_I^{-1}\tilde{\mathbf{S}}_{IJ}\mathbf{P}_J^{-1}\tilde{\mathbf{S}}_{JI}\right]},
\end{equation}
where $\mathbf{P}_I = \langle\delta\mathbf{t}_I\delta\mathbf{t}_I^{\rm T}\rangle$ is the autocovariance of the post-fit residuals in pulsar $I$, which we can write in terms of the matrices $\mathbf{T_I}$ and $\mathbf{\Psi_I}$ as:

\begin{equation}
\mathbf{P_I} = \mathbf{T_I}\mathbf{\Psi_I}\mathbf{T_I}^{\mathrm{T}},
\end{equation}
where the matrix $\mathbf{\Psi_I}$ is constructed from maximum-likelihood noise estimates obtained in previous single-pulsar analysis. Any GW signal will have been absorbed into the red-noise estimation during this previous analysis. The signal term $\tilde{\mathbf{S}}_{IJ}$ is defined such that $A^2 \tilde{\mathbf{S}}_{IJ} = \langle\delta\mathbf{t}_I\delta\mathbf{t}_J^{\rm T}\rangle = \mathbf{S}_{IJ}$, where we assume that no signal other than GWs induce cross-correlations between pulsar TOAs. The normalisation of $\hat{A}^2$ is chosen such that $\langle\hat{A}^2\rangle = A^2$.

The standard deviation of the statistic in the absence of a cross-correlated signal reduces to
\begin{equation}
\sigma_0 = \left(\sum_{IJ}{\rm tr}\left[\mathbf{P}_I^{-1}\tilde{\mathbf{S}}_{IJ}\mathbf{P}_J^{-1}\tilde{\mathbf{S}}_{JI}\right]\right)^{-1/2},
\end{equation}
which can be used as an approximation to the error on $\hat{A}^2$ in the weak-signal regime. Hence, for a particular signal and noise realisation where we have measured the optimal-statistic, the S/N of the power in the cross-correlated signal is given by
\begin{equation}
\rho = \frac{\hat{A}^2}{\sigma_0} = \frac{\sum_{IJ}\delta\mathbf{t}_I^{\rm T}\mathbf{P}_I^{-1}\tilde{\mathbf{S}}_{IJ}\mathbf{P}_J^{-1}\delta\mathbf{t}_J}{\left(\sum_{IJ}{\rm tr}\left[\mathbf{P}_I^{-1}\tilde{\mathbf{S}}_{IJ}\mathbf{P}_J^{-1}\tilde{\mathbf{S}}_{JI}\right]\right)^{1/2}},
\end{equation}
with an expectation over all realisations of
\begin{equation}
\langle\rho\rangle = A^2\left(\sum_{IJ}{\rm tr}\left[\mathbf{P}_I^{-1}\tilde{\mathbf{S}}_{IJ}\mathbf{P}_J^{-1}\tilde{\mathbf{S}}_{JI}\right]\right)^{1/2}.
\end{equation}
This S/N effectively measures how likely it is (in terms of number of standard deviations from zero) that we have found a cross-correlated signal in our data rather than an uncorrelated signal. The properties of the signal cross-term $\tilde{\mathbf{S}}_{IJ}$ are determined by a fixed input spectral shape, which in this case is a power-law with slope $\gamma  = 13/3$, matching the predicted spectral properties of the strain-spectrum resulting from a population of circular GW-driven SMBHBs.

To compute upper-limits with the optimal-statistic, we follow the procedure outlined in \citet{2009PhRvD..79h4030A}, where the distribution of $\hat{A}^2$ is assumed to be a Gaussian with mean $A^2$ and variance $\sigma_0^2$. The latter assumption is clearly only appropriate in the weak-signal regime, but serves as a useful approximation. We want to find $A^2_{\rm ul}$ such that, in some predetermined fraction of hypothetical experiments ($C$), the value of the optimal-statistic would exceed the actual measured value. Hence we can claim that $A^2\leq A^2_{\rm ul}$ to confidence $C$, otherwise we would have seen it exceed the measured value a fraction $C$ of the time. The solution is given by
\begin{equation}
A^2_{\rm ul} = \hat{A}^2 + \sqrt{2}\sigma_0{\rm erfc}^{-1}[2(1-C)].
\end{equation}


It was shown in \citet{2014arXiv1410.8256C} that the cross-correlation statistic of \citet{2013ApJ...762...94D} is identical to the aforementioned optimal-statistic, and in fact allows us to achieve a measure of the individual cross-power values between pulsars. In the high S/N limit one would expect these cross-power values to map out the Hellings and Downs curve when plotted as a function of pulsar angular separations. The cross-power values and their associated errors are given by
\begin{equation}
\chi_{IJ} = \frac{\delta\mathbf{t}_I^{\rm T}\mathbf{P}_I^{-1}\hat{\mathbf{S}}_{IJ}\mathbf{P}_J^{-1}\delta\mathbf{t}_J}{{\rm tr}\left[\mathbf{P}_I^{-1}\hat{\mathbf{S}}_{IJ}\mathbf{P}_J^{-1}\hat{\mathbf{S}}_{JI}\right]},\nonumber
\end{equation}
\begin{equation}
\sigma_{0,IJ} = \left({\rm tr}\left[\mathbf{P}_I^{-1}\hat{\mathbf{S}}_{IJ}\mathbf{P}_J^{-1}\hat{\mathbf{S}}_{JI}\right]\right)^{-1/2},
\end{equation}
where $A^2\Gamma_{IJ}\hat\mathbf{S}_{IJ} = \mathbf{S}_{IJ} = A^2 \tilde{\mathbf{S}}_{IJ}$, and $\Gamma_{IJ}$ are the Hellings and Downs cross-correlation values.


\section{The Dataset}
\label{Section:Dataset}

\begin{table*}
\scriptsize
\centering
\caption{Details of the 6 pulsars used for the isotropic stochastic background analysis.  Numbers in parentheses correspond to the maximum likelihood values from the 5-dimensional analysis described in Section \ref{Section:Dataset}.}
\begin{threeparttable}
\centering
\begin{tabular}{c c c c c c c}

\hline\hline
Pulsar 			& J0613$-$0200 & J1012+5307 & J1600$-$3053 &J1713+0747  &  J1744$-$1134 & J1909$-$3744 \\[0.5ex]
\hline
Dataspan (yr) 		& 16.05 & 16.83 & 7.66  & 17.66 & 17.25 & 9.38 \\
N$_{\mathrm{sys}}$\tnote{a}	& 14    & 15 & 4  & 14 & 9 & 3 \\
$\sigma (\mu$s) \tnote{b}	& 1.691 & 1.610 & 0.563 & 0.679 & 0.801 & 0.131\\
Log$_{10}$ A$_{\mathrm{SN}}$	&	-13.58	$\pm$	0.40	(-13.41)	&	-13.05	$\pm$	0.09	(-13.04)	&	-13.71	$\pm$	0.54	(-13.42)	&	-14.31	$\pm$	0.46	(-14.20)	&	-13.63	$\pm$	0.27	(-13.60)	&	-14.22	$\pm$	0.42	(-13.98)	\\
$\gamma_{\mathrm{SN}}$	&	2.50	$\pm$	0.99	(2.09)	&	1.56	$\pm$	0.37	(1.56)	&	1.91	$\pm$	1.05	(1.38)	&	3.50	$\pm$	1.16	(3.51)	&	2.21	$\pm$	0.82	(2.16)	&	2.23	$\pm$	0.89	(2.17)	\\
Log$_{10}$ A$_{\mathrm{DM}}$	&	-11.61	$\pm$	0.12	(-11.57)	&	-12.25	$\pm$	0.47	(-11.92)	&	-11.75	$\pm$	0.39	(-11.67)	&	-11.97	$\pm$	0.14	(-11.90)	&	-12.19	$\pm$	0.38	(-11.93)	&	-12.76	$\pm$	0.53	(-12.51)	\\
$\gamma_{\mathrm{DM}}$	&	1.36	$\pm$	0.48	(1.11)	&	1.26	$\pm$	0.97	(0.27)	&	1.64	$\pm$	0.80	(1.46)	&	2.03	$\pm$	0.55	(1.82)	&	1.41	$\pm$	1.09	(0.36)	&	2.23	$\pm$	1.07	(2.16)	\\
Global EFAC	&	1.01	$\pm$	0.02	(1.01)	&	0.98	$\pm$	0.02	(0.98)	&	1.03	$\pm$	0.04	(1.03)	&	1.00	$\pm$	0.02	(1.00)	&	1.01	$\pm$	0.03	(1.00)	&	1.02	$\pm$	0.04	(1.01)	\\

95\% upper limit \tnote{c}	& $9.7\times10^{-15}$ & $8.3\times10^{-15}$ & $2.1\times10^{-14}$ & $4.4\times10^{-15}$ & $7.0\times10^{-15}$ & $5.2\times10^{-15}$\\
\hline
\end{tabular}
\begin{tablenotes}
\item[a] Number of unique observing `systems' that make up the dataset for each pulsar.
\item[b] Weighted rms for the DM subtracted residuals for each pulsar (D15).
\item[c] Upper limit obtained from the 5-dimensional analysis described in Section  \ref{Section:Dataset}.
\end{tablenotes}
\end{threeparttable}
\label{Table:Pulsars}
\end{table*}

\begin{figure*}
\begin{center}$
\begin{array}{c}
\includegraphics[width=120mm]{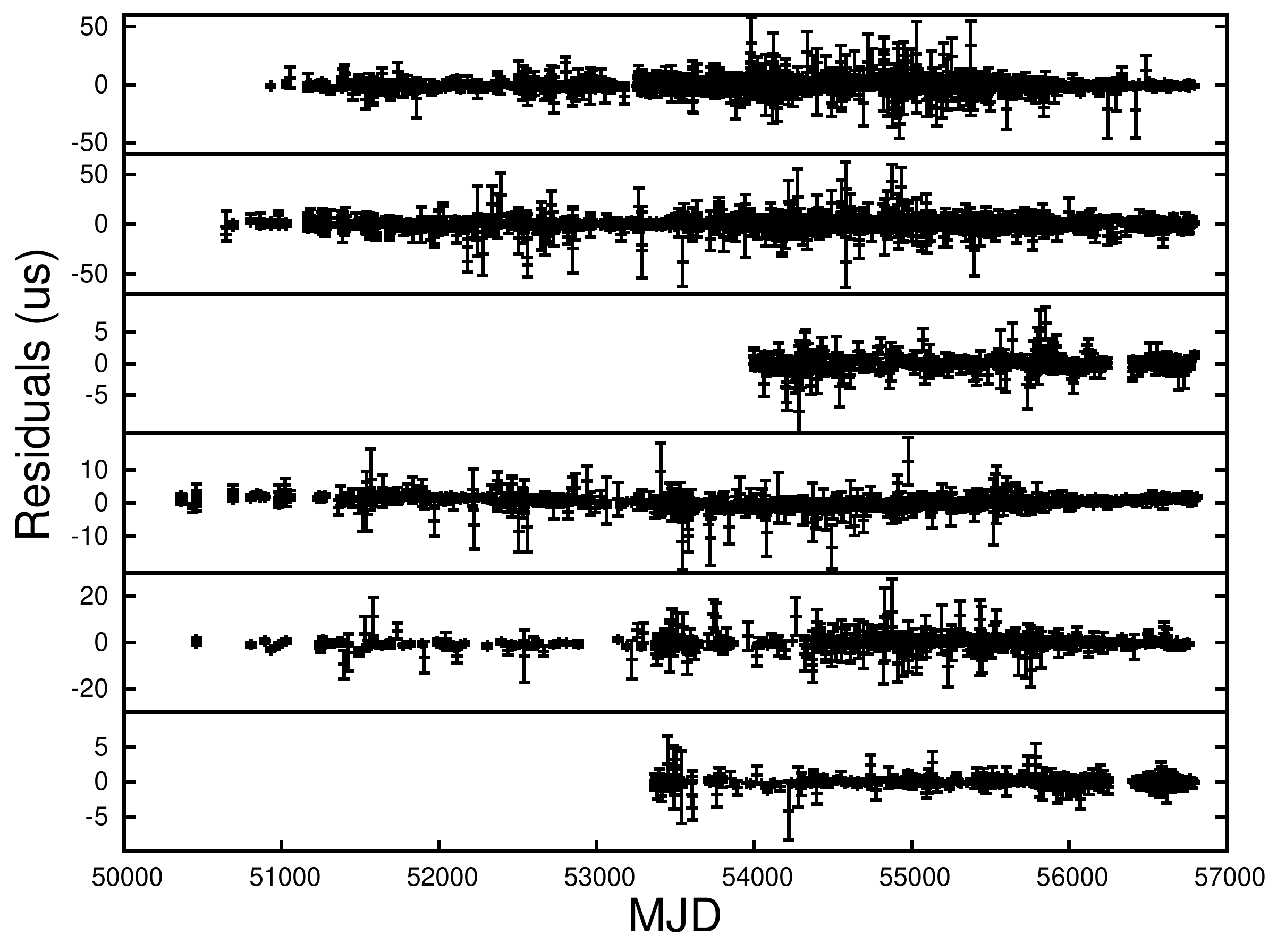} \\
\hspace{-0.5cm}
\includegraphics[width=125mm]{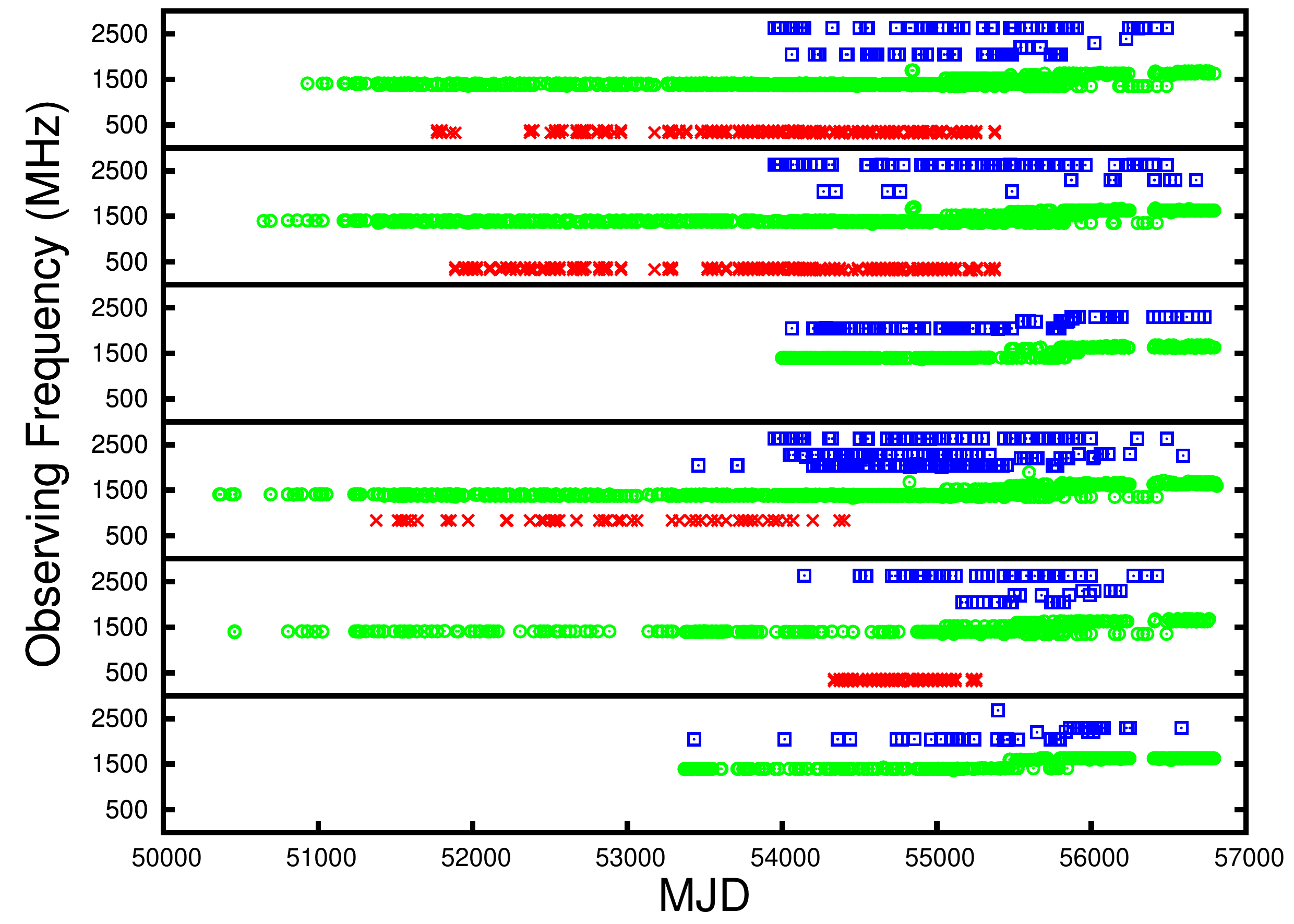} \\
\end{array}$
\end{center}
\caption{Top: timing residuals as a function of Modified Julian Date (MJD) for the 6 pulsars included in the stochastic GWB analysis presented in this work, after the maximum likelihood DM variations signal realisation has been subtracted.  From top to bottom these are PSRs: J0613$-$0200, J1012+5307, J1600$-$3053, J1713+0747,  J1744$-$1134, and J1909$-$3744. While the overall timing baseline for this dataset is $\sim 18$~yr, only four of the 6 pulsars have data that extends across the majority of this timespan, and in particular, PSR J1909$-$3744 contributes only to the latter half of the dataset, significantly reducing our overall sensitivity to signals at the lowest frequencies supported by the dataset. Bottom: Frequency coverage for the 6 pulsars included in the stochastic GWB analysis presented in this work.  The order of the pulsars is as in the top plot.  Colours indicate observing frequencies $< 1000$MHz (red crosses), between 1000 and 2000 MHz (green circles) and $>$ 2000 MHz (blue squares).  In addition to fewer pulsars extending across the full dataset, there is also less multi-frequency coverage in the early data.  This further decreases our sensitivity to a stochastic GWB at the lowest sampled frequencies as the signal becomes highly covariant with the DM variations for the individual pulsars in the first half of the dataset.}
\label{Fig:Dataplots}
\end{figure*}

\begin{figure}
\begin{center}$
\begin{array}{c}
\includegraphics[width=80mm]{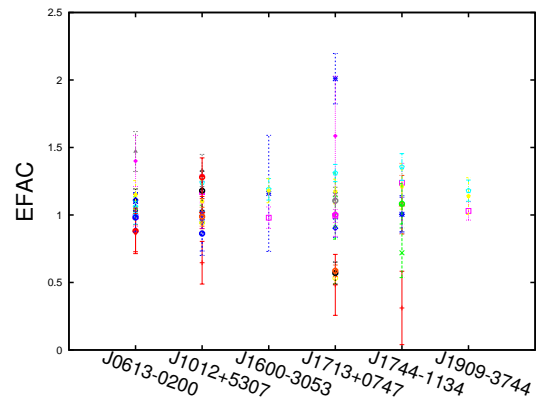} \\
\end{array}$
\end{center}
\caption{EFAC values obtained for all systems from the initial analysis performed for the 6 pulsars used in our analysis.  All EFACs are consistent with values equal to or greater than 1 within uncertainties, with the exception of the Westerbork 1380 MHz data in PSR J1713+0747 which have values consistent with $\sim 0.5$.  This could be the result of systematic effects that occur in the template forming phase, and is the subject of ongoing work.}
\label{Fig:EFACplot}
\end{figure}

Our limits for an isotropic stochastic background are obtained using a subset of the full 2015 EPTA data release described in Desvignes et al. (in prep.) (henceforth D15). 
In particular we use a set of 6 pulsars, listed in Table~\ref{Table:Pulsars}, that contribute 90$\%$ of the total S/N for this dataset (see Babak et al. (in prep.) for details).  We use this subset of the full 42 pulsar dataset in order to minimise the dimensionality of the problem, and thus enable accurate evidence calculations using \textsc{MultiNest}.  The pulsar that contributed next in terms of sensitivity, PSR J1640+2224, contributes at only the $\sim 2\%$ level, so even were we to add a small number of additional pulsars, the overall gain in sensitivity would be minimal.  The DM-subtracted residuals, as well as the frequency coverage as a function of time for these pulsars are shown in Fig. \ref{Fig:Dataplots} (left and right panel respectively).  For each of these pulsars a full timing analysis has been performed using the \textsc{TempoNest} plugin for the \textsc{Tempo2} pulsar timing package, which simultaneously includes the white noise modifiers EFAC and EQUAD for each observing system, as well as intrinsic red noise, and frequency dependent DM variations.  In Fig.~(\ref{Fig:EFACplot}) we show the mean parameter estimates and 1$\sigma$ uncertainties for the EFAC parameters obtained for each system from this initial analysis. We find all EFACs are consistent with values equal to or greater than 1 within their uncertainties, with the exception of the Westerbork 1380 MHz data in PSR J1713+0747.  This could be the result of systematic effects that occur in the template forming phase, and is the subject of ongoing work.  As these systems do not contribute a large fraction of the total weight in the dataset, however, it will not have a significant impact on the subsequent analysis. Further analysis of the white noise parameters will be presented in Caballero et al. (in prep.).  In the joint analysis presented in this work we use the linear approximation to the timing model.  As such, timing solutions obtained from the initial \textsc{TempoNest} analysis were checked for convergence, and the linear regime was found to be suitable in all cases.

The EPTA dataset contains observations from four of the largest radio telescopes in Europe: the Effelsberg Radio Telescope in Germany, the Lovell Radio Telescope at the Jodrell Bank Observatory in the UK, the Nancay Radio Telescope in France, and the Westerbork Synthesis Radio Telescope in The Netherlands.  Each of these telescopes operates at multiple observing frequencies, and so the number of unique `systems' present for any one pulsar can be as large as 15.   For the 6 pulsars in this dataset we list the number of observing systems present in each in Table ~\ref{Table:Pulsars}, which in combination results in 118 white noise parameters.  When accounting for the 4 spin-noise and DM variation power law parameters for each pulsar, we have a total of 142 intrinsic noise parameters before the addition of any correlated model components. In an effort to decrease the dimensionality we therefore use the mean estimates for the EFAC and EQUAD parameters from the individual timing analysis presented in D15, and fit a single global EFAC for each pulsar, reducing the number of intrinsic noise parameters to 30.

\begin{figure*}
\begin{center}$
\begin{array}{ccc}
\hspace{-0.5cm}
\includegraphics[width=60mm]{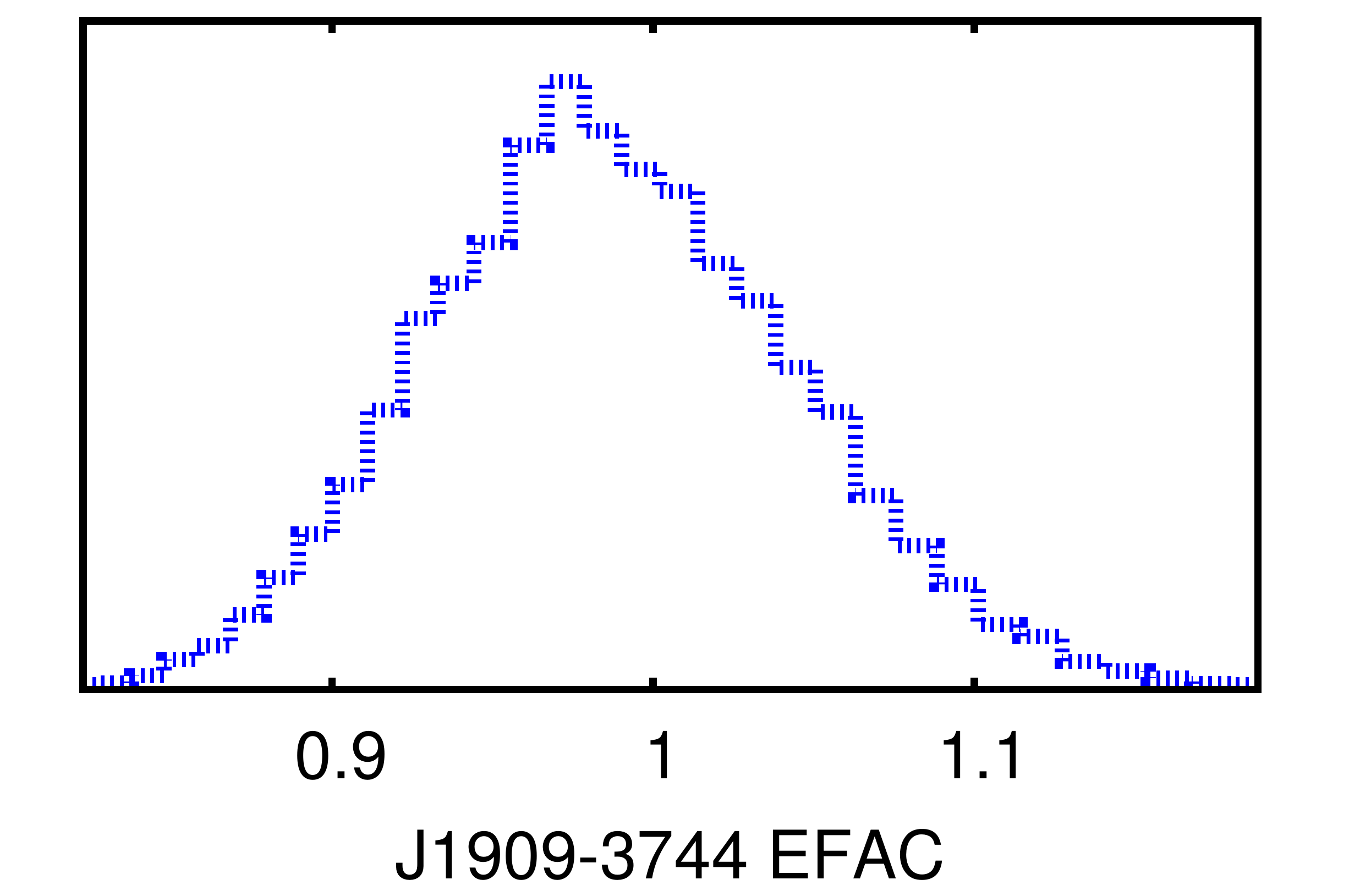} &
\includegraphics[width=60mm]{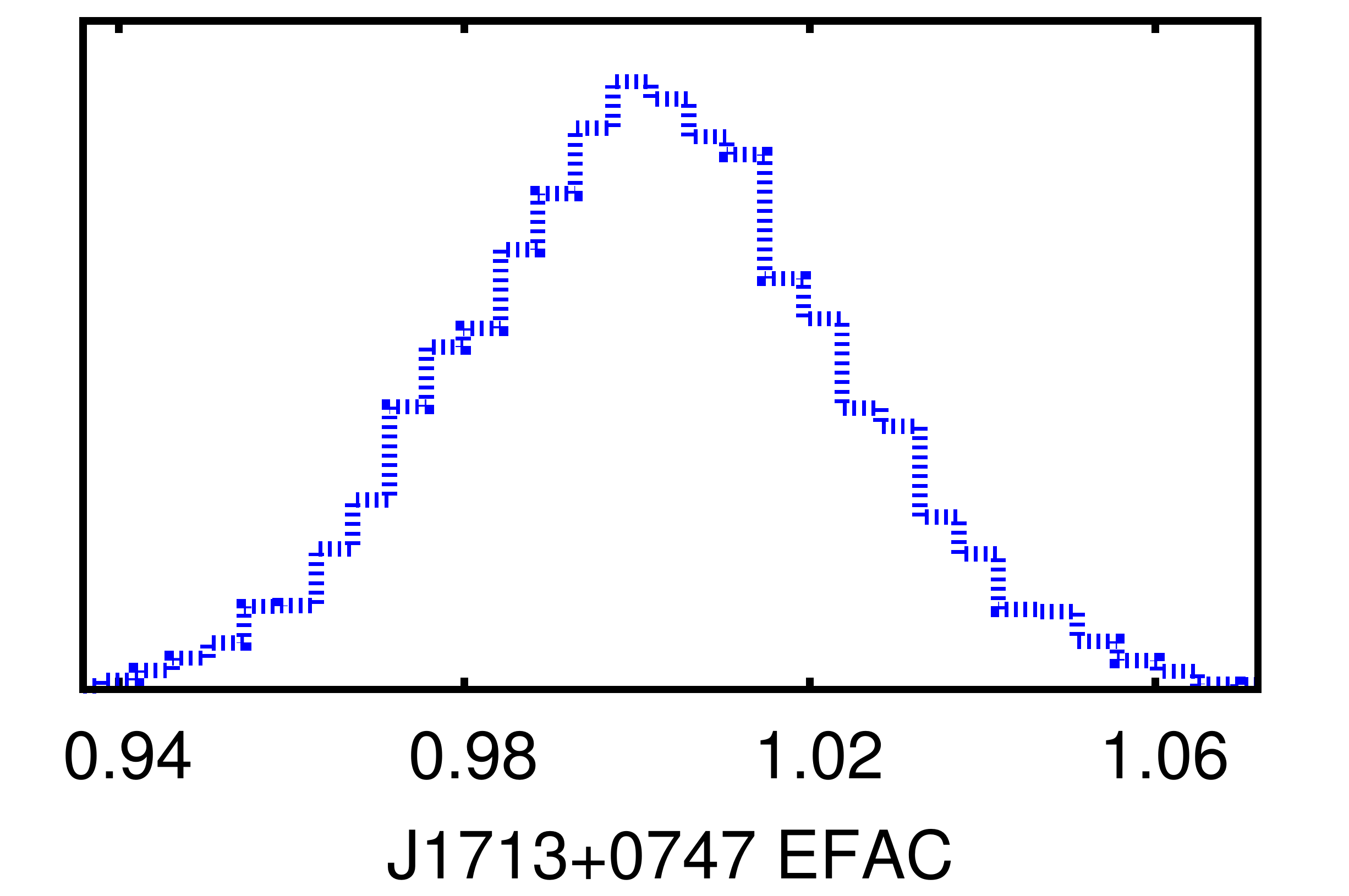} &
\includegraphics[width=60mm]{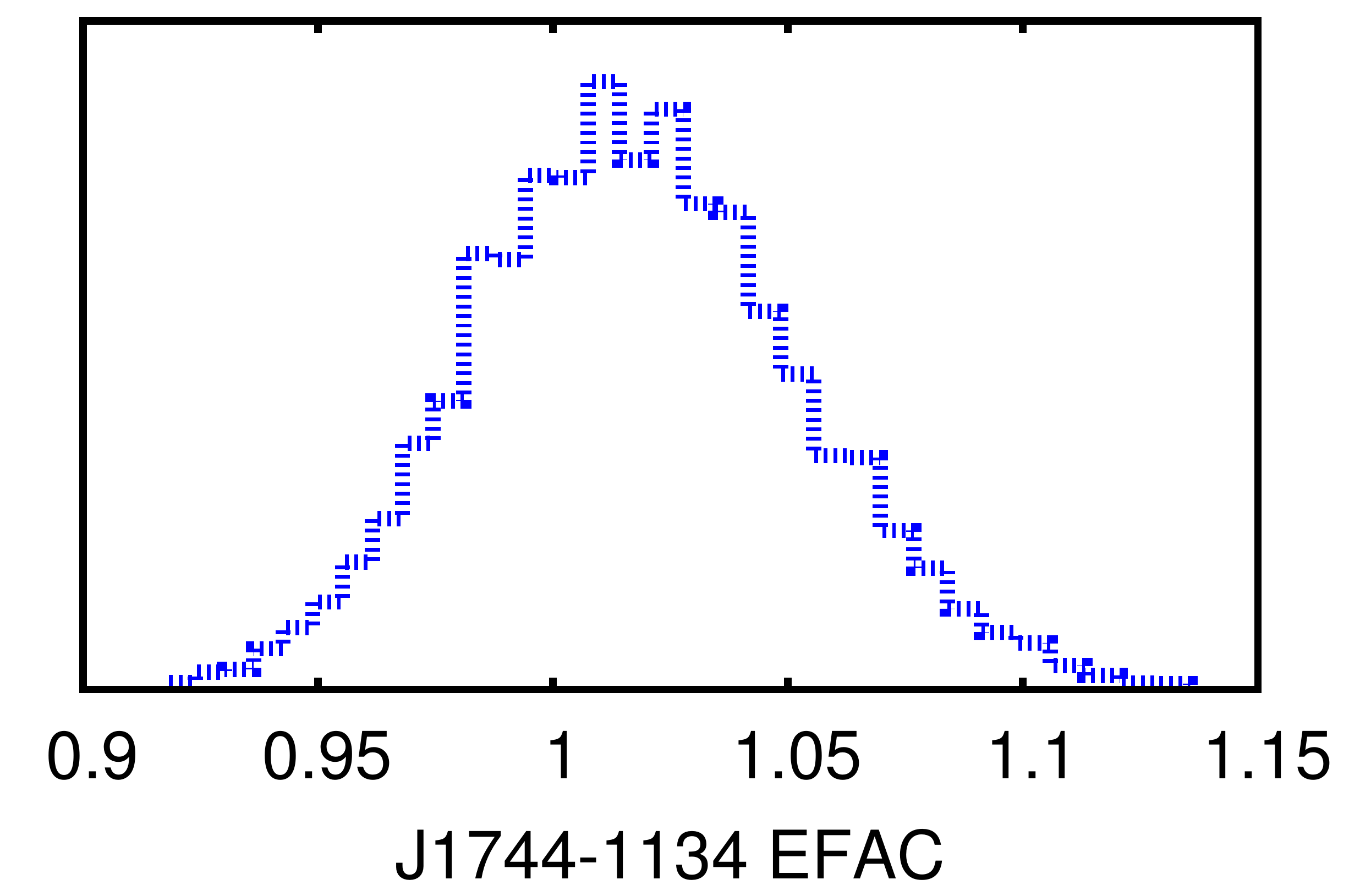} \\
\hspace{-0.5cm}
\includegraphics[width=60mm]{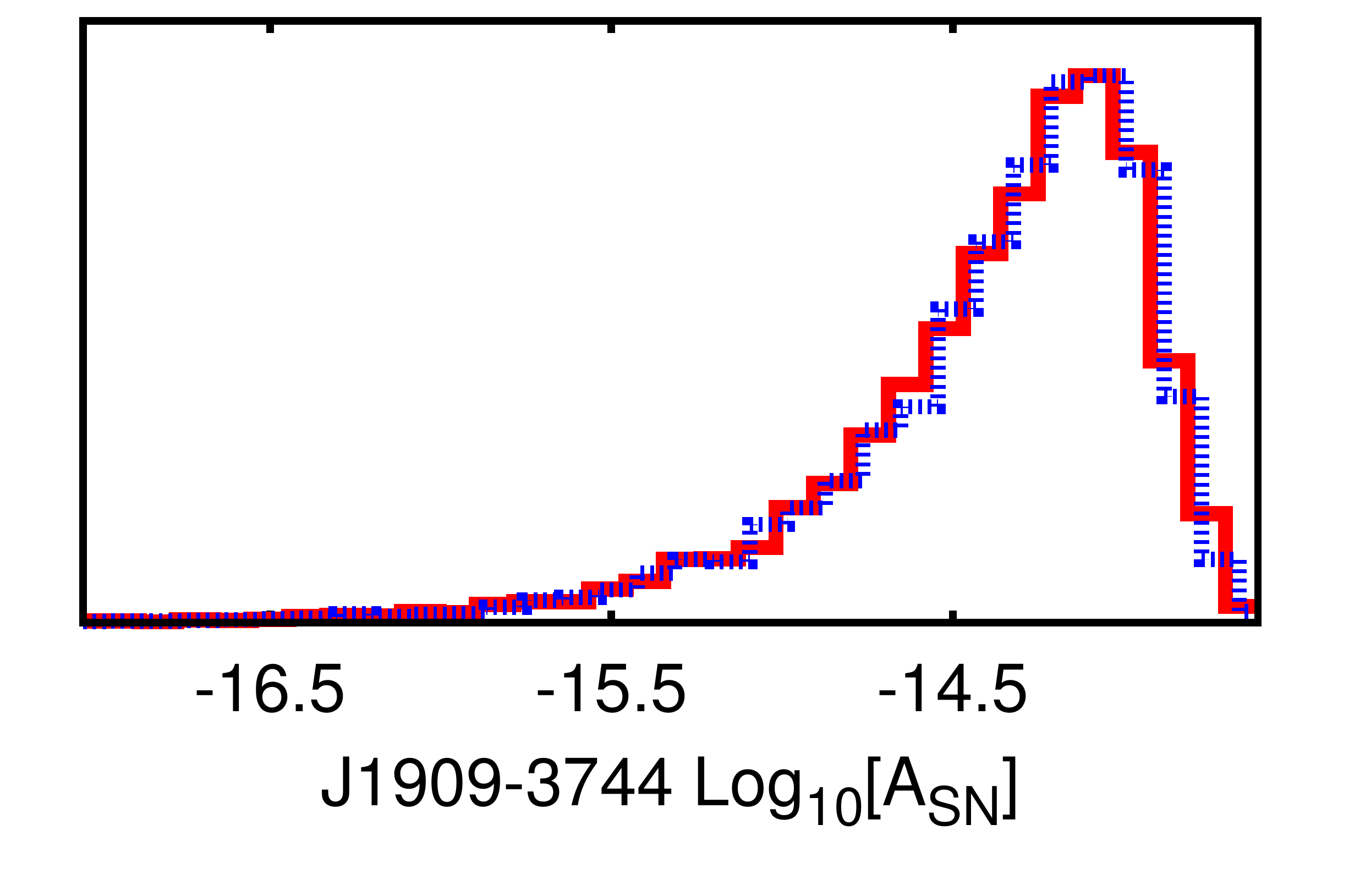} &
\includegraphics[width=60mm]{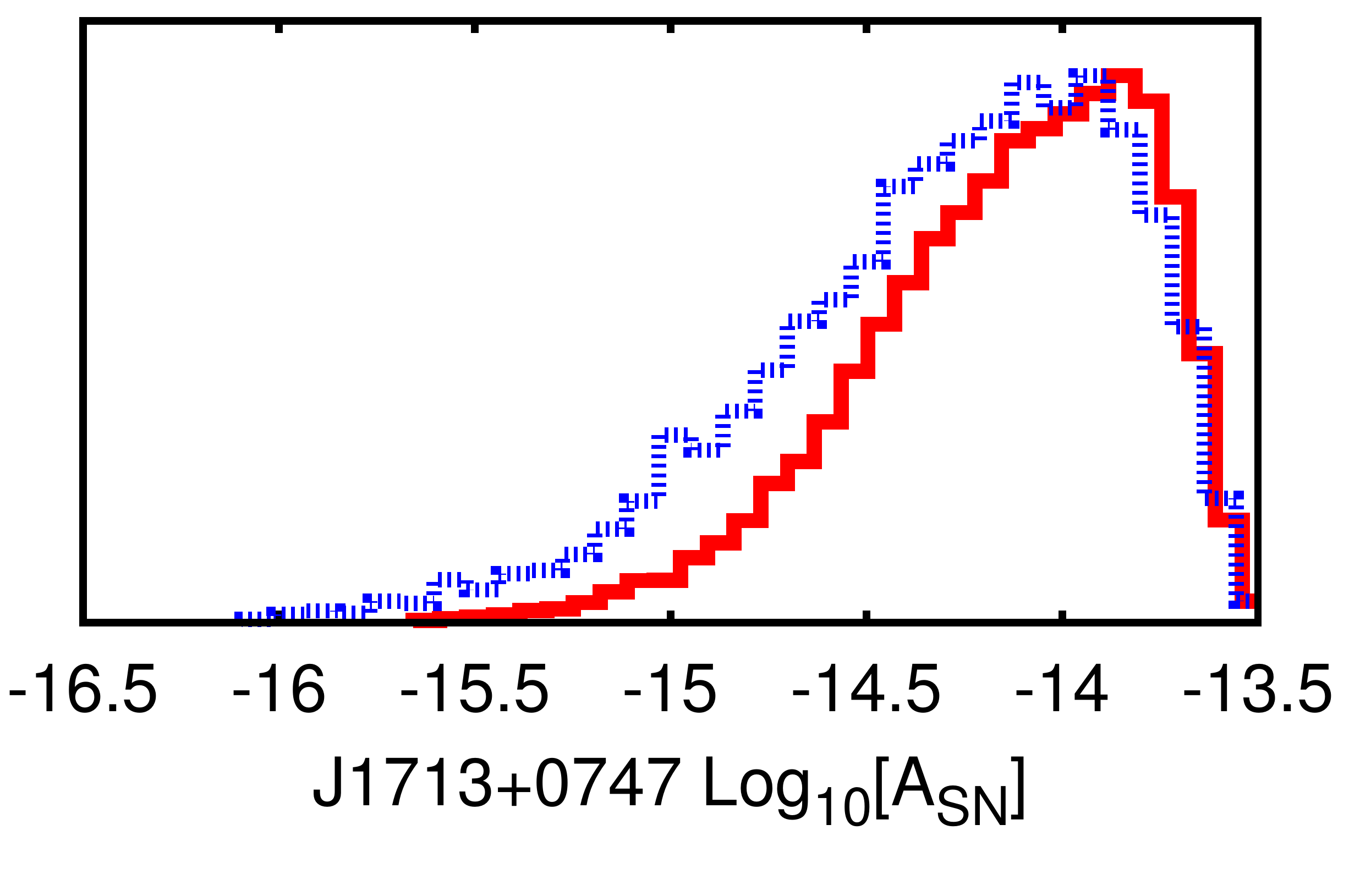} &
\includegraphics[width=60mm]{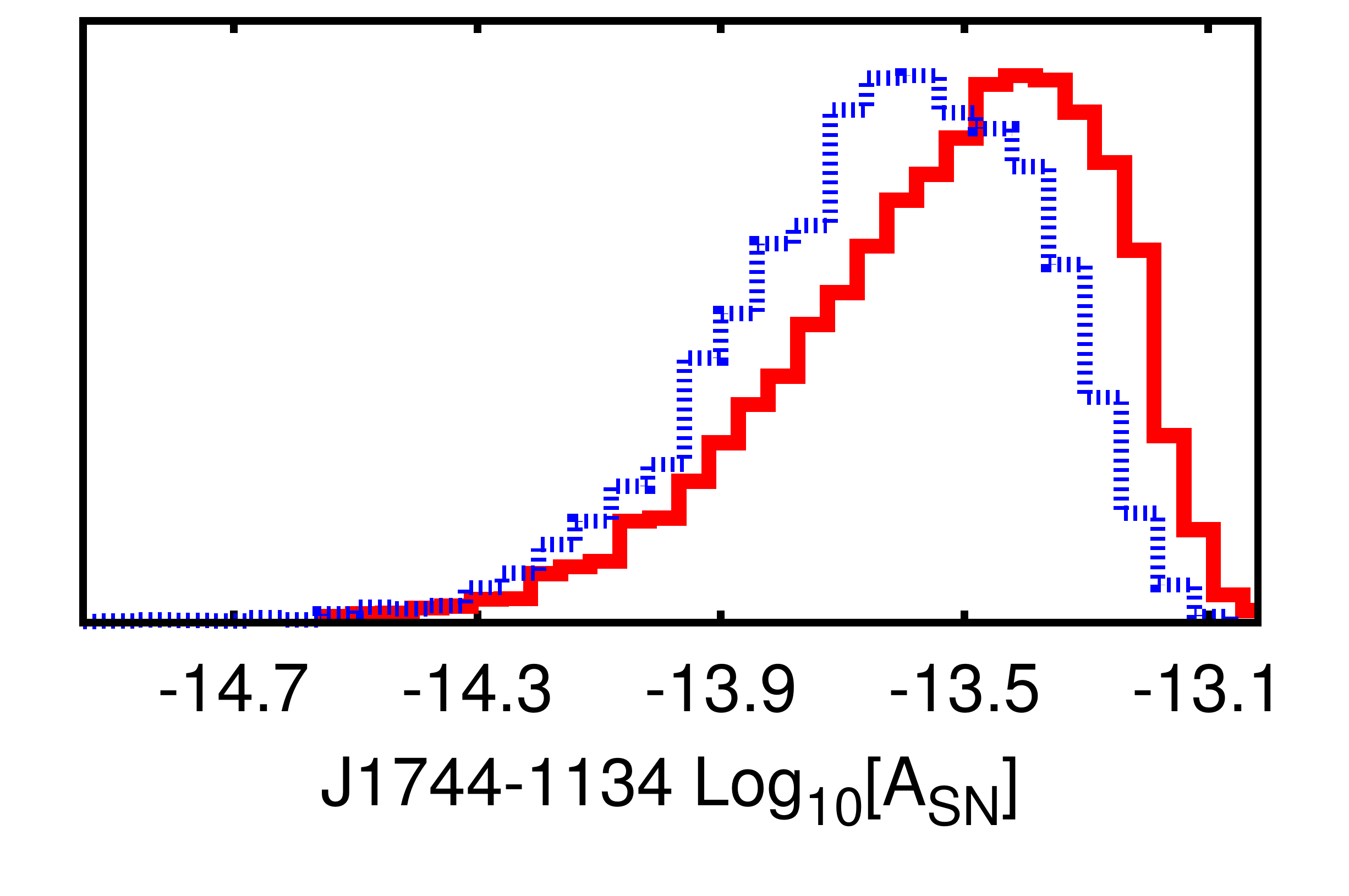} \\
\hspace{-0.5cm}
\includegraphics[width=60mm]{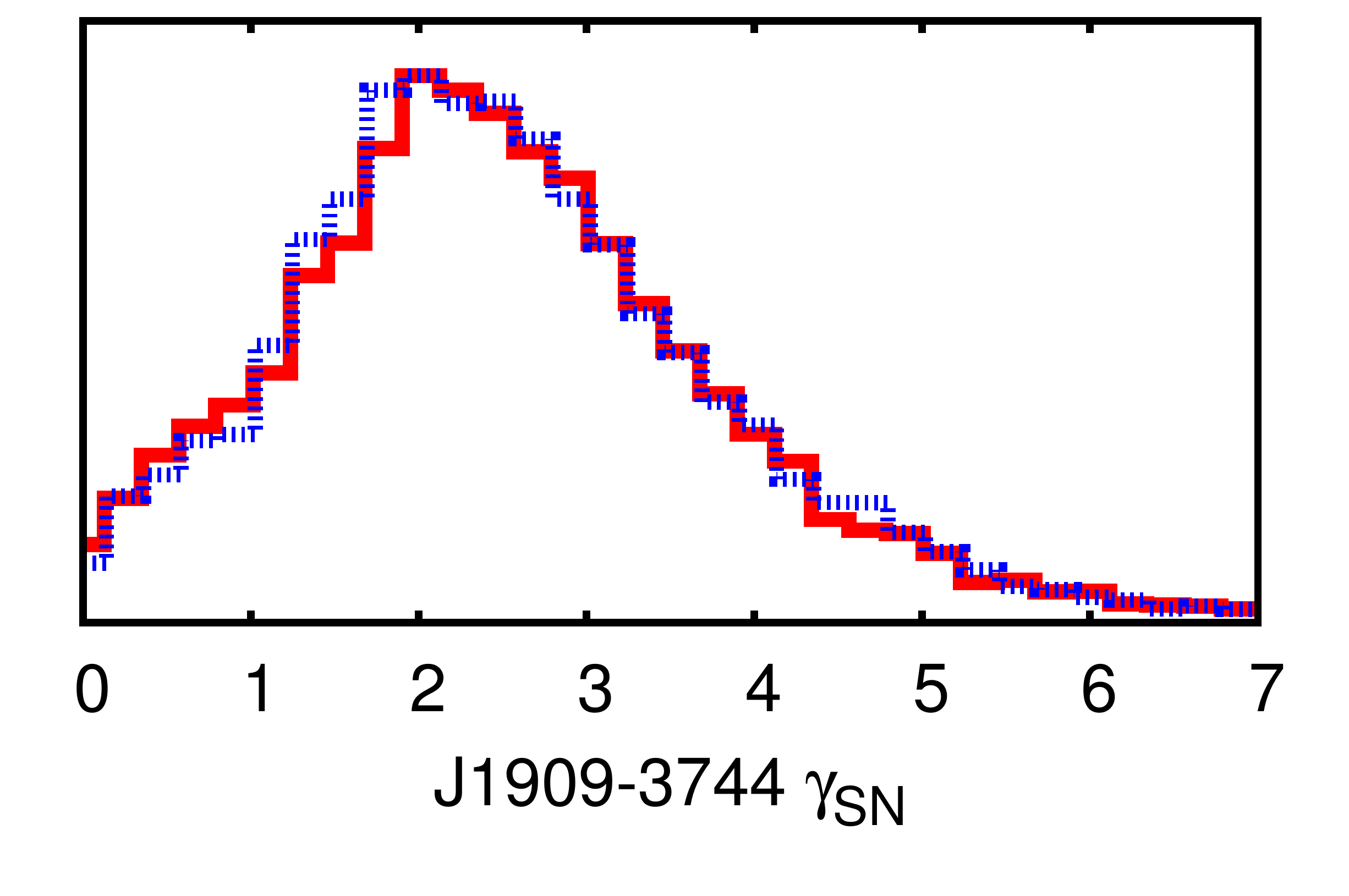} &
\includegraphics[width=60mm]{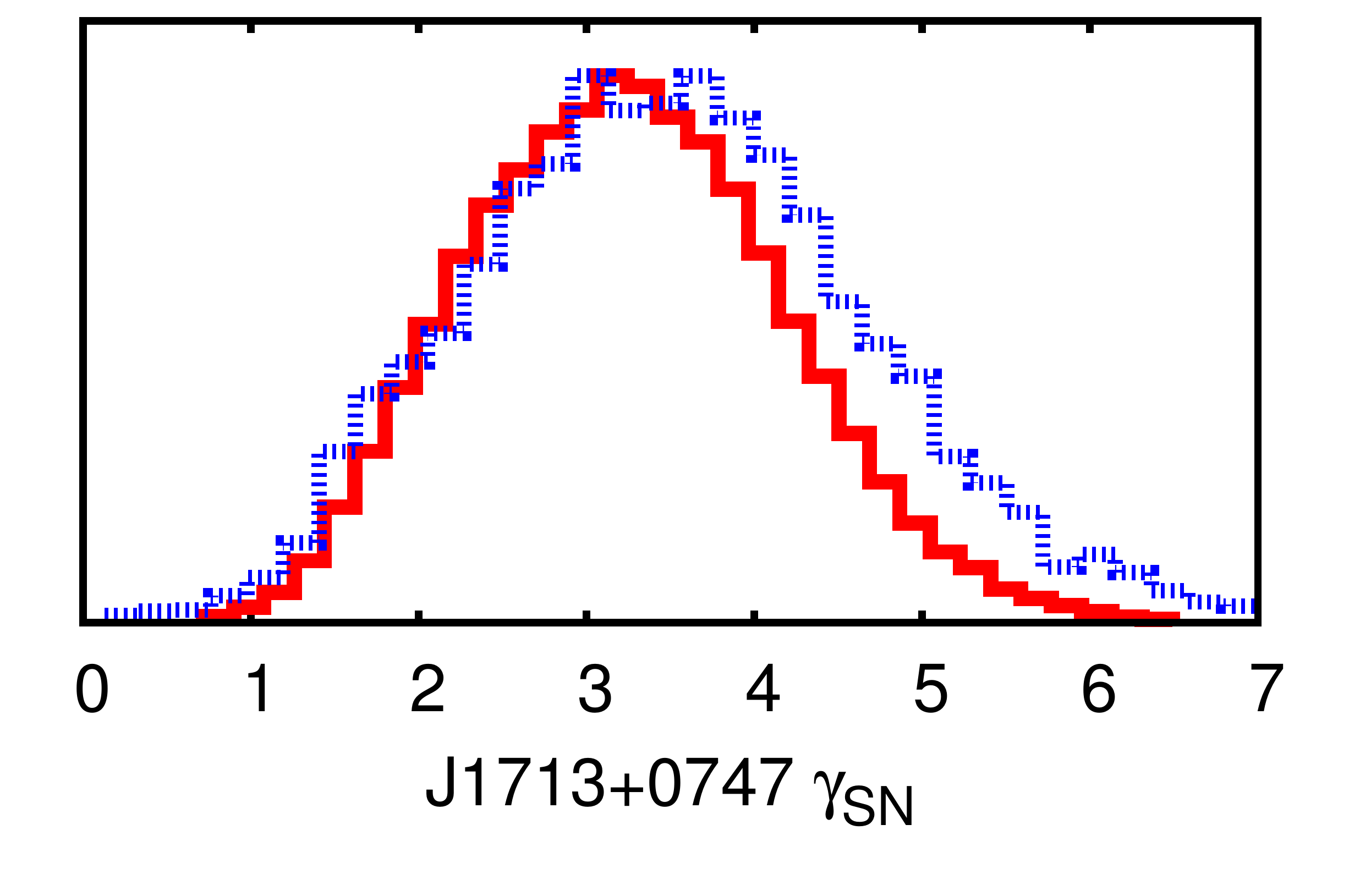} &
\includegraphics[width=60mm]{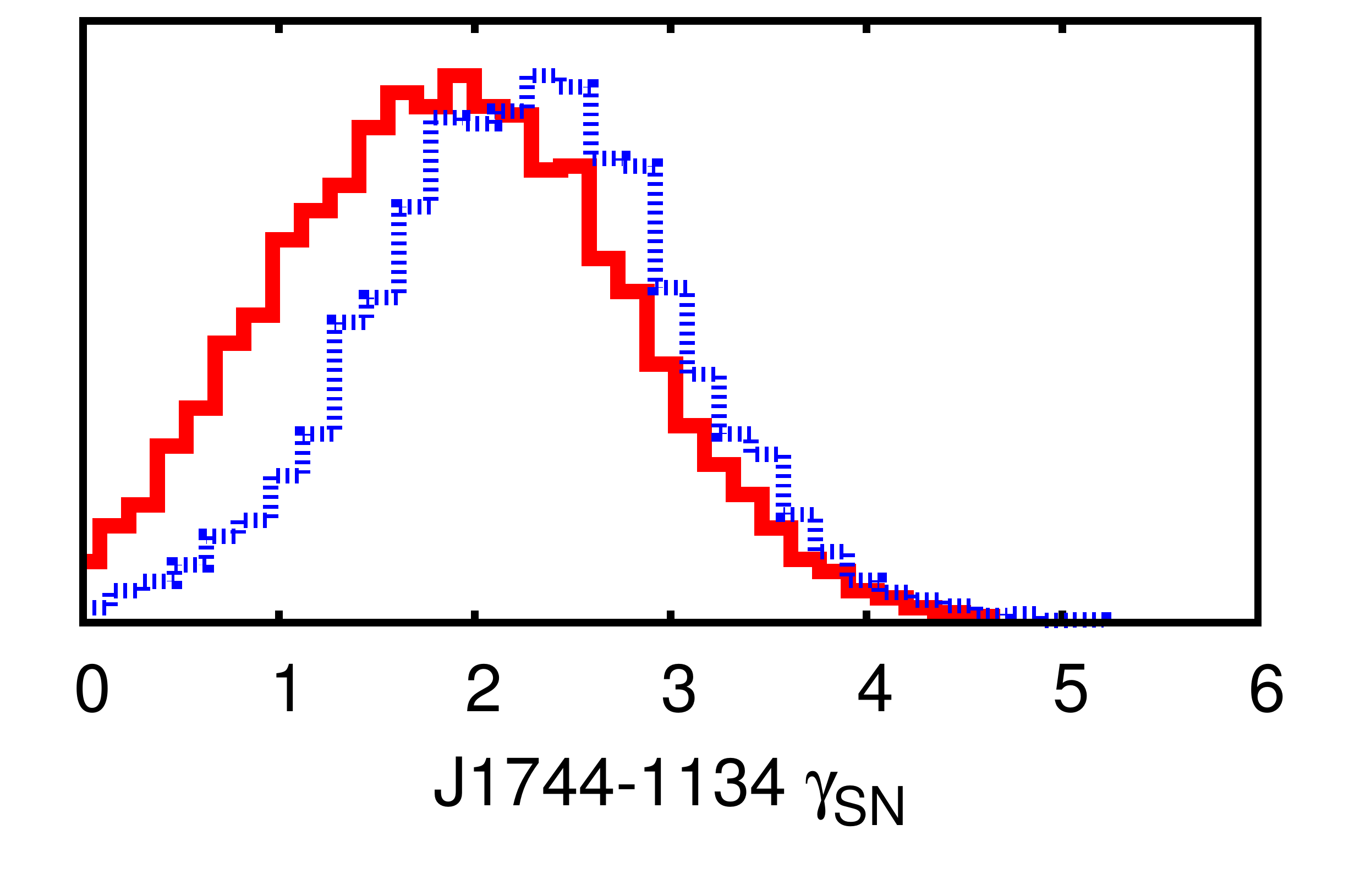} \\
\hspace{-0.5cm}
\includegraphics[width=60mm]{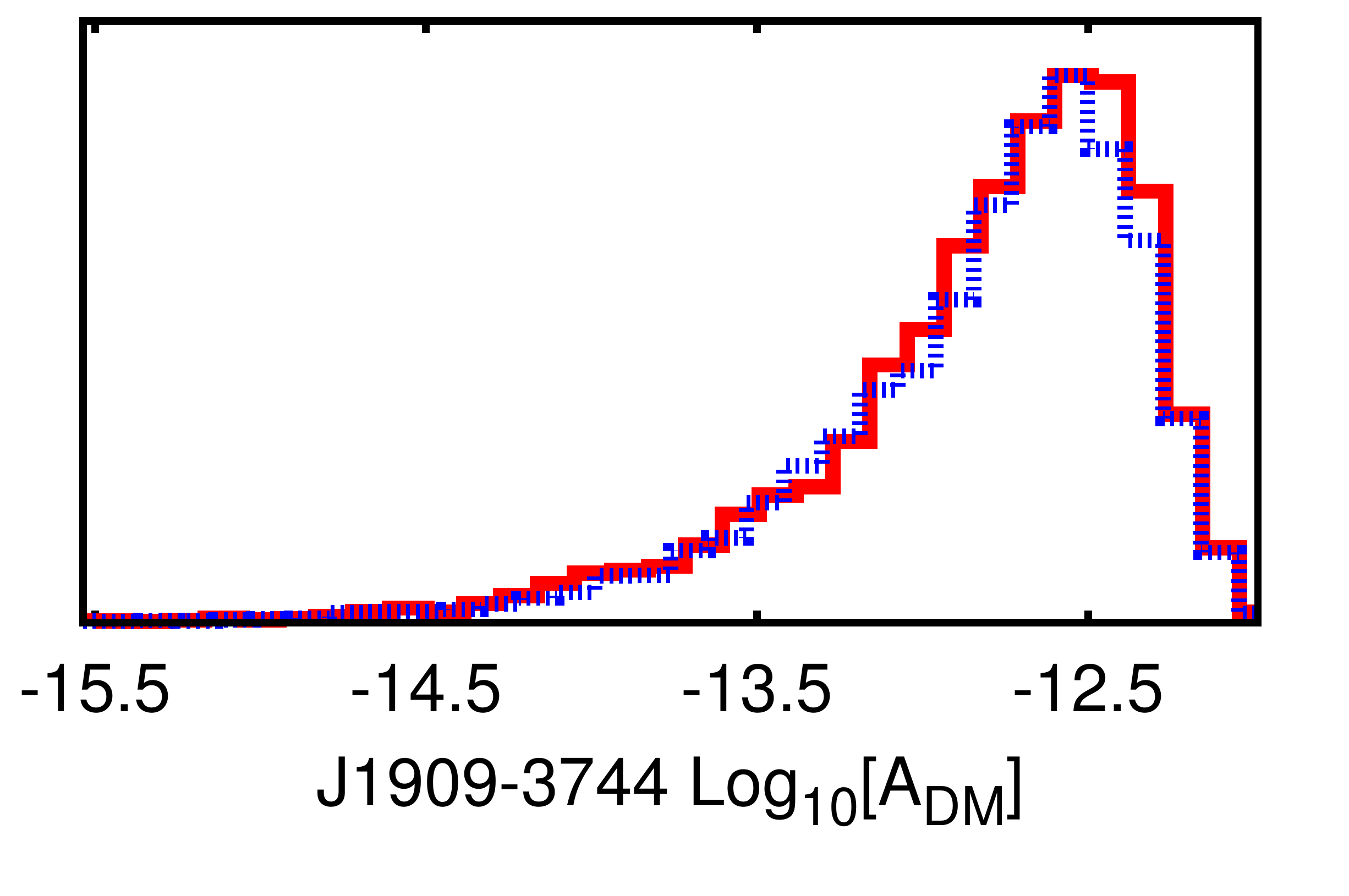}  &
\includegraphics[width=60mm]{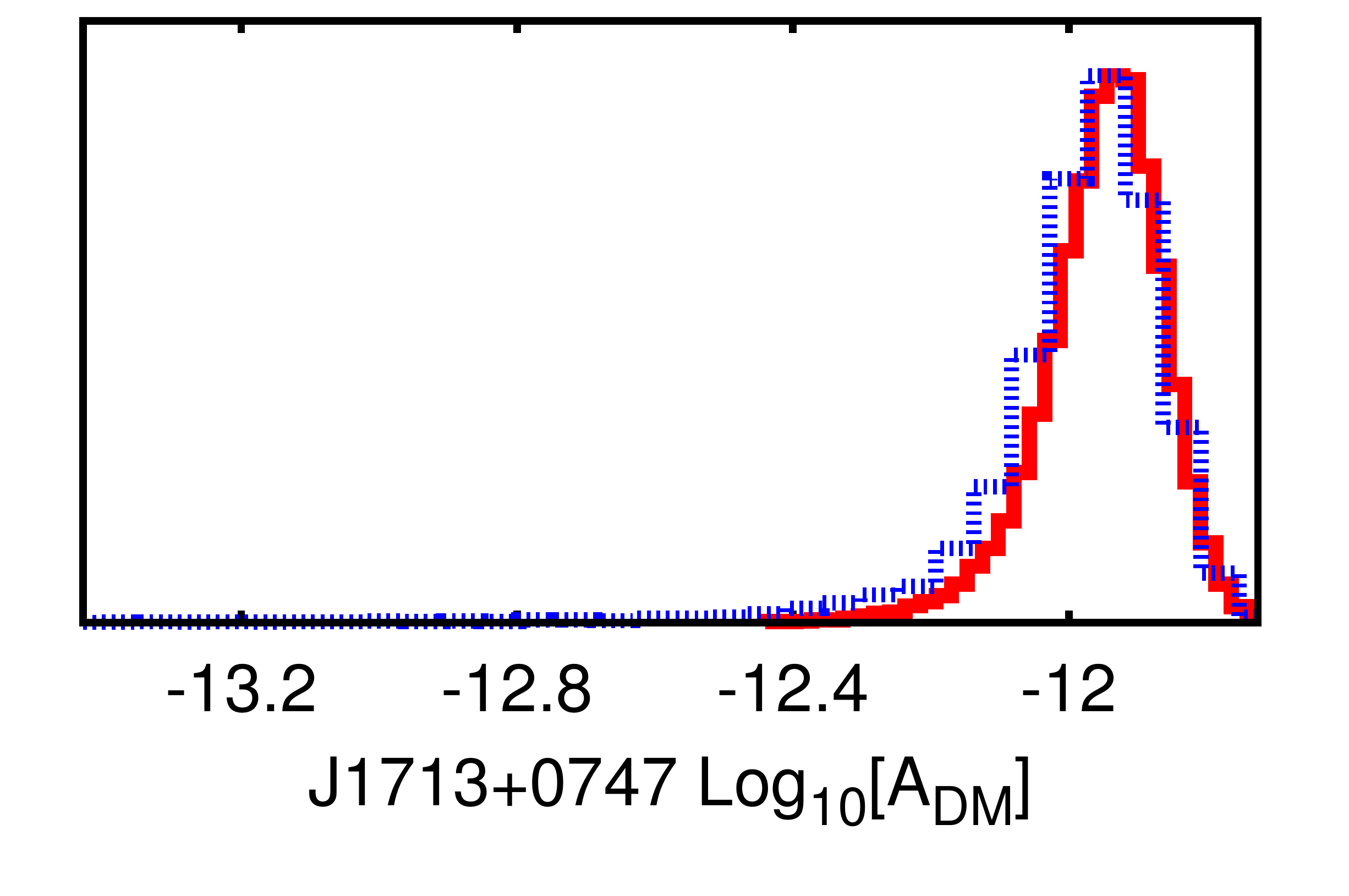} &
\includegraphics[width=60mm]{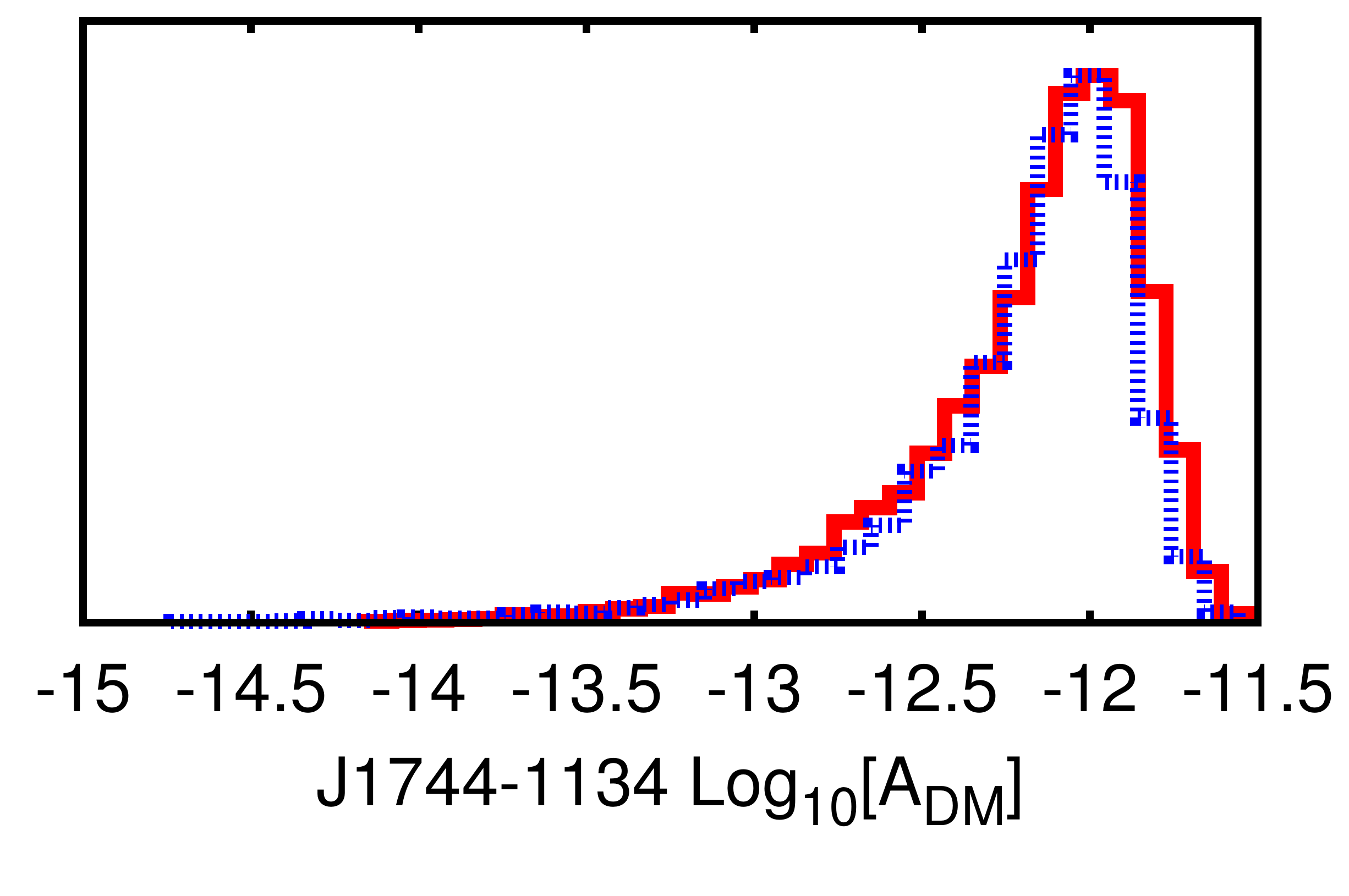} \\
\hspace{-0.5cm}
\includegraphics[width=60mm]{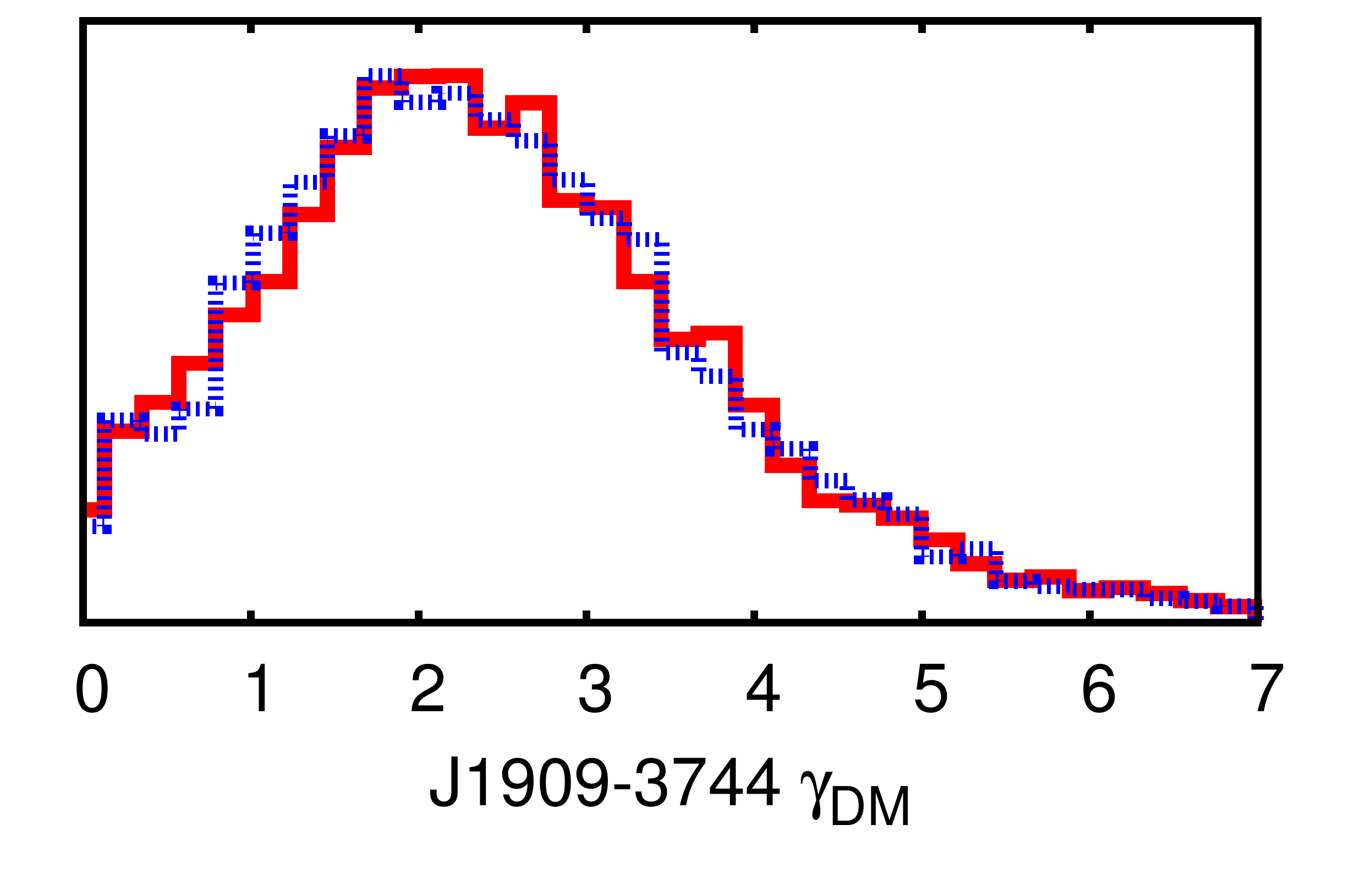}  &
\includegraphics[width=60mm]{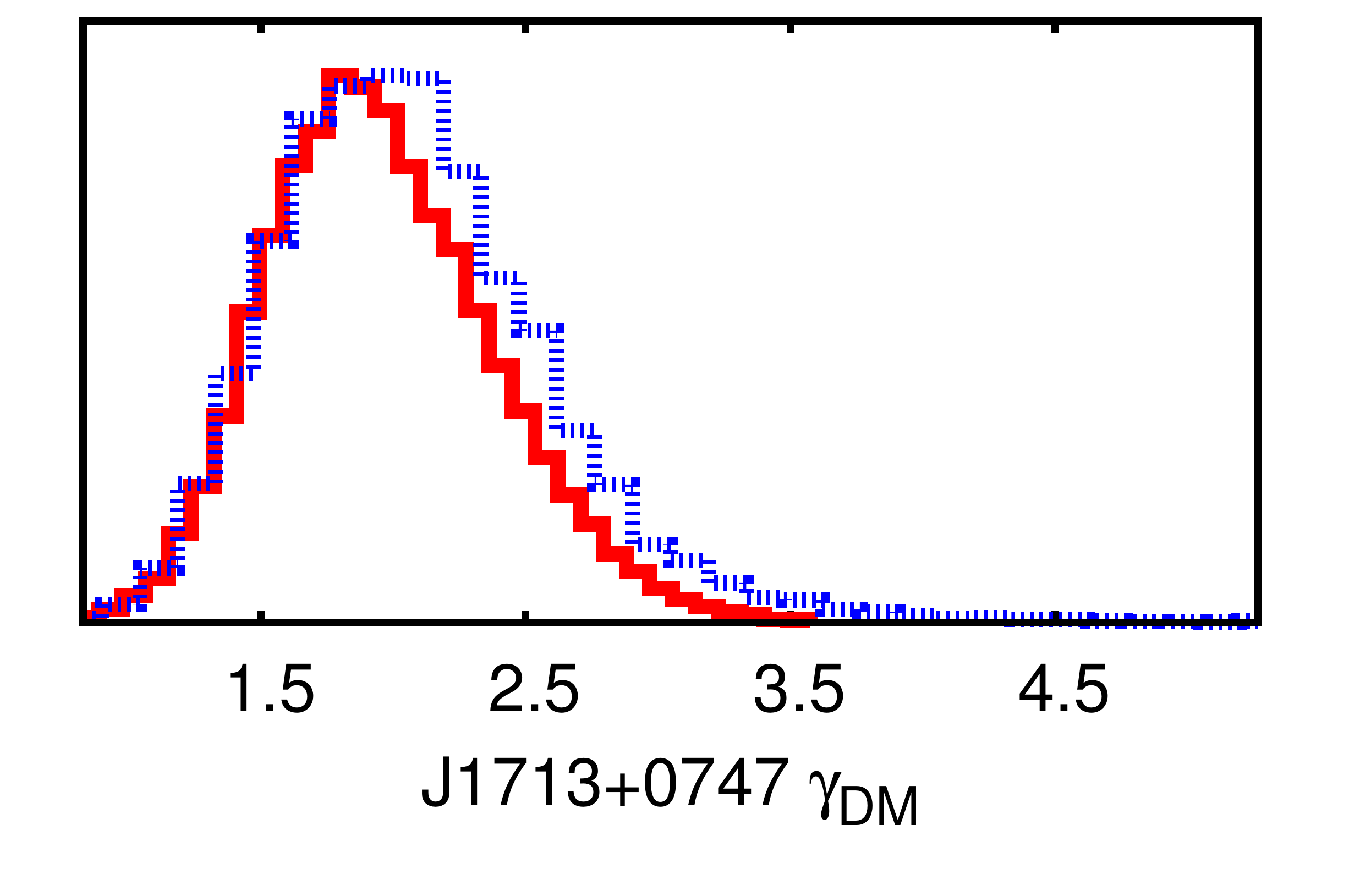}  &
\includegraphics[width=60mm]{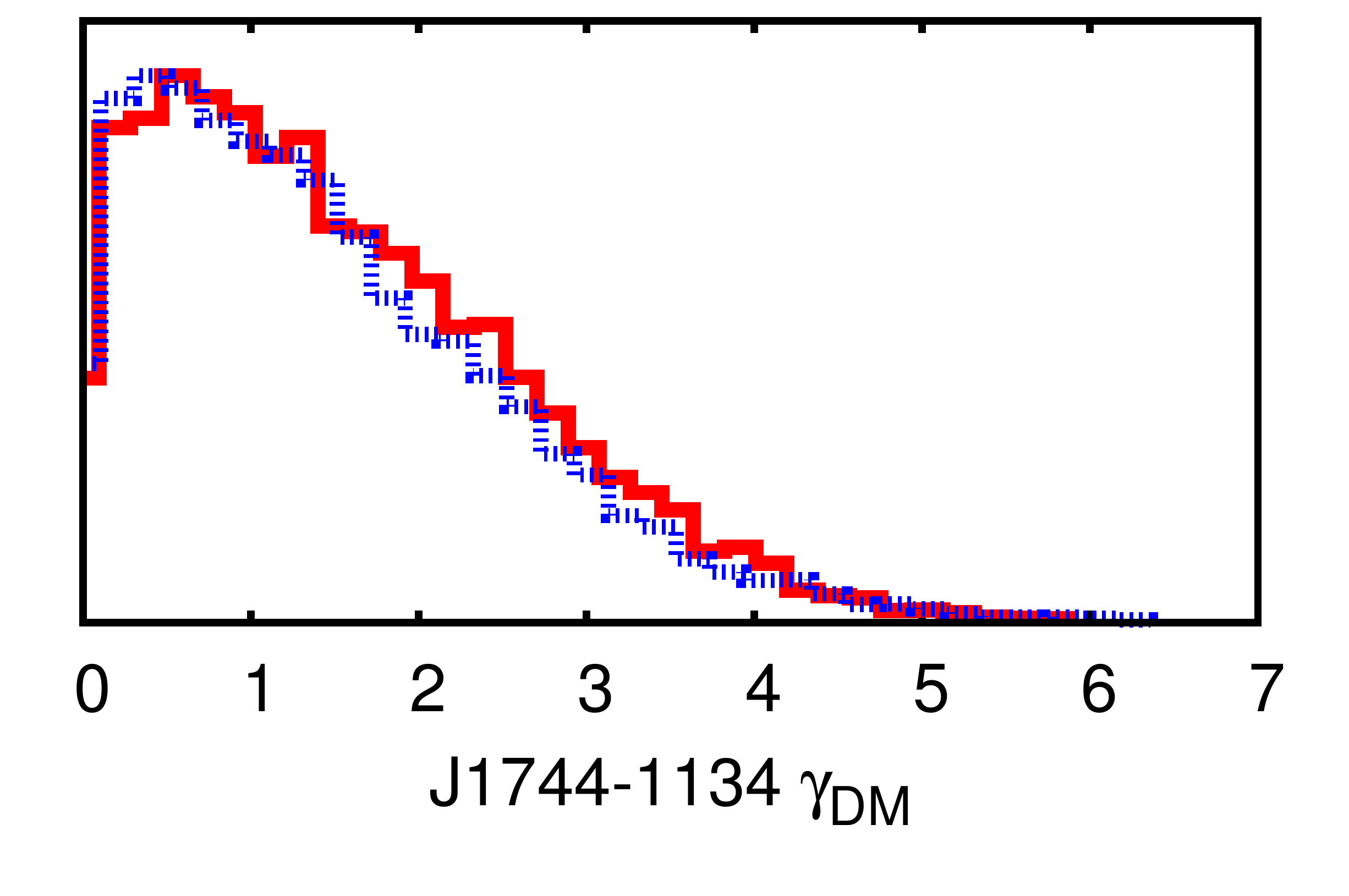}  \\
\end{array}$
\end{center}
\caption{Comparison of the 1-dimensional marginalised posterior probability distributions for PSRs (left to right) J1909$-$3744, J1713+0747, and J1744$-$1134. The y-axis in all plots represents probability. In each case we show the spin-noise and DM variation power law parameters for the full noise analysis (red solid lines) and 5-dimensional analysis where the TOA error bars have been pre scaled by the mean value of the EFAC/EQUAD parameters for each pulsar backend (blue dashed lines).  In both cases parameter estimates have been obtained using a uniform prior on the amplitude of the spin-noise and DM variations power law models. We also show the global EFAC parameters from the 5-dimensional analysis in each case.   We find the posteriors are consistent between the two sets of analysis.}
\label{Fig:5dcomp}
\end{figure*}

In order to check the validity of this simplification we performed a 5-dimensional analysis for each of the 6 pulsars in this dataset, fitting for power law intrinsic red noise and DM variations, in addition to a global EFAC parameter after adjusting the error bars using the mean values from D15.  The parameter estimates obtained are given in  Table ~\ref{Table:Pulsars}, and the 1-dimensional marginalised posteriors for J1909$-$3744, J1713+0747, and J1744$-$1134 from this analysis are shown in Fig. \ref{Fig:5dcomp}.  In each case we show the red noise and DM variation power law parameters for the full noise analysis (red) and 5-dimensional analysis (blue).  We also show the global EFAC parameter from the 5-dimensional analysis in each case.  We find the posteriors are consistent between the two sets of analysis.

Table ~\ref{Table:Pulsars} also lists the 95\% upper limit on a red noise process with a spectral index of $13/3$ at a reference frequency of $1\mathrm{yr^{-1}}$ for each of the 6 pulsars used in our analysis.  This limit was obtained when simultaneously fitting for the 5 intrinsic noise parameters for each pulsar in addition to the steep spectrum noise term.  The two pulsars with the most constraining upper limit are PSRs J1909$-$3744, and J1713+0747, consistent with the results obtained in Babak et al. (in prep.), with values of $\approx5\times10^{-15}$, and $\approx4\times10^{-15}$ respectively.

\section{Results}
\label{Section:Results}

\subsection{Limits on an Isotropic Stochastic GWB}
\label{Section:sgwblimits}
\subsubsection{Bayesian approach}
\label{sec:bayesian}

In Table \ref{Table:parameters} we list the complete set of free parameters that we include in the different models used in the analysis presented in this section, along with the prior ranges used for those parameters.

\begin{table*}
\scriptsize
\centering
\caption{Free parameters and prior ranges used in the Bayesian analysis.}
\centering
\begin{tabular}{l l l c}
\hline\hline
Parameter 	& Description		& 	Prior range	&	\\
\hline
White noise			&				&		&	\\
$\alpha$				& 	Global EFAC		&	uniform in [0.5 , 1.5]		&	1 parameter per pulsar (total 6) \\
\hline
Spin-noise				&	&	&	\\
$A_\mathrm{SN}$	&	Spin-noise power law amplitude	&	uniform in [$10^{-20}$ , $10^{-10}$]	&	1 parameter per pulsar (total 6)	\\
$\gamma_\mathrm{SN}$	&	Spin-noise power law spectral index	&	uniform in $[0 , 7]$ &	1 parameter per pulsar (total 6)	\\
\hline
DM variations	&	&	&	\\
$A_\mathrm{DM}$	&	DM variations power law  amplitude	&	uniform in [$10^{-20}$ , $10^{-10}$]	&	1 parameter per pulsar (total 6)	\\
$\gamma_\mathrm{DM}$	&	DM variations power law spectral index	&	uniform in $[0 , 7]$	&	1 parameter per pulsar (total 6)	\\
\hline
Common noise	&	&	&	\\
$A_\mathrm{CN}$	&	Uncorrelated common noise power law amplitude			&	uniform in [$10^{-20}$ , $10^{-10}$] &	1 parameter for the array	\\
$\gamma_\mathrm{CN}$	&	Uncorrelated common noise power law spectral index	&	uniform in $[0 , 7]$	&	1 parameter for the array	\\
$A_\mathrm{clk}$	&	Clock error power law amplitude			&	uniform in [$10^{-20}$ , $10^{-10}$] &	1 parameter for the array	\\
$\gamma_\mathrm{clk}$	&	Clock error power law  spectral index	&	uniform in $[0 , 7]$	&	1 parameter for the array	\\
$A_\mathrm{eph}$	&	Solar System ephemeris error power law amplitude			&	uniform in [$10^{-20}$ , $10^{-10}$]	&	3 parameters for the array	(x, y, z) \\
$\gamma_\mathrm{eph}$	&	Solar System ephemeris error power law spectral index	&	uniform in $[0 , 7]$	&	3 parameters for the array	(x, y, z)	\\
\hline
Stochastic GWB	&	&	&	\\
$A$						&	GWB power law amplitude		&	uniform in [$10^{-20}$ , $10^{-10}$] &	1 parameter for the array \\
$\gamma$				&	GWB power law spectral index		& 	uniform in $[0 , 7]$						 &	1 parameter for the array \\
$\rho_i$						&	GWB power spectrum coefficient at frequency $i/T$		&	uniform in [$10^{-20}$ , $10^{0}$] & 1 parameter for the array per frequency in \\
& & &unparameterised GWB power spectrum model	(total 20)\\
\hline
Stochastic background angular correlation function	&	&	&	\\
$c_{1\dots4}$						& Chebyshev polynomial coefficient 	&	uniform in $[-1 , 1]$					&	see Eq.~(\ref{e:gamma_Chebyshev})\\
$\Gamma_{IJ}$				& Correlation coefficient between pulsars (I,J) & uniform in $[-1 , 1]$	& 1 parameter for the array per unique pulsar pair (total 15)\\
\hline
\hline
\end{tabular}
\label{Table:parameters}
\end{table*}

\begin{table}
\centering
\caption{95\% upper limits on the amplitude of an isotropic stochastic GWB obtained for different models at a reference frequency of $1\mathrm{yr^{-1}}$.}
\centering
\begin{tabular}{l c}
\hline\hline
Model & 95\% upper limit\\[0.5ex]
	  &		$(\times10^{-15})$ \\
\hline
\hline
Bayesian Analysis & \\
\hline
Fixed Noise - Fixed Spectral Index & $1.7$\\
Varying Noise - Fixed Spectral Index & $3.0$\\
Additional Common Signals - Fixed Spectral Index & $3.0$\\
\hline
Fixed Noise - Varying Spectral Index & $8.0$\\
Varying Noise - Varying Spectral Index & $13$\\
Additional Common Signals - Varying Spectral Index & $13$\\
\hline
\hline
Frequentist Analysis & \\
\hline
Fixed Noise - Fixed Spectral Index & $2.1$\\
\hline
\hline
Simulations - Varying Spectral Index & \\
\hline
White Noise Only & $4.3$\\
White and intrinsic spin-noise & $7.2$\\
White and intrinsic spin-noise and DM variations & $12$\\
\hline
\hline
\end{tabular}
\label{Table:PLResults}
\end{table}
\begin{table}
\centering
\caption{95\% upper limits obtained for common noise terms at a reference frequency of $1\mathrm{yr^{-1}}$.}
\centering
\begin{tabular}{l c}
\hline\hline
Model & 95\% upper limit\\[0.5ex]
	  &		$(\times10^{-15})$ \\
\hline
\hline
Additional Common Signals - Varying Spectral Index  & \\
\hline
$A_\mathrm{CN}$ & $13$ \\
$A_\mathrm{clk}$ & $11$\\
$A_\mathrm{eph}$ (x) & $65$\\
$A_\mathrm{eph}$ (y) & $14$\\
$A_\mathrm{eph}$ (z) & $25$\\
\hline
\hline
\end{tabular}
\label{Table:CNResults}
\end{table}
When parameterising the stochastic GWB using the power law model in Eq.~(\ref{Sh}), we run two parallel sets of analyses: in the first set we fix $\gamma=13/3$, consistent with a stochastic GWB  dominated by SMBHBs; in the second set we allow $\gamma$ to vary freely within a prior range of [0,7]. In both cases we consider three different models: 

\begin{enumerate}
  \item with the intrinsic timing noise for each pulsar fixed at the maximum likelihood values given in Table \ref{Table:Pulsars};
  \item with the intrinsic timing noise for each pulsar allowed to vary;
  \item as in (ii), but including additional common uncorrelated red noise, a clock error, and errors in the Solar System ephemeris as discussed in Section \ref{Section:AdditionalNoise}.
\end{enumerate}


The $95\%$ upper limits for the amplitude of an isotropic stochastic GWB in the six different models are listed in Table \ref{Table:PLResults}.  In Table \ref{Table:CNResults} we then list the $95\%$ upper limits for the additional common noise terms that were included in model (iii), when allowing the spectral indicies to vary.  All upper limits in this section are reported at a reference frequency of $1\mathrm{yr^{-1}}$.
\begin{figure*}
  \begin{center}$
    \begin{array}{cc}
      \hspace{-1cm}
      \includegraphics[width=90mm]{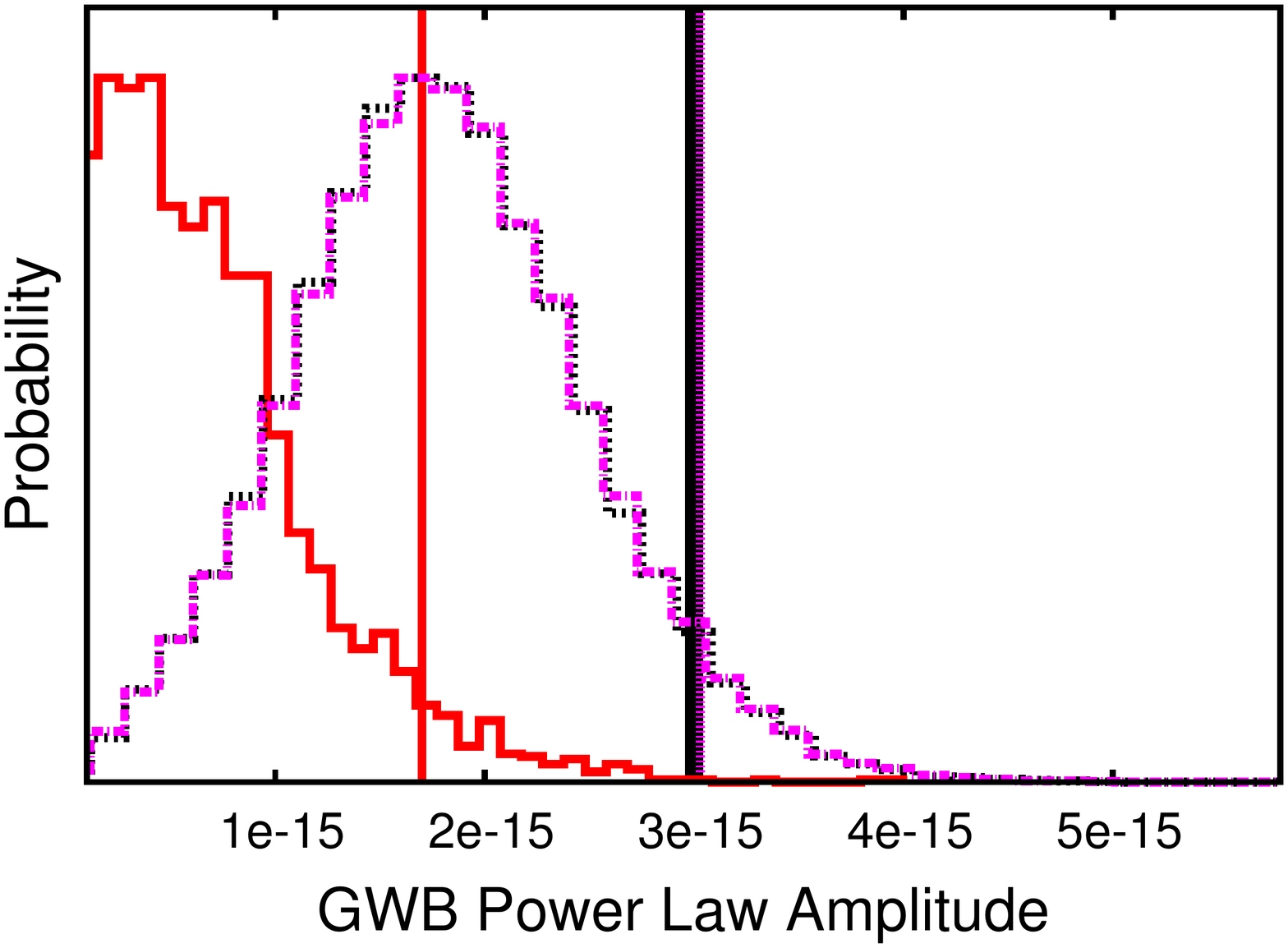}  &
      \hspace{-1cm}
      \includegraphics[width=90mm]{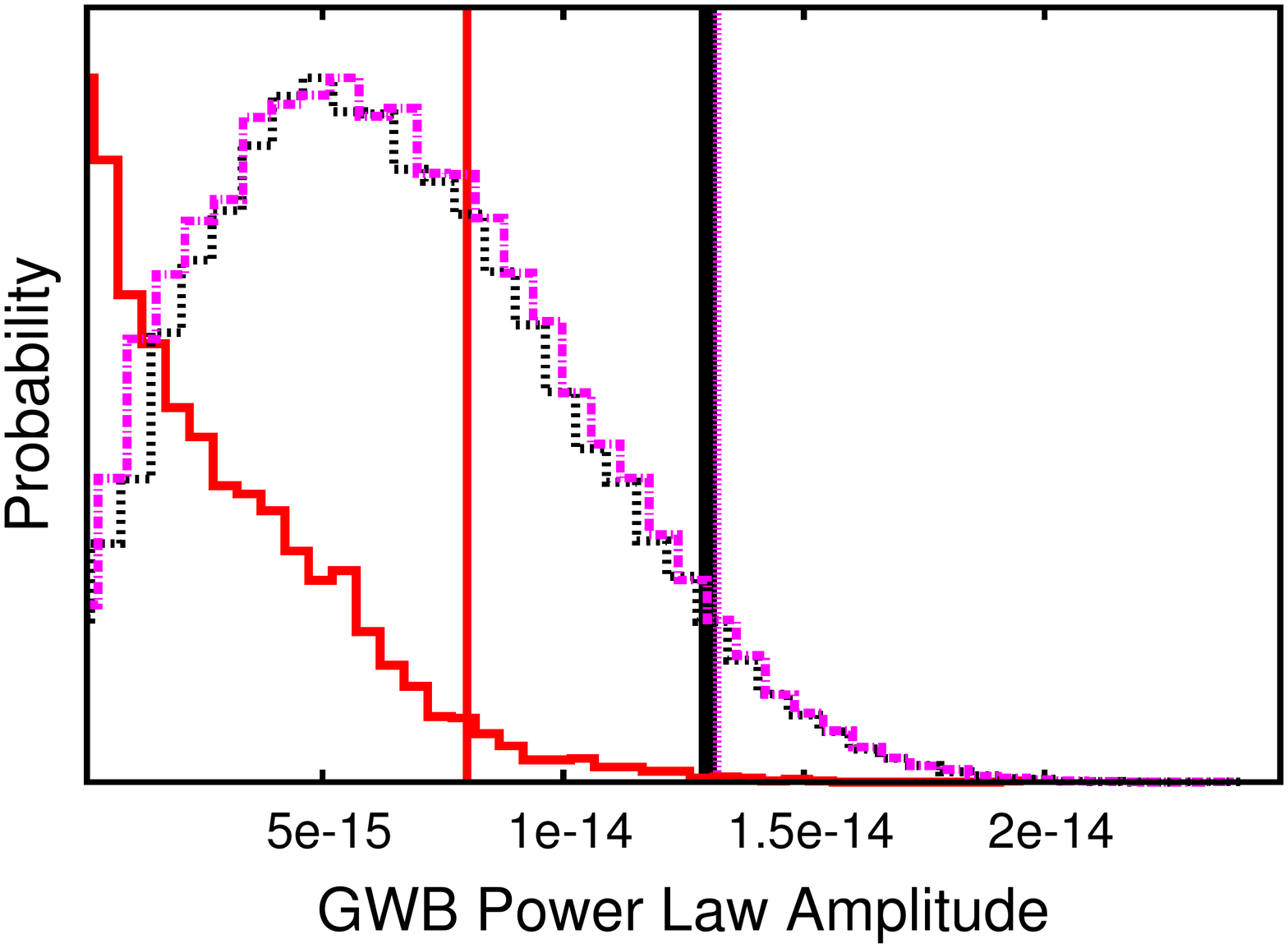}  \\
    \end{array}$
  \end{center}
  \vspace{-1cm}
  \caption{One-dimensional marginalised posterior parameter estimates for the amplitude of a correlated GWB in the 6 pulsar dataset presented in this paper when: (left) Fixing the spectral index of the power law to a value of $13/3$, consistent with a background dominated by a population of SMBHB, and (right) when marginalising over the spectral index given a prior range of [0,7].  In each case we show the posterior given: (red solid line) Fixed intrinsic noise parameters for each pulsar, where the values of the parameters are given by the maximum likelihood estimates listed in Table \ref{Table:Pulsars}, (black dotted line) varying noise parameters for each pulsar, and (magenta dashed line) varying noise parameters for each pulsar, and additionally including a common uncorrelated red noise process, clock errors, and errors in the Solar System ephemeris in the model.  Vertical lines in each case represent the $95\%$ upper limits for each model.}
  \label{figure:PowerLaws}
\end{figure*}

The one-dimensional marginalised posteriors for the amplitude of the GWB for each of these models are shown in Fig. \ref{figure:PowerLaws}.  We find that in both the fixed, and varying spectral index model for the GWB, limits placed under the assumption of fixed intrinsic timing noise are erroneously more stringent than when the noise is allowed to vary, by a factor $\sim 1.8$ and $\sim 1.6$ respectively. This is a direct result of using values for the intrinsic noise that have been obtained from analysis of the individual pulsars, in which the red spin-noise signal, and any potential GWB signal will be completely covariant.  The natural consequence is that fixing the properties of the intrinsic noise to those obtained from the single pulsar analysis will always push the upper limit for the GWB lower in a subsequent joint analysis.

Both of the most recent isotropic GWB limits from pulsar timing have been set using frequentist techniques, either by performing a fixed noise analysis \citep{2013ApJ...762...94D}, or using simulations \citep{2013Sci...342..334S}, obtaining 95\% upper limits of $7\times10^{-15}$ and  $2.4\times10^{-15}$ respectively.  In both cases, therefore, the analysis performed was fundamentally different to the Bayesian approach presented in this work.  As such it is difficult to compare our results directly, or to ascertain the effect of fixing the intrinsic noise parameters on limits obtained using those methods.


\begin{figure}
  \begin{center}$
    \begin{array}{c}
      \hspace{-1cm}\includegraphics[width=90mm]{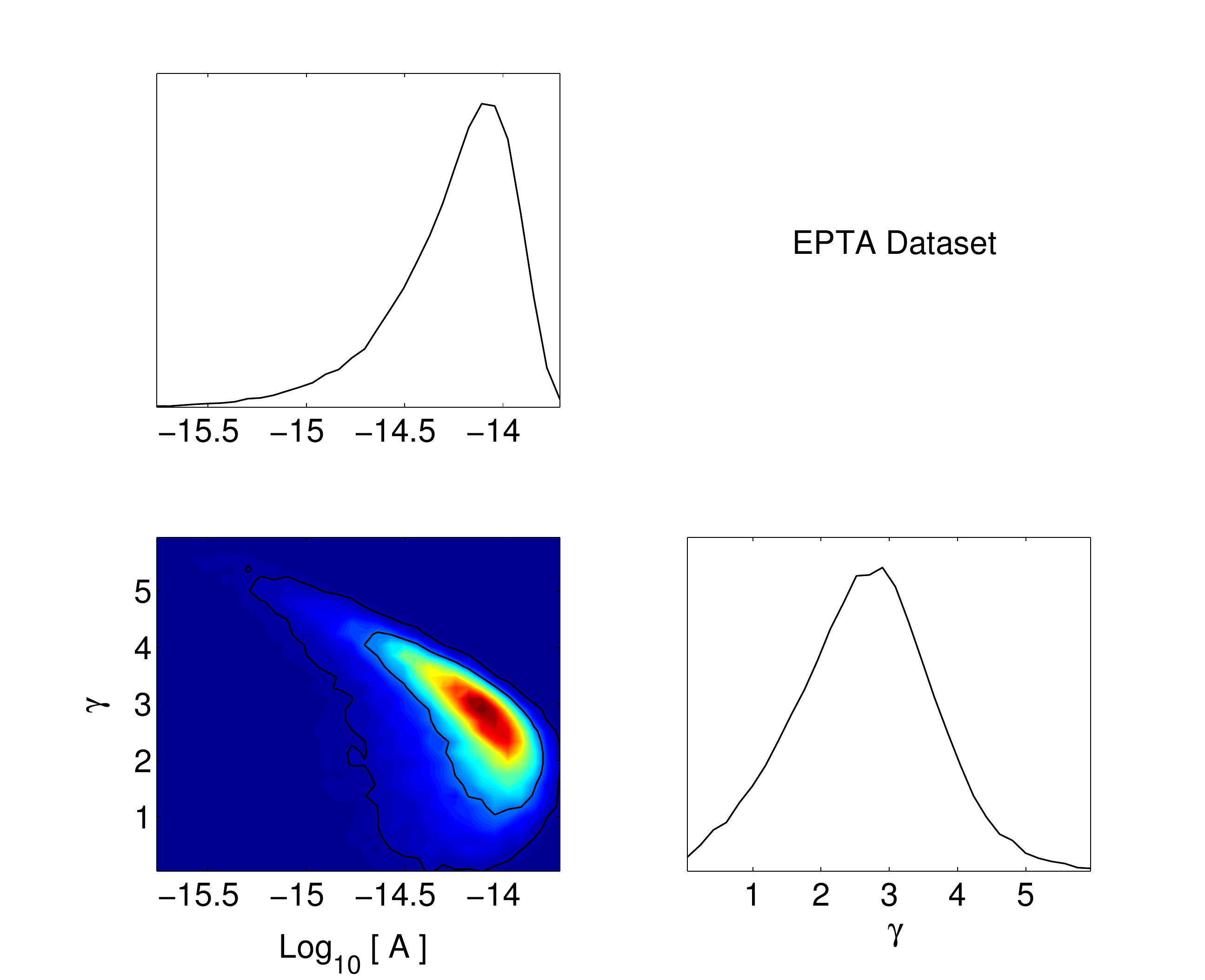}
    \end{array}$
  \end{center}
  \caption{One and two-dimensional marginalised posterior parameter estimates for the amplitude and spectral index of a correlated GWB in the 6 pulsar dataset presented in this paper when varying the intrinsic noise parameters for each pulsar.  The amplitude and spectral index are highly correlated, resulting in a significantly higher upper limit when allowing the spectral index to vary, as opposed to fixing it at a value of $13/3$.}
  \label{figure:2DGWB}
\end{figure}

\begin{figure}
  \begin{center}$
    \begin{array}{c}
      \hspace{-0.5cm}
      \includegraphics[width=80mm]{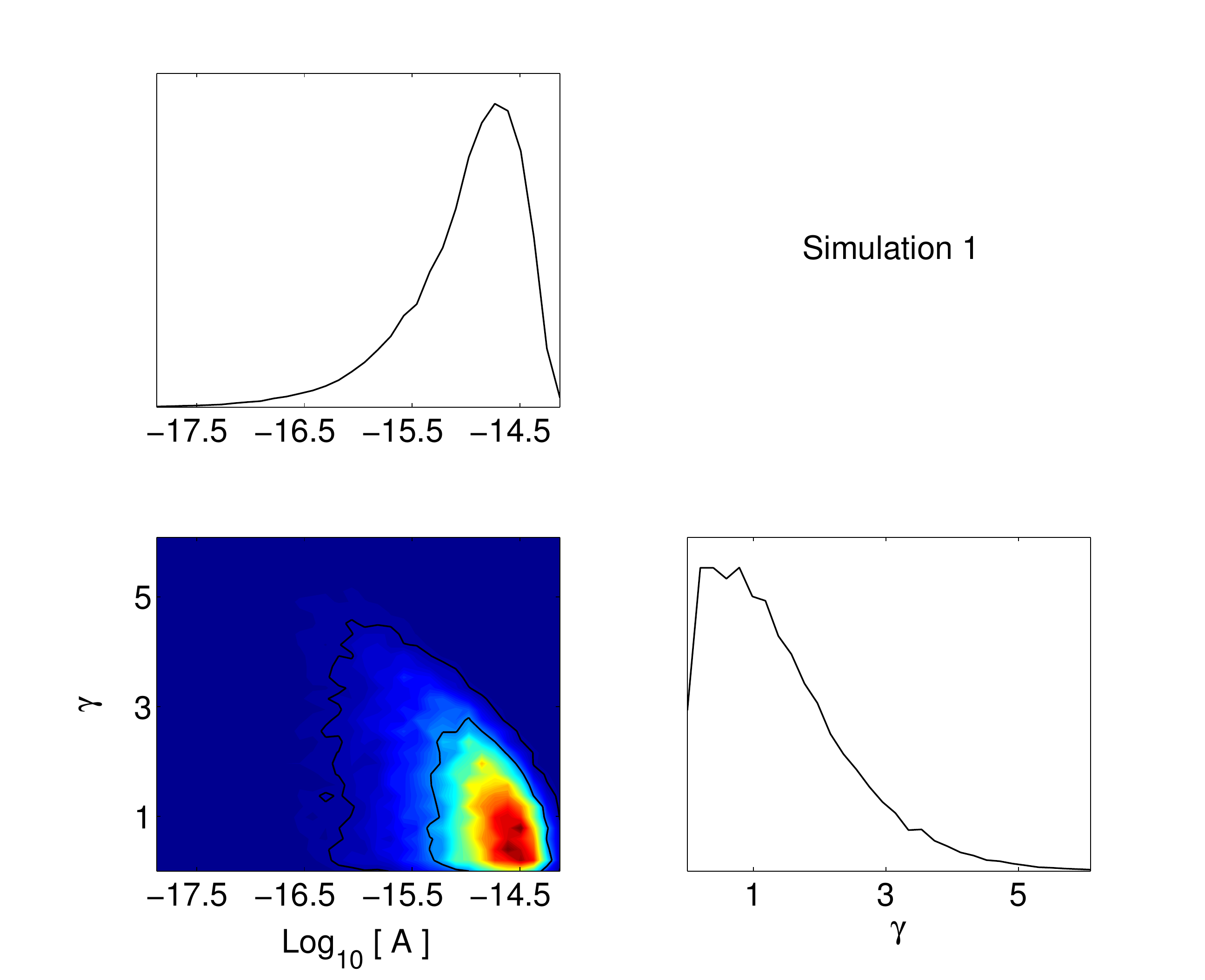}\\
      \hspace{-0.5cm}
      \includegraphics[width=80mm]{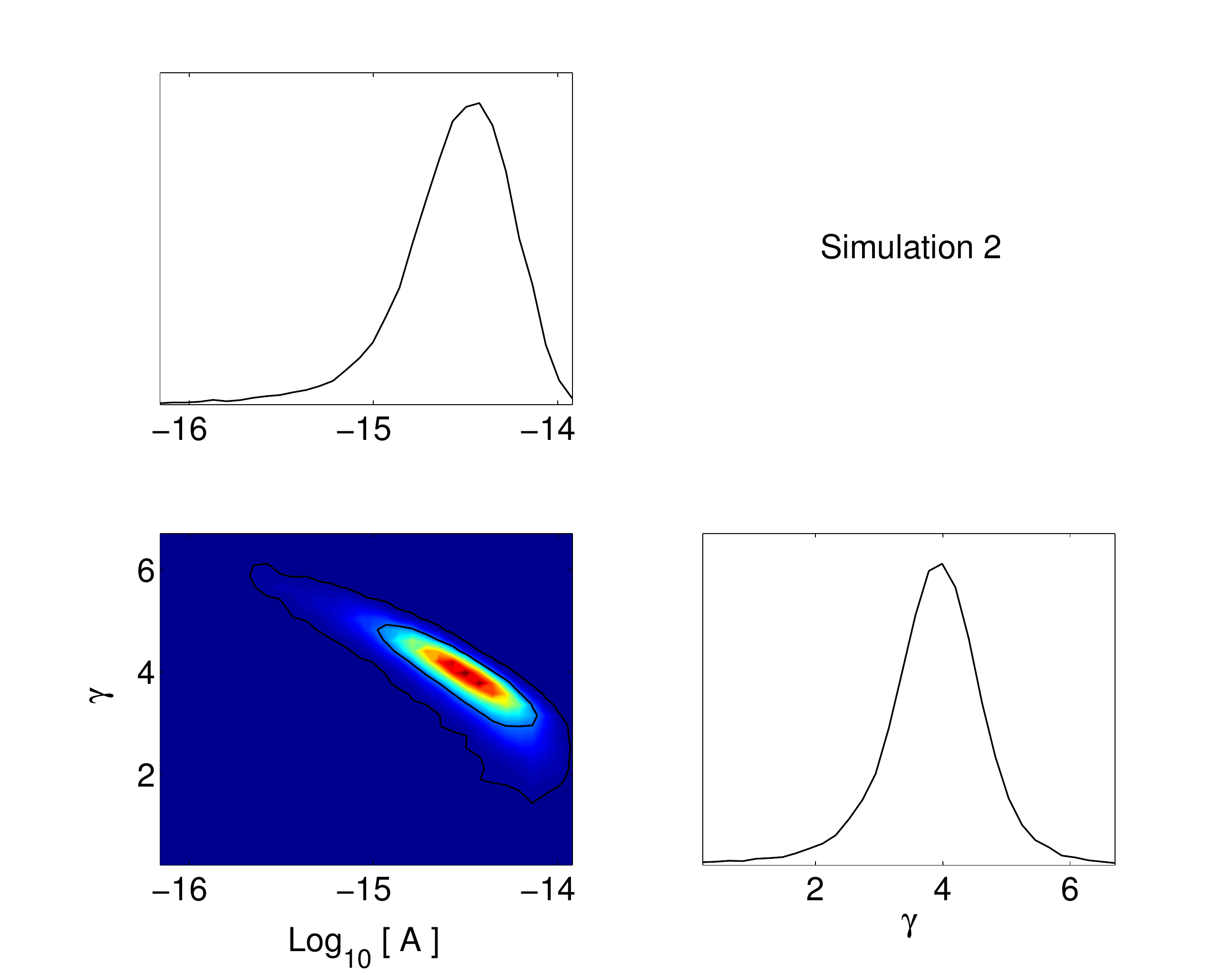}\\
      \hspace{-0.5cm}
      \includegraphics[width=80mm]{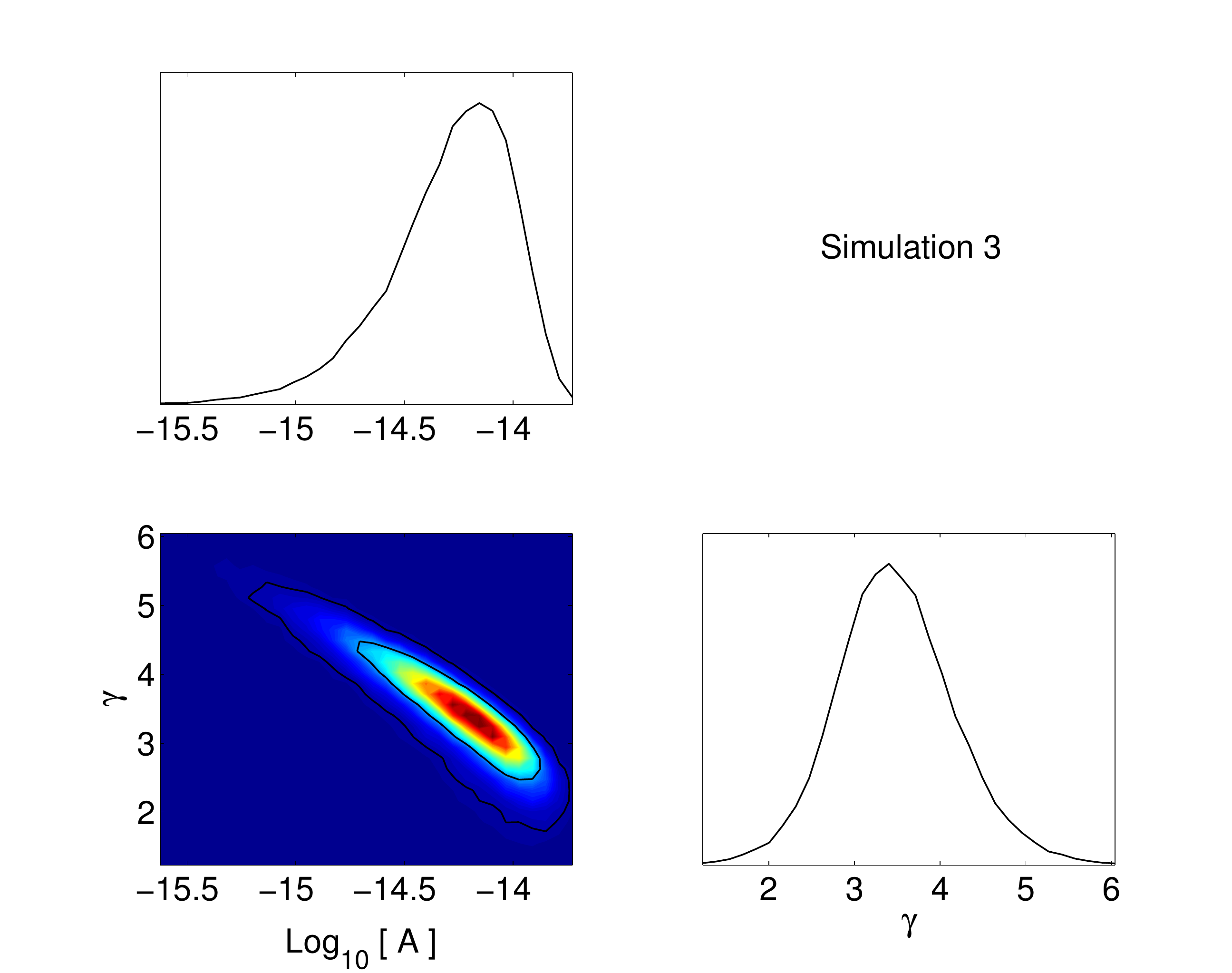}\\
    \end{array}$
  \end{center}
  \caption{One and two-dimensional marginalised posterior parameter estimates for the amplitude and spectral index of a correlated GWB in 3 simulated datasets, including (Top) White noise only, (Middle) White and intrinsic spin-noise only, and (Bottom) White noise, intrinsic spin-noise, and dispersion measure variations.  In each case we use the TOAs from the 6 pulsars used in the GWB analysis presented in this work, and use the maximum likelihood noise parameters from Table \ref{Table:Pulsars} when constructing the simulations.  The upper limits obtained in each case are $4.3\times 10^{-15}$, $7.2\times 10^{-15}$, and $1.2\times 10^{-14}$.  We find both the upper limit, and the form of the posterior to be consistent between the third simulation and the real dataset.  In both cases we are simply recovering our uniform prior on the amplitude of the GWB signal at small amplitudes, before the data begins to place constraints on the upper limit at large amplitudes. }
  \label{figure:NoGWBSims}
\end{figure}


The most recent limit placed when allowing the intrinsic noise parameters of the pulsars to vary is given by \cite{2011MNRAS.414.3117V}, in which a 95\% upper limit of $A=6\times10^{-15}$ was obtained at a spectral index of $13/3$.  Our model (ii) is most comparable to this analysis, in which we obtain a 95\% upper limit of $A=3\times10^{-15}$, an improvement of a factor of two.  This translates into a limit on $\Omega_\mathrm{gw}(f)h^2 = 1.1\times 10^{-9}$ at 2.8 nHz. We confirm this result by analysing the 2015 EPTA dataset with model (ii) using an independent code\footnote{https://github.com/stevertaylor/NX01}, which makes use of the PAL, parallel-tempered adaptive MCMC sampler\footnote{https://github.com/jellis18/PAL2} which explores the parameter space in a fundamentally different way to \textsc{MultiNest}, and obtain a consistent 95\% upper limit. 

Finally for model (iii) when including additional common or correlated terms in the analysis we find the extra parameters have a negligible impact on our sensitivity, with consistent upper limits obtained in both the fixed and varying spectral index models.

We find the upper limits for the uncorrelated common red noise model to be consistent with those obtained for the GWB, however we find the upper limit for the clock error signal to be slightly lower, with $A_\mathrm{clk} < 1.1\times10^{-14}$ compared to $A_\mathrm{CN} < 1.3\times10^{-14}$.  This is to be expected however, as the clock is handled coherently across all pulsars, whereas the GWB and common uncorrelated red noise signals are handled incoherently, as such we have greater sensitivity when searching for the clock signal and obtain a correspondingly lower limit for the amplitude.

The limits on the different errors originating in the Solar System emphemeris can be understood given the components of the unit vector from the SSB toward the two pulsars that contribute most to our analysis, PSRs J1713+0747 and J1909$-$3744, which are given by $(-0.20, -0.97, +0.14)$ and $(+0.24, -0.75, -0.61)$ respectively.  Both PSRs J1909$-$3744 and J1713+0747 contribute very little to the constraints on the ephemeris in the $x$ direction, and so here we see the greatest degradation in the limit on the amplitude, while in the $y$ direction PSR J1713+0747 contributes almost fully and so the limit we obtain is only slightly worse than that obtained for the GWB and uncorrelated common noise terms.

We consider model (iii) to be the most robust analysis presented in this paper, and so conclude that the 95\% upper limit provided by our dataset on a power law GWB is $A<3.0\times10^{-15}$ at $\gamma=13/3$ , and $A<1.3\times10^{-14}$ when marginalising over spectral index.

That the upper limit is considerably higher in the varying spectral index model can be understood from the two-dimensional posterior distribution for $A$ and $\gamma$ in Fig. \ref{figure:2DGWB}. Here we see the clear correlation between the two quantities; as we will see below, our PTA is most sensitive at frequencies $\ll1\,\mathrm{yr}^{-1}$, meaning that for a single power-law spectrum, the flatter the spectral index, the less stringent the limit on $A$.

As a consistency check on this result we perform a set of three simulations using the sampled time stamps present in the actual 6 pulsar dataset.  In the first simulation we include only a white noise component with an amplitude determined using the TOA uncertanties from the real dataset.  In the second simulation we then add an intrinsic spin-noise component, with amplitudes and spectral indices equal to the maximum likelihood values presented in Table \ref{Table:Pulsars}.  Finally in the third simulation we also include DM variations, where as with the intrinsic spin-noise we use the maximum likelihood values in Table \ref{Table:Pulsars} to set the amplitudes and spectral indices.  Critically in all the simulations we include no correlated GWB term, and so in each case we expect to recover only our uniform prior on the amplitude of the GWB included in our model.

The one and two-dimensional marginalised posterior parameter estimates for the amplitude and spectral index of the GWB from each of the three simulations are shown in Fig. \ref{figure:NoGWBSims}.  We obtain upper limits of $4.3\times 10^{-15}$, $7.2\times 10^{-15}$, and $1.2\times 10^{-14}$ in each case respectively. In all the one-dimensional posterior distributions for the amplitude of the signal we are simply recovering our prior, such that the probability is proportional to the amplitude of the signal below some limit set by the data.  In the case of the third simulation, which is most similar to the real dataset, both the upper limit, and the form of the posterior is consistent with the results presented in Fig. \ref{figure:2DGWB} and Table \ref{Table:PLResults}.


\begin{figure}
  \begin{center}$
    \begin{array}{c}
        \hspace{-1.8cm}\includegraphics[width=110mm]{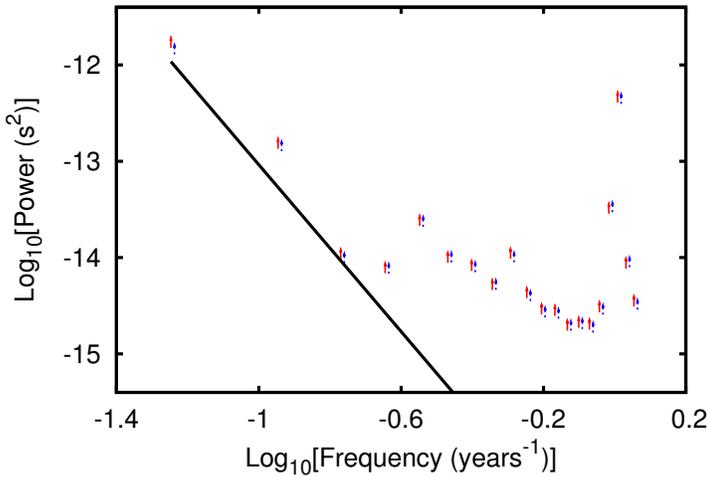}\\
    \end{array}$
  \end{center}
   \vspace{-1cm}
  \caption{(Top) 95\% upper limits from an unparameterised power spectrum analysis for a correlated GWB (red points), and uncorrelated common red noise process (blue points) for the 6 pulsar dataset described in Section \ref{Section:Dataset} obtained while varying the intrinsic noise parameters for each pulsar. The upper limit obtained from a power law analysis of the GWB at a spectral index of $13/3$ of $3\times10^{-15}$ is overplotted as a straight line.  Frequencies were included from $1/T$ up to $20/T$, with $T=17.66$~yr. Beyond these frequencies the data provides increasingly poorer constraints on a steep red noise process, and so we do not consider higher frequency terms in our model.  Both the correlated and uncorrelated power spectrum are completely consistent with one another at all frequencies.  The difference in the log Evidence between the correlated and uncorrelated models was $0.2\pm0.3$, indicating no support for the presence of a correlated signal in the dataset.   }
  \label{figure:PowerSpectrum}
\end{figure}

\begin{figure}
  \begin{center}$
    \begin{array}{c}
      \hspace{-1cm}\includegraphics[width=90mm]{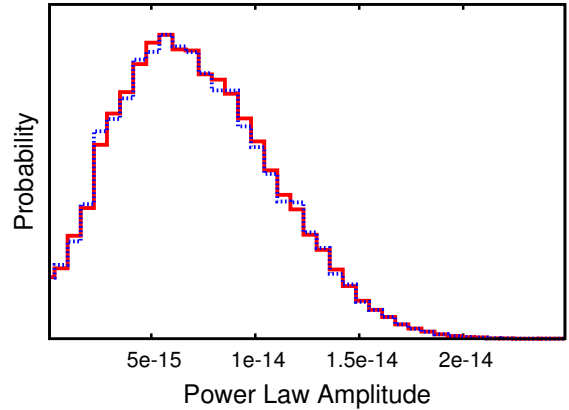}  \\
    \end{array}$
  \end{center}
   \vspace{-1cm}
  \caption{One-dimensional marginalised posterior parameter estimates for the amplitude of a correlated GWB (red solid line) and an uncorrelated common red noise model (blue dashed line) in the 6 pulsar dataset presented in this paper when marginalising over the spectral index given a prior range of [0,7]. Posteriors were obtained when varying the intrinsic noise parameters, and including only either the GWB, or common uncorrelated terms.  We find the upper limits to be completely consistent with one another, and obtain a change in the log evidence of $-1.0 \pm 0.5$ for the GWB model over the uncorrelated common red noise model, suggesting no strong support for either.}
  \label{figure:CommonGWBComp}
\end{figure}
\begin{figure}
  \begin{center}$
    \begin{array}{c}
      \hspace{-1cm}\includegraphics[width=90mm]{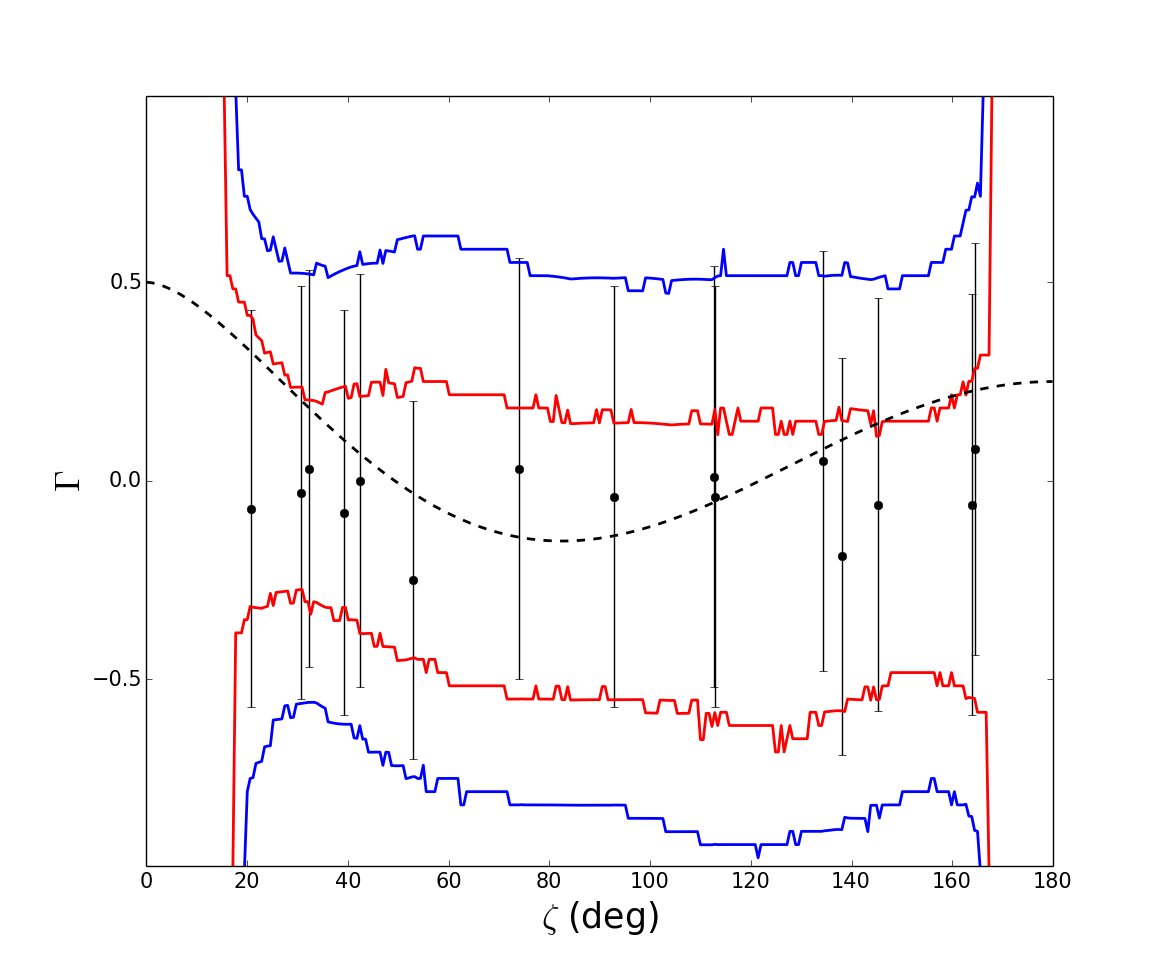}\\
    \end{array}$
  \end{center}
  \caption{The recovered correlation between pulsars as a function of angular separation on the sky for a power law noise process. The red and blue lines represent the $68\%$ and $95\%$ confidence intervals for the correlation function when modelled by the lowest 4 Chebyshev polynomials, while the individual points are the mean correlation coefficient with $1\sigma$ uncertainty for each pulsar pair when fitting without assuming a smooth model.  The Hellings-Downs correlation is represented by the dotted line.  For both models the correlation of a common power law model between pulsars is consistent with effectively all possible values (the range [-1,1]).}
  \label{figure:Correlation}
\end{figure}

\begin{figure*}
\begin{center}$
\begin{array}{cc}
\hspace{-1.0cm}
\includegraphics[width=90mm]{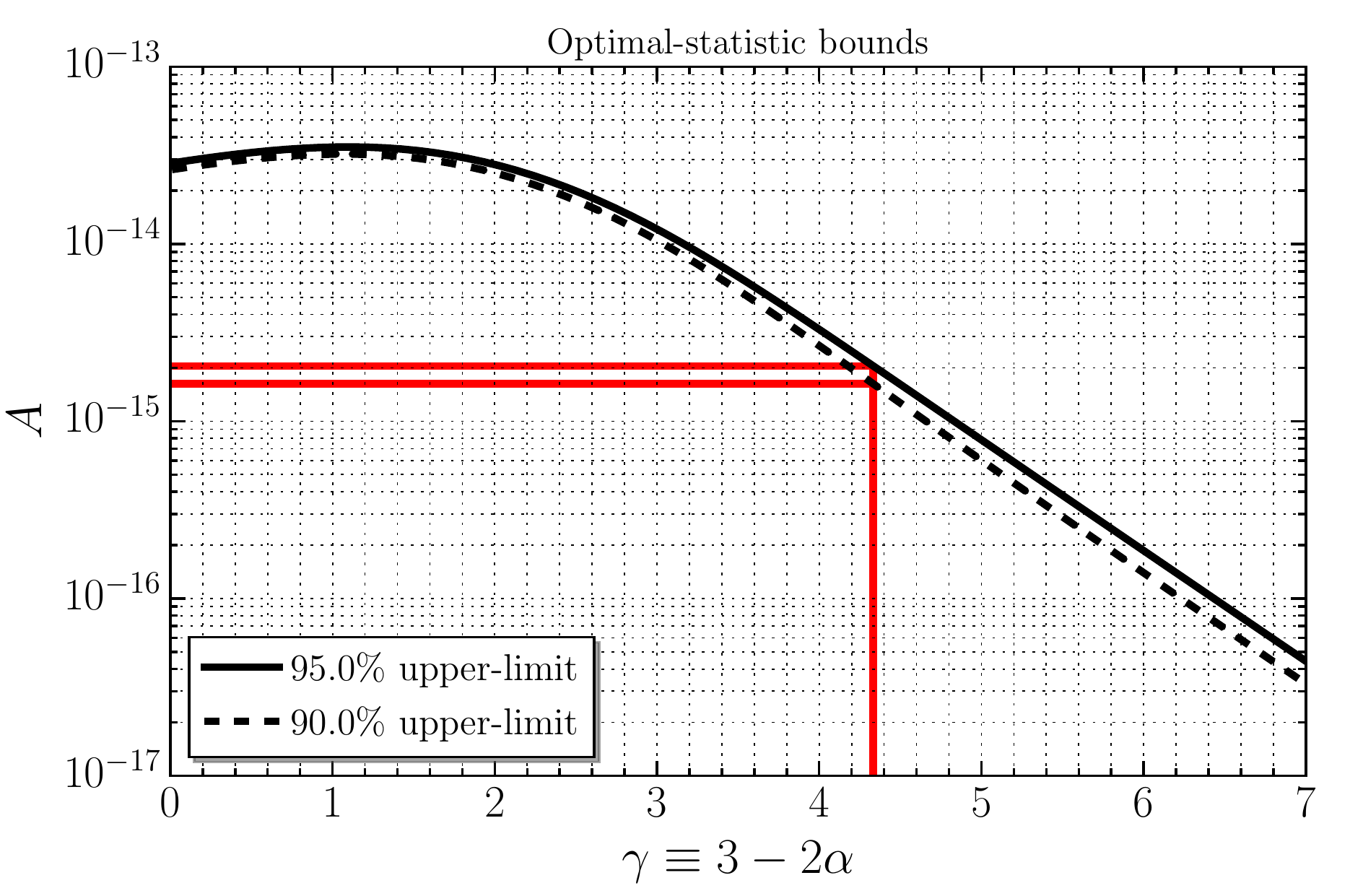}  &
\hspace{-0.5cm}
\includegraphics[width=90mm]{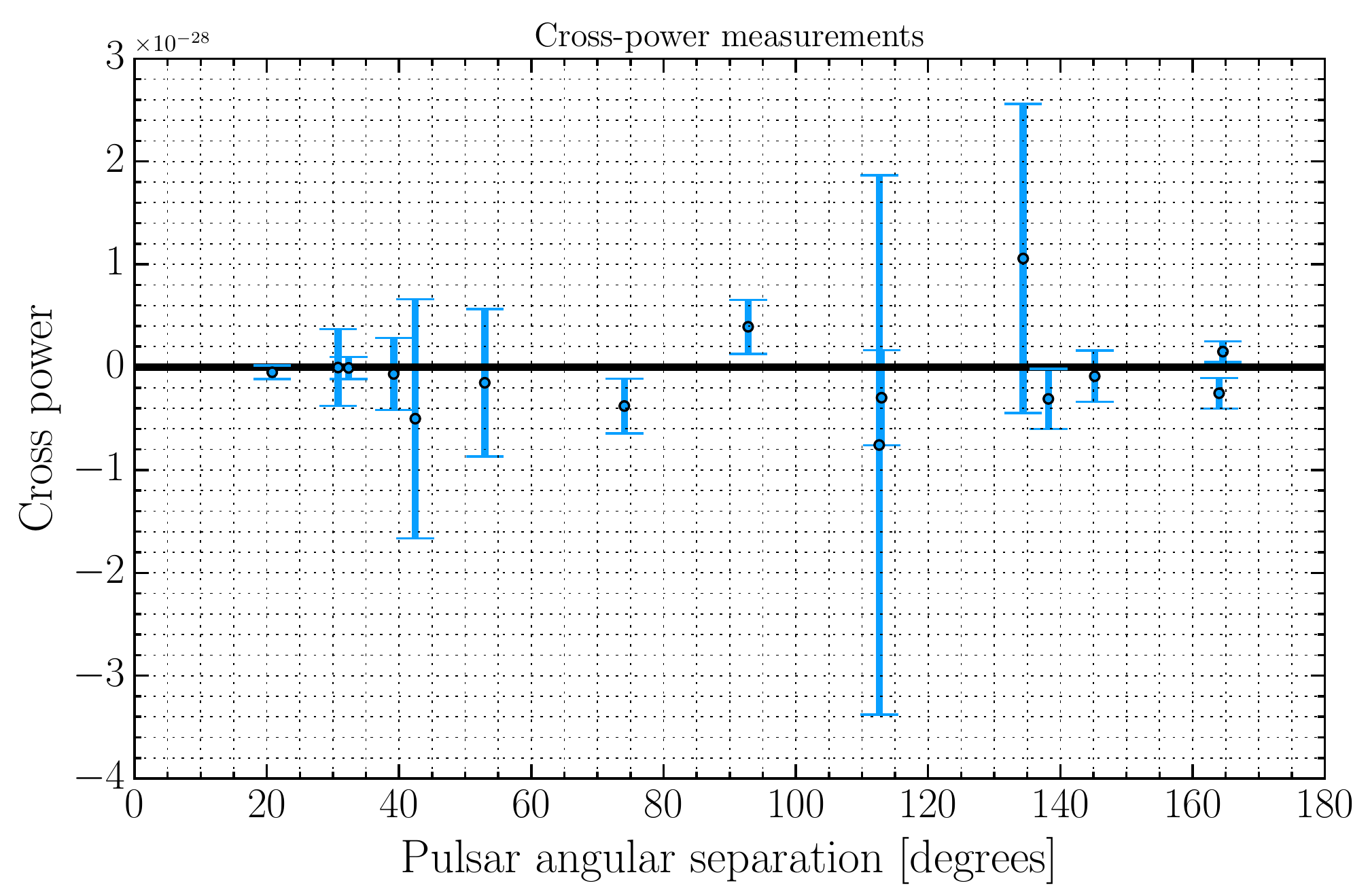}  \\
\end{array}$
\end{center}
\caption{(left) Frequentist upper-limits on the strain amplitude of the stochastic GWB obtained via the optimal-statistic for our 6 pulsar dataset. Red lines indicate $90\%$ and $95\%$ upper-limits at the fiducial slope of the strain-spectrum of $-2/3$, which corresponds to a slope of the residual PSD of $-13/3$. (right) The individual cross-power values are shown for our 6 pulsar dataset. All values are consistent with zero cross-correlation.}
\label{figure:OptimalStatistic}
\end{figure*}

In Fig. \ref{figure:PowerSpectrum}, we show the 95\% upper limits from a power spectrum analysis that does not assume a power law model, but allows the power at each frequency included in the model to vary separately.  We perform this analysis separately both for a correlated GWB (red points), and uncorrelated common red noise process (blue points) while varying the intrinsic noise parameters for each pulsar, however we do not include any additional common terms. The upper limit obtained from the equivalent power law analysis of the GWB at a spectral index of $13/3$ of $3\times10^{-15}$ is overplotted as a straight line. Frequencies were included from $1/T$ up to $20/T$, with $T=17.66$~yr, beyond which we do not expect the data to provide significant constraints on a steep red noise process. We find that our limit at a spectral index of $13/3$ is most heavily constrained by the $3/T$ term, corresponding to  $f\approx5\times10^{-9}$Hz. This is likely a combination of the lack of multifrequency data in the early data for PSR J1713+0747 as shown in Fig. \ref{Fig:Dataplots}, which significantly impacts our ability to disentangle DM variations from frequency-independent red noise at the lowest frequencies, and that our PSR J1909$-$3744 dataset is only $\sim 9$~yr in length.  Despite these limitations, the long timing baseline of the EPTA dataset used in this work is still critical for placing limits on the lowest frequencies in our analysis. Both the correlated and uncorrelated power spectrum are completely consistent with one another at all frequencies.  The difference in the log Evidence between the correlated and uncorrelated models was $0.2\pm0.3$, indicating no support for the presence of a correlated signal in the dataset.

In Fig. \ref{figure:CommonGWBComp} we further assess the impact of including, or not, the Hellings-Downs correlation on the upper limit obtained in our power law model (ii) for the varying spectral index case.  We show the one-dimensional marginalised posterior for both the amplitude of the GWB, which includes the correlation between pulsars (red solid line), and for the amplitude of an uncorrelated common red noise power law process (blue dashed line).  We find the upper limits to be completely consistent with one another, and obtain a change in the log evidence of $-1.0 \pm 0.5$ for the GWB model over the uncorrelated common red noise model, indicating no strong support for either model in the data.
We confirm this result by obtaining constraints on the correlation between pulsars as a function of angular separation for a common power law model, where the amplitude and spectral index of the power law are free to vary, and we fit simultaneously for the intrinsic noise parameters for the individual pulsars.  We fit for the correlation using the two methods described in Section \ref{Section:Correlation}, using either the four lowest order Chebyshev polynomials, or fitting for the correlation coefficient directly.  In Fig. \ref{figure:Correlation} the red and blue lines represent the $68\%$ and $95\%$ confidence intervals for the correlation function when modelled by the lowest 4 Chebyshev polynomials, while the individual points are the mean correlation coefficient with $1\sigma$ uncertainty for each pulsar pair when fitting directly.  In both cases the correlation is consistent with effectively all possible values (the range [-1,1]).

\subsubsection{Frequentist approach}

Applying the optimal-statistic introduced in Section \ref{Section:Frequentist} to our reduced $6$ pulsar dataset, and testing for a strain-spectrum slope of $-2/3$, gives $\hat{A}^2 = (-2.86\pm 4.29)\times 10^{-30}$, with an associated ${\rm S/N} = -0.67$. This is clearly a non-detection, however the $95\%$ upper-limit on $A$ is $2.05\times 10^{-15}$, which is more constraining than the best published limit of \citet{2013Sci...342..334S}, and is consistent with the fixed noise Bayesian limit of $1.7\times10^{-15}$. The optimal-statistic limits as a function of slope are shown in the left panel of Fig.\ \ref{figure:OptimalStatistic}, with limits at the fiducial slope value marked in red.

The computed cross-power values for our 6 pulsar dataset are shown in the right panel of Fig.\ \ref{figure:OptimalStatistic}, and are all consistent with zero correlation, as expected.

\section{Discussion}
\label{Section:Discussion}
\subsection{Implications for SMBHB astrophysics}
\label{Section:implicationsSMBHB}
As discused previously, the most promising astrophysical source of GWs in the nHz regime relevant to PTA observations is a cosmological population of adiabatically in-spiralling SMBHBs. 

%

\begin{figure}
\includegraphics[width=8.6cm]{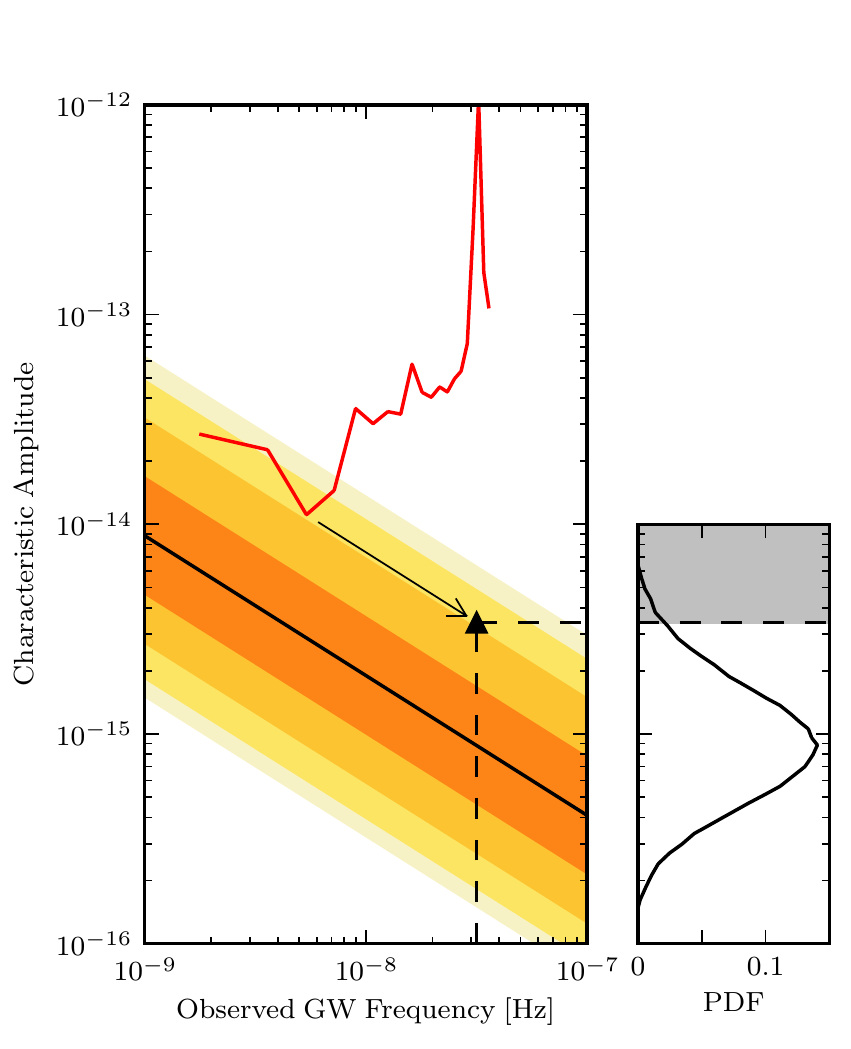}
\caption{Comparison between the expected GWB amplitude from a cosmological population of SMBHBs and the 95\% upper limit obtained with our PTA experiment.  Shaded areas represent the central 68\%, 95\%, 99.7\% and 100\% confidence interval of the predicted signal according to \protect\cite{2013MNRAS.433L...1S}, whereas the red curve is the 95\% upper limit presented in this paper, obtained by converting the unparametrised power spectrum shown in Fig. \ref{figure:PowerSpectrum} to characteristic amplitude. The black triangle is the extrapolated 95\% upper limit on $A$ at $f=1$yr$^{-1}$. The right panel shows the PDF of the predicted $h_c$ at $f=1$yr$^{-1}$, and the shaded area marks the region excluded at 95\% confidence by our limit (less than 5\% of the distribution).}
\label{Fig:astrolimit}
\end{figure}

In this section we will consider the implications of the upper-limit obtained in Section \ref{Section:Results} on the gravitational-wave signal for models of astrophysical populations of SMBHBs. If we assume that a SMBHB evolves purely due to gravitational radiation reaction, and that all SMBHBs are in circular orbits -- we will return to these assumptions at the end of the section -- the characteristic amplitude, Eq.~(\ref{hcA}), is given by:
\begin{equation}
h_c^2(f) =\frac{4f^{-4/3}}{3\pi^{1/3}}\int \int dzd{\cal M} \, \frac{d^2n}{dzd{\cal M}}{1\over{(1+z)^{1/3}}}{\cal M}^{5/3},
\label{hcirc}
\end{equation}
so that
\begin{equation}
A = \frac{2\,f_{\rm 1yr}^{-2/3}}{\sqrt{3}\pi^{1/6}}\,
\left[\int \int dzd{\cal M} \, \frac{d^2n}{dzd{\cal M}}{1\over{(1+z)^{1/3}}}{\cal M}^{5/3}\right]^{1/2}\,.
\end{equation}
A limit on the amplitude $A$, as given by Eq.~(\ref{hcirc}) therefore places constraints on $d^2n/(dzd{\cal M})$, i.e., the number density of SMBHB mergers per unit redshift and unit chirp mass across cosmic history. Although the dominant contribution to the signal comes from relatively massive (${\cal M}>10^{8}\msun$), low redshift ($z<2$) systems \citep{2008MNRAS.390..192S}, their merger rate is still poorly constrained, resulting in a fairly wide range of possible signal amplitudes. \cite{2008MNRAS.390..192S} exploited semi-analytical merger trees from \cite{2003ApJ...582..559V} to estimate a plausible range for the amplitude of a GWB of $5\times10^{-16}<A<3\times10^{-15}$. Merger rates extracted from cosmological simulations like the Millennium \citep{2005Natur.435..629S} and Massive Black \citep{2014arXiv1402.0888K} simulations, coupled with different prescriptions for the SMBH-galaxy relation results in a compatable range of $4\times10^{-16}<A<2\times10^{-15}$ \citep{2009MNRAS.394.2255S,2012ApJ...761...84R}. Recently, \cite{2013MNRAS.433L...1S} constrained the expected range of $A$ by building a set of phenomenological models based on the observed properties of interacting galaxies. Thousands of models fulfilling all relevant observational constraints were assembled by combining different estimates of the galaxy mass function, pair fractions, estimated merger times and galaxy-SMBH relations. The central 90\% of the probability distribution function (PDF) in the amplitude $A$ lies in the range $3\times10^{-16}<A<3\times10^{-15}$, as shown in the right panel of Fig. \ref{Fig:astrolimit}. A similar approach was employed by \cite{2014arXiv1406.5297R}, yielding consistent results.

In Fig. \ref{Fig:astrolimit} we compare the expected range in $h_c$ predicted by the phenomenological models presented in \cite{2013MNRAS.433L...1S} to the 95\% upper limit obtained in Section \ref{Section:sgwblimits}. Shaded areas represent the central 68\%, 95\%, 99.7\% and 100\% confidence intervals for the predicted signal. The red curve is derived by converting the 95\% limit on the unparametrised power spectrum shown in Fig. \ref{figure:PowerSpectrum} into $h_c$ as
\begin{equation}
  h_c=({\rm power}\times12\pi^2f^3T)^{1/2},
  \label{eq:powerhc}
\end{equation}
where $T$ is the total observation time. Eq.~(\ref{eq:powerhc}) can be calculated directly from Eq.~(\ref{Sh}) by noting that the power at each frequency is the integral of $S(f)$ over a frequency bin $\Delta{f}=1/T$. Fig. \ref{Fig:astrolimit} instructively shows how the limit on $A$ that is usually quoted in the literature is extrapolated from the actual sensitivity of the PTA. Our dataset is most sensitive at $f\approx5\times10^{-9}$Hz, where the 95\% limit on $h_c$ is $\sim 1.1\times10^{-14}$. This is then extrapolated to $f=1$yr$^{-1}$ \emph{assuming} a $f^{-2/3}$ power-law, to get a 95\% upper limit of $A=3.0\times10^{-15}$. The right panel of Fig. \ref{Fig:astrolimit} shows the probability distribution of $A$ inferred from the theoretical models of \cite{2013MNRAS.433L...1S}
together with the region excluded at 95\% confidence by our analysis. Only about 5\% of the distribution is excluded, meaning that our limit does not place severe restrictions on the cosmic SMBHB population. Note that $A$ values obtained from merger trees and cosmological simulations quoted above are generally  in the range $4\times10^{-16}<A<3\times 10^{-15}$, and therefore also consistent with this limit.

We caution that Fig. \ref{Fig:astrolimit} shows the expected range of the GW signal given circular GW driven SMBHBs, with negligible coupling to the environment. It has been shown \citep[see, e.g.][]{2007PThPh.117..241E,2011MNRAS.411.1467K,2013CQGra..30v4014S,2014ApJ...789..156M,2014MNRAS.442...56R} that both high eccentricities and strong environmental coupling might cause a significant suppression of the GW signal at $f<10^{-8}$Hz. If this is the case, our $1.1\times10^{-14}$ limit at $f\approx5\times10^{-9}$Hz cannot be extrapolated to $f=1$yr$^{-1}$ using a simple power-law model. Even a much more stringent limit obtained at such low frequencies might not have significant implications for the cosmic SMBHB merger rate, because the suppression at low frequencies  would invalidate the $f^{-2/3}$ extrapolation to higher frequencies \citep[see Fig. 2 in][]{2013CQGra..30v4014S}.
Since the detailed dynamical evolution of SMBHBs coupled with their environment is still poorly understood, current PTA limits do not allow us to formulate strong astrophysical statements about the cosmological population of SMBHBs.

\subsection{Limits on the cosmic (super)string tension}
\label{Section:implicationsStrings}

The limits computed in Section \ref{Section:sgwblimits} can be converted into upper limits on the linear energy density of a cosmic (super)string network, $\mu$, or tension in the Nambu-Goto approximation, usually described by the dimensionless quantity $G\mu/c^2$, where $G$ is Newton's constant and $c$ the speed of light. Field theory cosmic strings \citep{1976JPhA....9.1387K,2003PhRvD..68j3514J}, are one-dimensional topological defects, relics of an early, more symmetric state of the Universe, created through the mechanism of spontaneous symmetry breaking during the various phase transitions that the early Universe underwent. Their formation is a generic property of supersymmetric hybrid inflation scenarios \citep{2003PhRvD..68j3514J}, whereas the creation of their superstring theory counterparts, usually referred to as cosmic superstrings, are also a natural by-product of brane inflation scenarios \citep[e.g.][]{2002PhLB..536..185S,2003PhLB..563....6J}.

A cosmic string network consists of ``infinite'' (larger than the particle horizon) strings and cosmic string loops. A cosmic string network grows along the expansion of the Universe and is expected to settle in a scaling regime, were all the fundamental properties of the network grow proportionally with cosmic time, something achieved with the creation of loops; the main energy loss mechanism of the network. When two cosmic strings intersect, they ``intercommute'' (exchange partners) with a characteristic probability, and form loops. Cosmic string loops, once created, oscillate and decay emitting all of their energy in various forms of radiation, with the dominant form thought to be GWs \citep{1981PhLB..107...47V}. The stochastic GWB created by a cosmic string network is broadband (from \finetilde$10^{-16}$\,Hz, to higher than $10^{9}$\,Hz, depending on the size of the loops created), a characteristic feature of primordial GW sources, and is potentially detectable by any present or future GW detector. The cosmic string GW spectrum consists of a flat part at high frequencies, originating from loops decaying in the radiation era, and a broad peak at lower frequencies originating from loops decaying in the matter era. PTAs are in the privileged position to typically probe the GW emission originating from loops decaying either in the matter era or in the radiation-to-matter era transition, making them excellent instruments to detect the stochastic GWB of a cosmic string network. In the case of a non-detection, the upper limits on the GWB can be used to constrain the energy scale of cosmic strings.

There are numerous investigations on the cosmic string GWB in the literature \citep[e.g.][]{2005PhRvD..71f3510D,2010PhRvD..81j4028O, 2007PhRvD..75l5006D}, depending on various assumptions concerning the GW emission mechanisms and the computation of the loop number density. The cosmic string GWB spectra used in this paper are those computed in \cite{2013ApJ...764..108S}, which are based on an updated version of the modelling presented in \cite{2012PhRvD..85l2003}. The main difference of these two investigations is the inclusion of the effects of massive particle annihilation, which reduces the amplitude of the spectrum at higher frequencies. Both investigations, however, do not make any assumptions about the values of the fundamental model parameters used to compute the GW spectrum, granting characteristic robustness to the results presented here. \cite{2012PhRvD..85l2003} presented a generic way to compute the GWB based on the widely accepted one-scale model for cosmic strings. The basic parameters of this model are the string tension $G\mu/c^2$, the birth-scale of loops relative to the horizon, $\alpha_{\rm cs}$, and the intercommutation probability, $p$. Based on this modelling, and assuming that the cosmic string network maintains a scaling evolution along the expansion of the Universe, one can compute the loop number density for all the parameter combinations that create a GWB at the frequency regime probed by PTAs. Additionally, no assumption is made concerning the dominant GW emission mechanism from the strings (i.e., kinks or cusps), which is modelled using two additional parameters, a spectral index, $q$, and a cut-off, $n_*$, on the number of emission harmonics $n$.

\begin{figure}
\includegraphics[width=8.6cm]{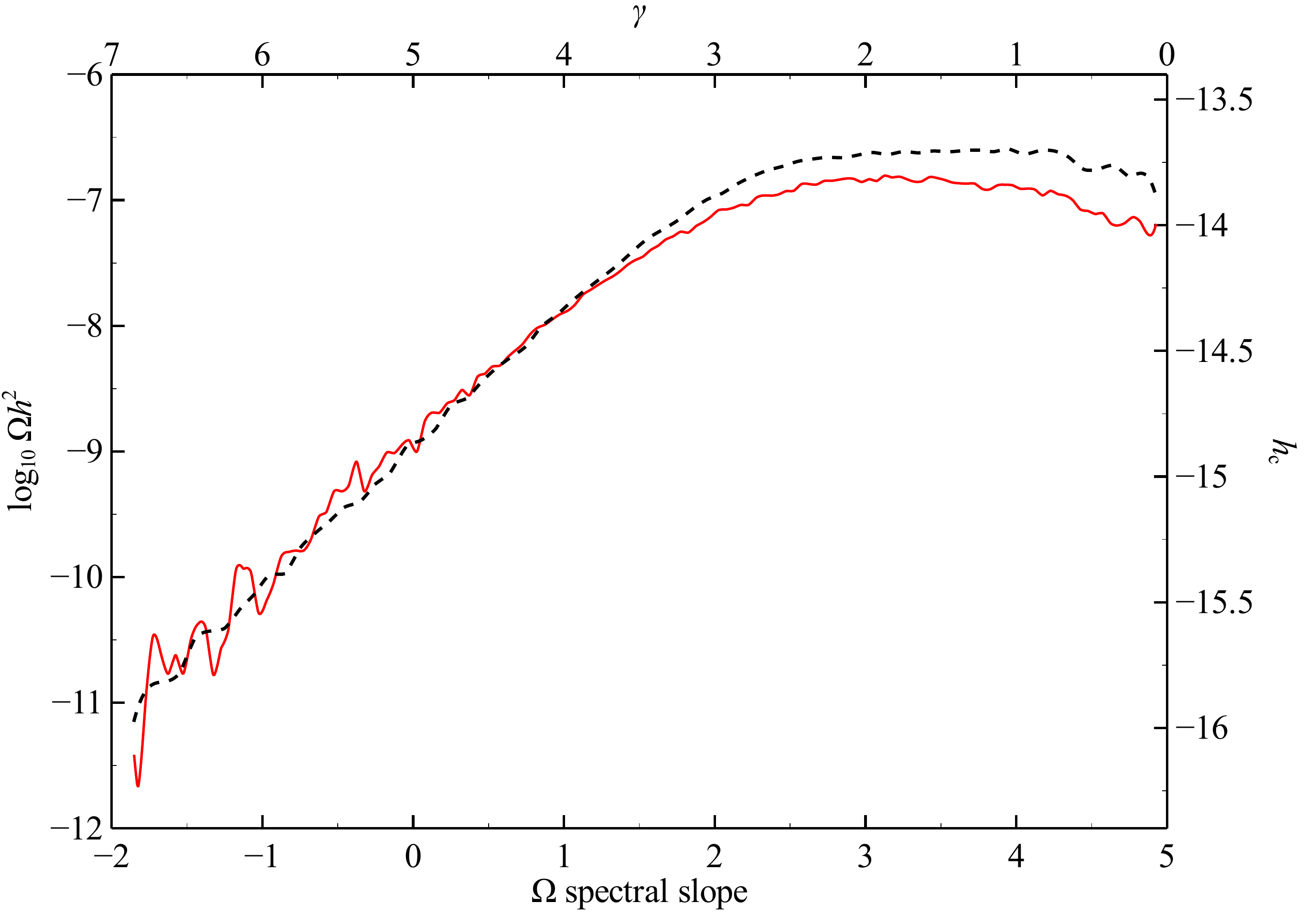}
\caption{The $95\%$ confidence upper limits on the amplitude of an isotropic, stochastic GWB as a function of the local spectral index at the frequency of $1\,{\rm yr}^{-1}$, imposed by models (ii) (dashed black) and (iii) (solid red) of Section \ref{Section:sgwblimits}. The limits are expressed in terms of the strain of the GWB, $h_{\rm c}$,  and the dimensionless spectral energy density of GWs, $\Omega h^2$, where $H_0=100h\,\rm{km\, s^{-1}\, Mpc^{-1}}$, as a function of their respective spectral indices.\label{Fig:gwblimits}}
\end{figure}
\begin{figure}
  \includegraphics[width=8.0cm]{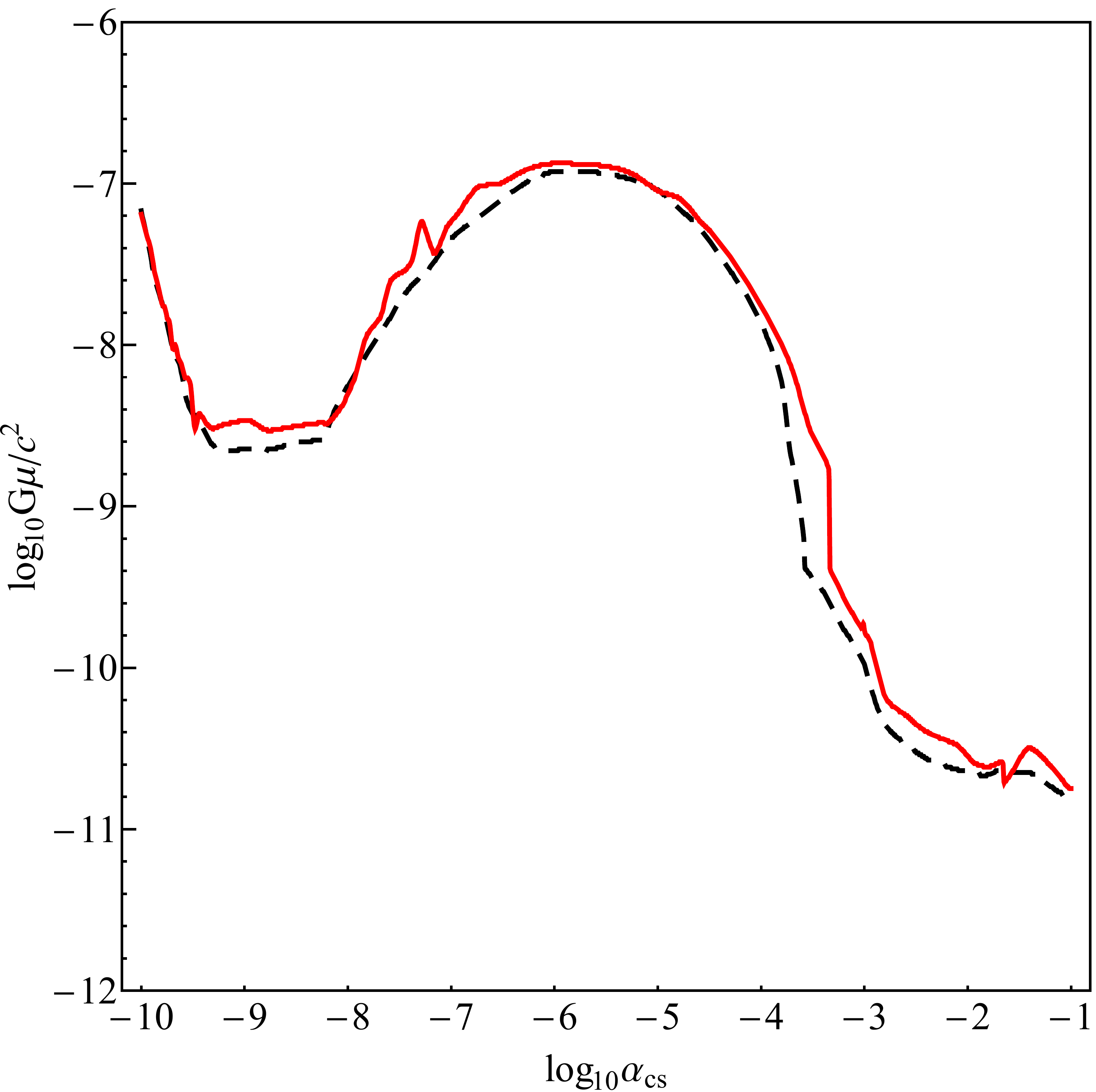}
  \caption{Exclusion limits for different cosmic string network configurations with the same $\Omega h^2$ value at a frequency $f=1\,{\rm yr}^{-1}$ in the $G\mu/c^2-\alpha_{\rm cs}$ parameter space, for $p=1$. Both the amplitude and spectral slope information of the GWB limits were used to construct the limits. The dashed black and solid red curves, are based on the results of models (ii) and (iii) presented in Fig.~\ref{Fig:gwblimits} respectively. In the mid-$\alpha_{\rm cs}$ region, the tension upper limits are provided by the $n_*=1$ networks, whereas in the regime of large and small loops, by the $n_*=10^4$ networks. \label{Fig:tension_limits}}
\end{figure}

In this section we present updated constraints on the cosmic string tension, for various values of the intercommutation probability, but independent of all the rest model parameters. The GWB upper limits we used are those obtained in the analysis for the models (ii) and (iii) of Section \ref{Section:sgwblimits}, where the spectral index $\gamma$ was a free parameter, as these are presented in Fig.~\ref{Fig:gwblimits}. Additionally, we used the GW sensitivity curve in Fig.~\ref{Fig:astrolimit} to investigate tension constraints across the probed frequency range. In the first case, we used the information of both the amplitude and the local spectral index of the GWB in order to create the tension exclusion curves in the cosmic string model parameter space, whereas in the second case we used only the amplitude information. In Fig.~\ref{Fig:tension_limits}, we present the cosmic string tension exclusion curves in the parameter space $G\mu/c^2-\alpha_{\rm cs}$ accessed by the PTAs, for field theory strings ($p=1$). These exclusion curves were constructed from the combination of the $n_*=1$ and $n_*=10^4$ exclusion curves, depending on which one provided the highest tension value. The limits are provided by the $n_*=1$ networks in the mid-$\alpha_{\rm cs}$ region ($-8\lesssim\log_{10}\alpha_{\rm cs}\lesssim -3$), whereas in the large ($\log_{10}\alpha_{\rm cs}\gtrsim-3$) and small ($-8\lesssim\log_{10}\alpha_{\rm cs}$) loop regimes, by the $n_*=10^4$  networks. As discussed in \cite{2012PhRvD..85l2003}, the $n_*=1$ and $n_*=10^4$ cases will always provide the largest tension values for fixed values of the rest cosmic string model parameters. The limits from the $n_*=1$ networks are independent of the GW emission mechanism since the power emitted per emission mode is $\propto n^{-q}$. For the limits provided by the $n_*=10^4$ networks, we used a spectral index $q=4/3$, corresponding to cusp emission, which always provides larger tension values than the $q=2$ case which corresponds to emission from kinks. The cosmic string tension upper limit for model (ii), where the intrinsic timing noise of each pulsar is allowed to vary, is
\begin{equation}
G\mu/c^2<1.2\times10^{-7} (95\%\,{\rm confidence}),
\end{equation}
whereas the tension upper limit for model (iii) where the common uncorrelated red noise, clock and Solar System ephemeris errors are included is
\begin{equation}
G\mu/c^2<1.3\times10^{-7} (95\%\,{\rm confidence}).
\end{equation}

For the particular case of large loop production with $\alpha=0.05$, as suggested by the most recent Nambu-Goto cosmic string evolution simulations \citep{2011PhRvD..83h3514B,2014PhRvD..89b3512B}, we get an upper limit $G\mu/c^2<3.0\times10^{-11}$. The tension constraint that was obtained based on that loop number density and the previous EPTA GWB limit was $G\mu/c^2<2.8\times10^{-9}$ \citep{2014PhRvD..89b3512B}. Assuming a $\finetilde 1.7$ times improvement on the value of the $\Omega_{\rm gw}$ used in that work, one would expect approximately a constraint $G\mu/c^2\lesssim 9.7\times10^{-10}$. This order of magnitude difference stems mostly from the normalization imposed on the produced number of loops. Whereas in our model we assume that all of the energy lost by the cosmic string network in order to attain scaling is channeled in loops of just one size, the loop production in \citep{2014PhRvD..89b3512B} takes place on a much wider range of large loops (wider range of $\alpha$). The difference in these two constraints is expected, since as we have demonstrated in \cite{2012PhRvD..85l2003}, where we also assumed a log-normal distribution for the loop birth scale, a wider range for the values of $\alpha$ has as an effect the lowering of the matter era peak GWB amplitude and its broadening. However, a direct comparison of these two results is not straightforward, not only because of the loop number density calculation differences, but also due to differences in computing the stochastic GWB.

Additionally, we computed the tension upper limits in the case of $p\ne 1$. In general, the intercommutation probability for cosmic superstrings can acquire values in the range $p\in[10^{-3},1]$, depending on their nature (F- or D- cosmic superstrings) \citep{2005JHEP...10..013J}. The effect of $p\ne 1$ is the increase of the amplitude of the stochastic GWB, without affecting the shape of the GW spectrum, $\Omega_{\rm gw}\propto p^{-k}$, where $k$ is the dependence of the scaling law that describes the effects of the intercommutation probability on the energy density of the infinite cosmic strings ($\rho_{\infty}\propto p^{-k}$). 

\begin{figure}
   \includegraphics[width=8.0cm]{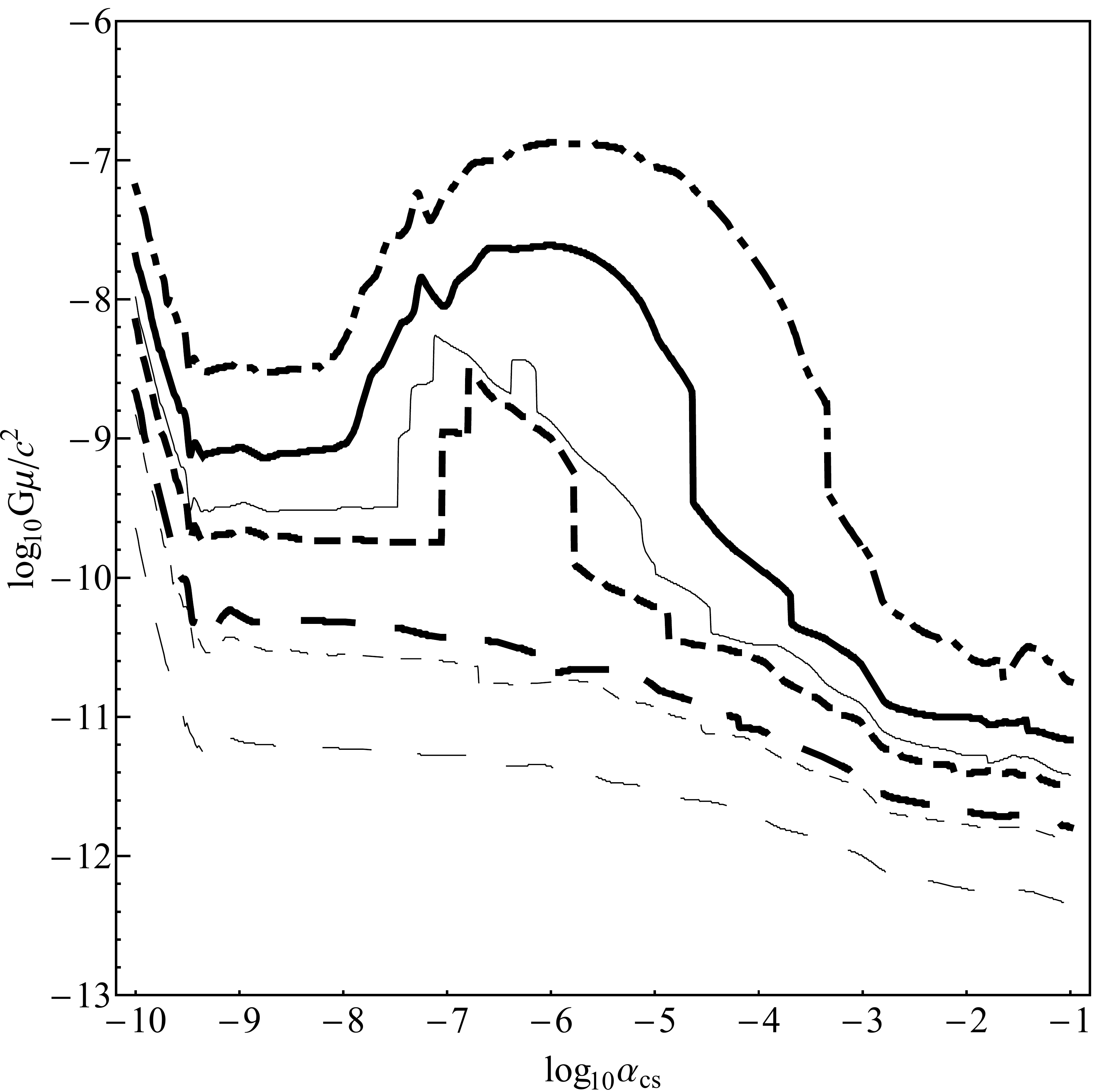}
   \caption{Exclusion curves for a set of network configurations with $p\ne 1$. With thick lines are the configurations with $k=0.6$ and with thin lines the configurations with $k=1$. We present exclusion curves for $p=0.1$ (solid lines), $p=10^{-2}$ (short dashed lines) and $p=10^{-3}$ (long dashed lines). The dot-dashed curve is the exclusion curve for $p=1$, for reference purposes. For all the results, we used the stochastic GWB limit of model (iii), placed at a frequency $f=1\,{\rm yr}^{-1}$. \label{Fig:pne1excurv}}
\end{figure}

There is no general consensus on the value of $k$, with different investigations suggesting a value of $k=0.6$ \citep{2005PhRvD..71l3513A} or $k=1$ \citep{2005JCAP...04..003S}. The reduced intercommutation probability, has as a result an increased number of intercommutations in order to maintain the scaling evolution of the network, and therefore, an increased number of loops which give GWBs of higher amplitude. In Table ~\ref{Table:TensionPLimits}, we present the tension upper limits for various cosmic superstring configurations, covering the whole possible range of $p$ and $k$ values. These upper limits, which are linked with small tension values, are provided by the networks with the smallest loop size accessible \citep[$\alpha_{\rm cs}\sim6\times10^{-11}$; see discussion in][]{2013ApJ...764..108S}. In Fig.~\ref{Fig:pne1excurv} we present a set of such exclusion curves to demonstrate the change in the shape of the exclusion curves as we probe lower tension values. The exclusion curves in the region $G\mu/c^2\lesssim10^{-10}$ are always provided by the $n_*=10^4$ networks. The apparent discontinuities in some of these exclusion curves are a combinatory result of the abrupt local changes in the GWB upper limit curve (more evident in the slope region [-2,0] for the $\Omega$ in Fig.~\ref{Fig:gwblimits}), and our requirement for a matching in the spectral slope of the GWB limit and the spectral slope of the cosmic string GW spectrum at a frequency\footnote{This has been verified by using smoother GWB sensitivity curves. The regions of the $G\mu/c^2-\alpha_{\rm cs}$ parameter space where there is also a significant change in the local spectral slope of the cosmic string GW spectrum, might also create such discontinuities, but such were not observed in the various tests we have conducted to investigate this effect. If one neglects the requirement for a matching local spectral index, no such artifacts are observed.} $f=1\,{\rm yr}^{-1}$.

\begin{table}
\centering
\caption{Upper limits on the cosmic superstring tension $G\mu/c^2$ for $p\ne 1$ and the two scaling laws proposed in the literature, using the GWB limits placed by models (ii) and (iii).}
\centering
\begin{tabular}{l c c c c}
\hline\hline
 & \multicolumn{2}{c}{Scenario ii}&\multicolumn{2}{c}{Scenario iii}\\
Model & \multicolumn{2}{c}{(varying spectral index,}&\multicolumn{2}{c}{(varying spectral index,}\\
 & \multicolumn{2}{c}{varying noise)}&\multicolumn{2}{c}{additional common noise)}\\
\hline\hline
Scaling law& k=0.6 & k=1 & k=0.6 & k=1\\
\hline\hline
$p=10^{-1}$ & $2.2\times10^{-8}$ & $1.1\times10^{-8}$ & $2.4\times10^{-8}$ & $1.0\times10^{-8}$ \\
$p=10^{-2}$ & $7.3\times10^{-9}$ & $1.6\times10^{-9}$ & $6.9\times10^{-9}$ & $1.5\times10^{-9}$ \\
$p=10^{-3}$ & $2.3\times10^{-9}$ & $2.8\times10^{-10}$ & $2.1\times10^{-9}$ & $2.2\times10^{-10}$ \\
\hline
\end{tabular}
\label{Table:TensionPLimits}
\end{table}

\begin{figure}
   \includegraphics[width=8.0cm]{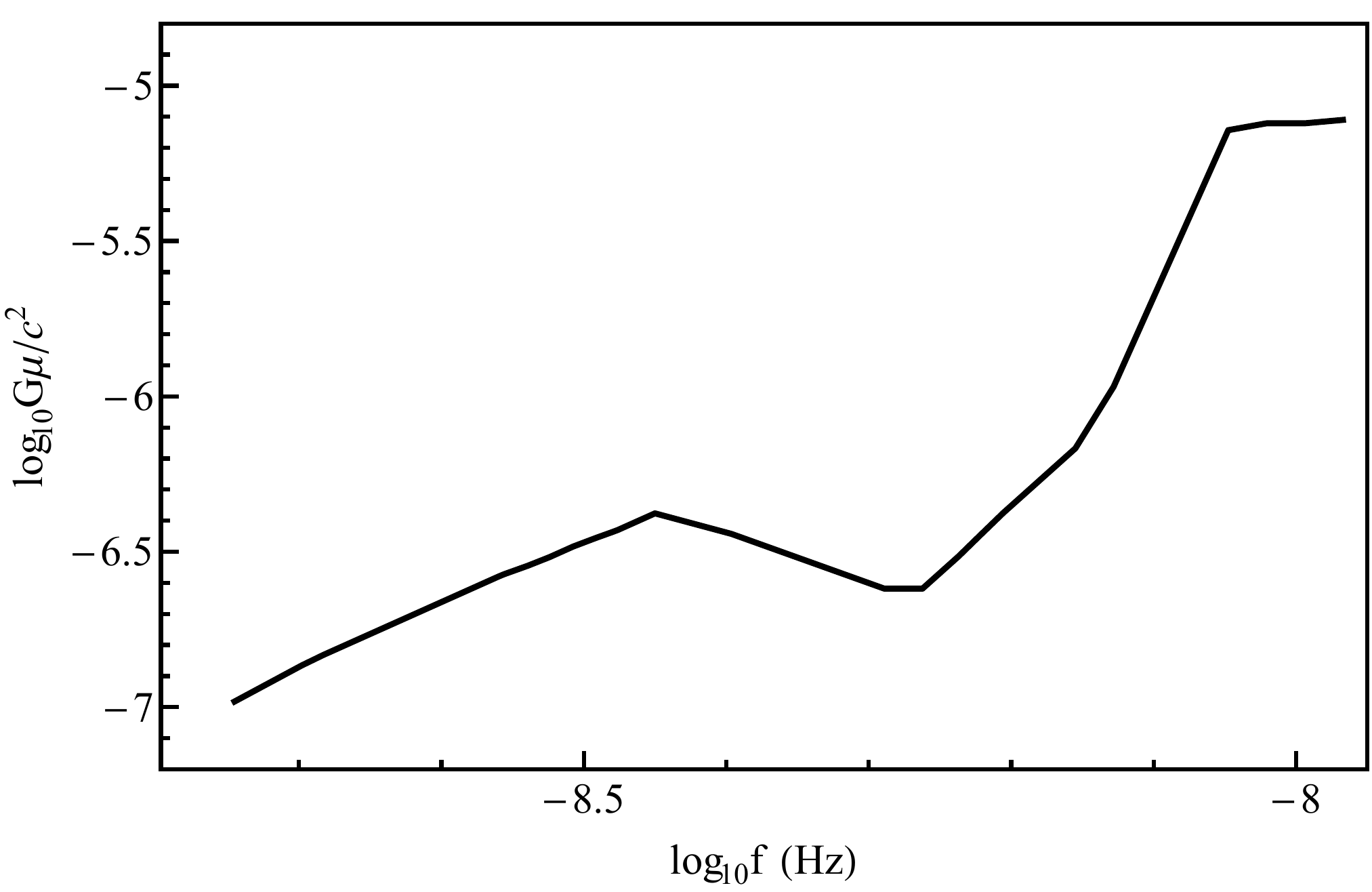}
   \caption{The cosmic string tension upper limits in the case of $p=1$, obtained from the sensitivity curve presented in Fig.~\ref{Fig:astrolimit} for a range of the probed frequencies. For these limits, only the information about the amplitude of the GWB has been used. The most stringent constraint is given by the lowest frequency probed, whereas results for frequencies higher than $\sim 10^{-8},\,{\rm Hz}$ have been omitted, since already for frequencies $\gtrsim7\times10^{-9}\,{\rm Hz}$ the tension upper limits are unphysically high.\label{Fig:tensioncs}}
\end{figure}

In Fig.~\ref{Fig:tensioncs}, we present the tension upper limits created from the sensitivity curve of Fig.~\ref{Fig:astrolimit} in the frequency range $f\in[1.7\times10^{-9},10^{-8}]$ in the case where $p=1$. We did not produce results for the whole frequency range of Fig.~\ref{Fig:astrolimit}, since already for frequencies $\gtrsim7\times 10^{-9}\, {\rm Hz}$ the tension upper limits are incompatible to the large scale structure of the Universe as we observe it (i.e., CMB anisotropies at large scales and galaxy distribution at small scales). For its computation, we used only the amplitude of the GWB per frequency bin, making those results less robust than those which also incorporate the information of the local GWB spectral index. The most stringent constraint is provided by the lowest frequency and is $G\mu/c^2<1.1\times10^{-7}$.

Our upper limit on the string tension for Nambu-Goto strings is slightly better than the one obtained in the extensive analysis performed by the {\it Planck} Collaboration \citep{2014A&A...571A..25P} using {\it Planck} data only ($G\mu/c^2<1.5\times10^{-7}$), and identical to that obtained when {\it Planck} data were combined with high-$\ell$ data from the Atacama Cosmology Telescope (ACT) and the South Pole Telescope (SPT). We note the significant improvement of the EPTA result in comparison to that presented in \cite{2012PhRvD..85l2003}, which was \finetilde3 times less constraining than the best available limit at that time, set by the WMAP 7-year+ACT results.

In terms of robustness, the CMB results are inherently more reliable than any GW-derived result because they depend only on the large scale properties of the cosmic string network (infinite strings), and not on the much more complicated details concerning the GW emission mechanism and the real cosmic string loop population. Our results, however, can be considered as quite reliable upper limits, for several reasons. First, the approach we used to compute the GW spectra does not make any assumption about the fundamental cosmic string model parameters used, and is only subject to a few fundamental assumptions, such as the validity of the one-scale model and the scale-invariant evolution of the cosmic string network. In \cite{2013PhRvD..88b3516S}, the authors presented results based on a possible delay of the onset of the scaling evolution as the network transverses from the radiation to the matter era evolution. Such a scenario leads to an increase in the amplitude of the resulting stochastic GWB, with a consequent strengthening of the upper limits, in comparison to our results where we assumed that the scaling evolution is always maintained and used a simple linear transition between the values of characteristic parameters of the cosmic string network (i.e., the number of infinite strings within our horizon) from their radiation era to their matter era values. Therefore, our upper limits remain robust even if such a possibility is true. Second, we have not included the GW emission from kinks on the infinite strings \citep{1990PhLB..251...28H,1990PhRvD..42..354S,1992PhRvD..45.1898A,2010PhRvD..81j3523K} or the emission of GWs due to the scaling evolution of the cosmic string network in the radiation era {\it per se} \citep{2013PhRvL.110j1302F}. These mechanisms, even though they contribute to the general string GWB, can be a few orders of magnitude smaller than the loop emission and omitting them from our calculations will not affect our upper limits. Note, however, that if the GWB originating from loops is not detectable by PTAs (i.e., in the case of very small loop creation by the network), the GWB created by the aforementioned mechanisms is the only one that can be detected is this frequency window.

\subsection{Relic Gravitational Waves}
\label{sec:relicGWs}
Quantum fluctuations of the gravitational field in the early Universe, amplified by an inflationary phase, are expected to produce a stochastic relic GWB \citep[see e.g.][]{g76, g77, s80, l82}. Observations of this {radiation} 
would provide a unique insight into poorly understood processes in the very-early Universe, at energy scales $\sim 10^{16}$~GeV and cosmic times $\sim 10^{-32}$~s \citep{bpa+15, aab+14}. At long wavelengths, gravitational-waves generated during an inflationary epoch produce 
a characteristic signature in the polarization of the CMB radiation, as well as CMB temperature anisotropies \citep{2005PhyU...48.1235G}. At shorter wavelengths, such as the PTA observational window, this radiation manifests itself as a contribution to the present day energy density spectrum $\Omega_{\mathrm{gw}}(f)$. The background spectrum is directly related to the primordial tensor spectral index $n_t$ and the equation of state of the early-Universe~\citep[see e.g.][]{z11, zzy+13}.

For standard single-field inflationary models, $n_t$ is related to the scalar-to-tensor ratio $r$ by $n_t = -r/8$, \citep{1993PhRvL..71..219C}. If we further \emph{assume} that at the end of inflation the Universe is not characterised by a `stiff' equation of state but enters a radiation dominated era, $w = 1/3$, the spectrum of the background can be written as in \cite{zzy+13}:
%

\begin{equation}
\Omega^{\mathrm{relic}}_{\mathrm{gw}}(f) \approx 1.3 \times 10^{-16} 10^{10 n_t}\left(\frac{r}{0.12}\right)\,\left(\frac{f}{\,\mathrm{yr}^{-1}}\right)^{n_t}\, ,
\label{e:Omega_infl}
\end{equation}
where it was implicitly assumed that $h=0.6711$.
Current results from \cite{bpa+15} set a limit of $r < 0.12$, and therefore the background spectrum is almost flat over a wide frequency range.  
By comparing the frequency dependency with our model we can therefore see that $\gamma \approx 5$.

We can now use the Bayesian analysis methods reported in Sec.~\ref{sec:bayesian} to perform an analysis where we vary the intrinsic noise parameters for the pulsars, and fix the spectral index of the correlated GWB term to $\gamma=5$, which implicitly assumes $n_t=0$, see e.g. \cite{zzy+13}. This yields a 95\% upper limit of $A<1.4\times 10^{-15}$ which translates to 
\begin{equation}
\label{eq:omega_lim}
\Omega^{\mathrm{relic}}_\mathrm{gw}(f)h^2 < 1.2 \times 10^{-9}\, ,
\end{equation}
a factor of 9 improvement from previously reported limits in \cite{2013ApJ...762...94D}, and a factor of 16 more constraining than \cite{2006ApJ...653.1571J}. Eq.~(\ref{e:Omega_infl}) can be inverted to yield a limit on the scalar-to-tensor ratio of $r<2.5\times 10^6$ (with $n_t=0$). While standard inflationary scenarios assume $n_t\leq0$, string-gas cosmology predicts a positive tilt in the primordial tensor spectrum (see e.g. \citet{PhysRevD.90.067301, 2007PhRvL..98w1302B}) which we cannot yet rule out. Future searches can set limits on $n_t$, providing the cosmology community independent measurements of this value, as well as independent constraints on $r$.

The limit on $\Omega^\mathrm{relic}_\mathrm{gw}(f)< 1.2 \times 10^{-9}$ is a factor $\sim 10^6$ from the predicted value reported in Eq.~(\ref{e:Omega_infl}), which may even be beyond the capabilities of a future PTA using the Square Kilometer Array \citep[SKA;][]{2015arXiv150100127J}, but is significantly more stringent than the indirect Big-Bang-Nucleosynthesis limit $\int \Omega_\mathrm{gw}(f)\mathrm{d}(\ln f) < 1.1\times 10^{-5}(N_\nu-3)$, where $N_\nu$ is the effective number of neutrino species at the time of Big-Bang-Nucleosynthesis, see e.g. \citet{1997rggr.conf..373A, 2000PhR...331..283M}. Current \cite{2015arXiv150201589P} limits place a limit on $N_\nu<3.7$. CMB experiments also set limits of a similar order of magnitude, $h^2\int \Omega_\mathrm{gw}(f)\mathrm{d}(\ln f) < 8.7\times 10^{-6}$, see \citet{2012PhRvD..85l3002S, 2006PhRvL..97b1301S}, whereas ground-based interferometers, which measure the relic GWB at specific frequency intervals, have recently been able to do better. Indeed, LIGO and Virgo reported new constraints in four different frequency bands, the most stringent being at $f=41.5-169.25$~Hz, where $\Omega_\mathrm{gw}(f)=5.6\times 10^{-6}$ at $95\%$ confidence, with an assumed $H_0=68$~km/s/Mpc, \citet{2014PhRvL.113w1101A}. To make a direct comparison to our result, we set $h=0.68$ in Eq.~(\ref{eq:omega_lim}) and find $\Omega_\mathrm{gw}(f) = 2.6\times10^{-9}$, over three orders of magnitude more constraining. 

We want to stress, however, 
that models such as those described in \cite{2005PhyU...48.1235G, PhysRevD.90.067301, 2007PhRvL..98w1302B} with values of $\gamma\in[4.6,5]$ with $n_t\in[0,0.9]$, may lead to much larger values $\Omega^\mathrm{relic}_\mathrm{gw}(f)h^2\sim 10^{-14}$, which may be within the reach of the SKA, see \cite{zzy+13} for further details.

\section{Conclusions}
\label{Section:Conclusions}

In this paper we have used a 6 pulsar subset of the recent EPTA data release presented in D15 to set a robust limit on the amplitude of a stochastic GWB using several models. When considering a power law model for the background we obtain a limit of $A=3.0\times10^{-15}$ at a spectral index of $\gamma=13/3$, consistent with a GWB dominated by SMBHBs, equivalent to $\Omega_\mathrm{gw}(f)h^2 = 1.1\times 10^{-9}$ at 2.8 nHz. When allowing the spectral index to vary freely over a prior range from $0\to7$,  $A=1.3\times10^{-14}$. This limit was obtained using a Bayesian analysis, in which we fit simultaneously for the intrinsic spin-noise and DM variation parameters for each pulsar, along with the GWB and additional common signals, including clock and Solar System ephemeris errors.  We stress that the simultaneous fit of the GWB signal with the individual pulsar noise parameters, and additional sources of common noise is crucial to obtain a robust limit. Fixing the intrinsic pulsar noise to the maximum likelihood values obtained by the single pulsar analysis and searching for a correlated signal {\it a posteriori} erroneously leads to an upper limit which is a factor of two more stringent. A series of simulations, and a parallel frequentist pipeline employing the optimal statistic yields consistent results, corroborating the robustness of our analysis. We also present a more general analysis, where we do not use a power law model for the background, but obtain limits on the correlated power spectrum at a series of discrete frequencies, and show that our sensitivity is greatest at a frequency of $\sim 5\times10^{-9}$~Hz.

In both cases we performed model selection using the Bayesian evidence for models that include a common red noise process that is either correlated between pulsars according to the isotropic overlap reduction function, or that is uncorrelated between the pulsars in the dataset. We obtained a difference in the logarithm of the Bayesian evidence of  $-1.0 \pm 0.5$ for the power law, and $0.2\pm0.3$ for the more general model, indicating that the dataset is not able to differentiate between these two cases.  We confirm this result by obtaining confidence intervals for the correlation coefficients between pulsars as a function of their angular separation on the sky and find them to be consistent both with zero correlation, and the Hellings-Downs curve.

Finally, we discussed the implications of our analysis on the astrophysics of SMBHBs, and derived upper limits on the string tension of a cosmic (super)string network and for a relic GWB. Our upper limit of $A=3.0\times10^{-15}$ at a spectral index of $\gamma=13/3$ skims the region of the expected GWB predicted by recent astrophysical models, but is still too high to place stringent constraints on the cosmological SMBHB population. An improvement of a factor 2-3 would place our sensitivity at the heart of the expected signal range for the included models, placing considerable constraints on possible populations of merging supermassive black holes. In the case of a Nambu-Goto field theory cosmic string network, the upper limit on the string tension was evaluated to be $G\mu/c^2<1.3\times10^{-7}$, identical to the best so far result from CMB investigations; the result presented by the {\it Planck} Collaboration combining data from {\it Planck}, SPT and ACT. {\it Planck} has managed to measure the temperature anisotropies of the CMB to an unprecedented detail, and it is expected that the string tension limits will not be improved significantly in the future unless there is an inclusion of CMB polarisation data. On the other hand, the constraints from PTAs will continue to improve significantly as longer and more precise datasets are obtained, since $\mu\propto\Omega^{1/2}$. Therefore, the PTA constraints on the string tension, as long as there is a careful treatment of all the involved uncertainties, have the potential to be the most stringent in the coming years, until full-sky CMB polarisation instruments become a reality (i.e., COrE+\footnote{https://hangar.iasfbo.inaf.it/core/}). Our limit on the relic GWB, $\Omega^\mathrm{relic}_\mathrm{gw}(f)h^2=1.2 \times10^{-9}$, is a factor of 9 more constraining than the previous NANOGrav limit reported in \cite{2013ApJ...762...94D}, and 16 times more constraining than the last PPTA limit, see \cite{2006ApJ...653.1571J}. Although the expected level of the relic GW energy density is $\Omega^\mathrm{relic}_\mathrm{gw}(f)h^2\sim 10^{-15}$ in the PTA band, this number may increase by an order of magnitude for models with non-flat primordial spectra (i.e. with nonzero tensor indices, $n_t$), described in \cite{2005PhyU...48.1235G}. Such models describe a relic GWB which may just be within the grasp of the SKA according to studies by \cite{zzy+13}.

\section{Acknowledgements}

The European Pulsar Timing Array (EPTA) is a collaboration between European institutes namely ASTRON (NL), INAF/Osservatorio di Cagliari (IT), Max-Planck-Institut für Radioastronomie (GER), Nançay/Paris Observatory (FRA), the University of Manchester (UK), the University of Birmingham (UK), the University of Cambridge (UK) and the University of Bielefeld (GER), with the aim to provide high precision pulsar timing to work towards the direct detection of low-frequency gravitational waves. An Advanced Grant of the European Research Council to implement the Large European Array for Pulsars (LEAP) also provides funding.

Part of this work is based on observations with the 100-m telescope of the Max-Planck-Institut f{\"u}r Radioastronomie (MPIfR) at Effelsberg. The Nan{\c c}ay radio Observatory is operated by the Paris Observatory, associated to the French Centre National de la Recherche Scientifique (CNRS). We acknowledge financial support from 'Programme National de Cosmologie and Galaxies' (PNCG) of CNRS/INSU, France.   Pulsar research at the Jodrell Bank Centre for Astrophysics and the observations using the Lovell Telescope is supported by a consolidated grant from the STFC in the UK. We thank A. G. Lyne and C. A. Jordan for carrying out the pulsar observations at JBCA. The Westerbork Synthesis Radio Telescope is operated by the Netherlands Institute for Radio Astronomy (ASTRON) with support from The Netherlands Foundation for Scientific Research NWO. 

This research was performed using the Darwin Supercomputer of the University of Cambridge High Performance Computing Service (http://www.hpc.cam.ac.uk/), provided by Dell Inc. using Strategic Research Infrastructure Funding from the Higher Education Funding Council for England and funding from the Science and Technology Facilities Council, and the Zwicky computer cluster at Caltech is supported by the NSF under MRI-R2 award no. PHY-0960291 and by the Sherman Fairchild Foundation.

LL was supported by a Junior Research Fellowship at Trinity Hall College, Cambridge University.  ST was supported by appointment to the NASA Postdoctoral Program at the Jet Propulsion Laboratory, administered by Oak Ridge Associated Universities through a contract with NASA. CMFM was supported by a Marie Curie International Outgoing Fellowship within the 7th European Community Framework Programme and would like to thank Paul Lasky for useful discussions.  AS and JG are supported by the Royal Society.  SAS would like to thank Richard Battye for various discussions and comments regarding the cosmic string section in this paper, and acknowledges funding from an NWO Vidi fellowship (PI JWTH).  RNC acknowledges the support of the International Max Planck Research School Bonn/Cologne and the Bonn-Cologne Graduate School.  KJL is supported by the National Natural Science Foundation of China (Grant No.11373011).  RvH is supported by NASA Einstein Fellowship grant PF3-140116.  JWTH acknowledges funding from an NWO Vidi fellowship and ERC Starting Grant `DRAGNET' (337062).   PL acknowledges the support of the International Max Planck Research School Bonn/Cologne. KL acknowledges the financial support by the European Research Council for the ERC Synergy Grant BlackHoleCam under contract no. 610058. SO is supported by the Alexander von Humboldt Foundation.




\bibliographystyle{mn2e}
\bibliography{references}

 \end{document}